%% file: lumo-phd.tex
\documentclass{ruthesis-smallmargins}
\usepackage{bbm,epsfig}
\newfont{\frak}{eufm10 scaled 1200}
\newcommand{\mfrak}[1]{\mbox{\frak #1}}
\newfont{\Bbb}{msbm10 scaled 1200}     
\DeclareSymbolFont{AMSa}{U}{msa}{m}{n}
\DeclareSymbolFont{AMSb}{U}{msb}{m}{n}
\let\Box\relax
\DeclareMathSymbol{\Box}{\mathord}{AMSa}{"03}

\begin{document}


    \include{title}
    \afterpreface 
    \include{chap1-intro}

    \include{chap2-screw}
    \include{chap3-hetm}
    \include{chap4-knit}

    \include{chap5-nonsusy}

    \include{chap6-bfm}

    \include{chap7-bfmhetero}


\input{references}

\include{vita}

\end{document}

%% file: title.tex

    \phd
    \title{\bf Nonperturbative Formulations\\
of Superstring Theory} 

\author{Lubo\v{s} Motl}
\campus{New Brunswick} \program{Physics and Astronomy}
\director{Thomas Banks} \approvals{5}
    \submissionmonth{October}
    \submissionyear{2001}
    \figurespage
    \tablespage
\input{mymacros.tex}

    \abstract{\input{abstract}}  
\beforepreface 

   \acknowledgements{\input{thanks}}
    \dedication{
For the victims of 20$^{\mbox{\small th}}$ century
totalitarian regimes, those who have been
executed, jailed or dismissed from schools by fascists and
communists.
%
}

%% file: mymacros.tex
\newcommand{\href}[2]{\bgroup #2\egroup}

\def\@spires#1{\href{http://www-spires.slac.stanford.edu/spires/find/hep/www?j=#1}}
\catcode`\%=12
\newcommand\adp[3]   {\@spires{ADPHA
                {{\it Adv.\ Phys.\ }{\bf #1} (#2) #3}}
\newcommand\ap[3]    {\@spires{APNYA
                {{\it Ann.\ Phys.\ (NY) }{\bf #1} (#2) #3}}
\newcommand\app[3]   {\@spires{APHYE
                {{\it Astropart.\ Phys.\ }{\bf #1} (#2) #3}}
\newcommand\appol[3] {\@spires{APPOL
                {{\it Acta Phys.\ Polon.\ }{\bf #1} (#2) #3}}
\newcommand\arnps[3] {\@spires{ARNUA
                {{\it Ann.\ Rev.\ Nucl.\ Part.\ Sci.\ }{\bf #1} (#2) #3}}
\newcommand\atmp[3] {\@spires{00203
                {{\it Adv.\ Theor.\ Math.\ Phys.\ }{\bf #1} (#2) #3}}
\newcommand\cpc[3]   {\@spires{CPHCB
                {{\it Comput.\ Phys.\ Commun.\ }{\bf #1} (#2) #3}}
\newcommand\cmp[3]   {\@spires{CMPHA
                {{\it Comm.\ Math.\ Phys.\ }{\bf #1} (#2) #3}}
\newcommand\dmj[3]   {\@spires{DUMJA
                {{\it Duke Math.\ J. }{\bf #1} (#2) #3}}
\newcommand\epjc[3]  {\@spires{EPHJA
                {{\it Eur.\ Phys.\ J. }{\bf C #1} (#2) #3}}
\newcommand\jmp[3]   {\@spires{JMAPA
                {{\it J.\ Math.\ Phys.\ }{\bf #1} (#2) #3}}
\newcommand\jgp[3]   {\@spires{JGPHE
                {{\it J.\ Geom.\ Phys.\ }{\bf #1} (#2) #3}}
\newcommand\jphg[3]   {\@spires{JPAGB
                {{\it J. Phys.\ }{\bf G #1} (#2) #3}}
\newcommand\cqg[3]   {\@spires{CQGRD
                {{\it Class.\ and Quant.\ Grav.\ }{\bf #1} (#2) #3}}
\newcommand\hpa[3]   {\@spires{HPACA
                {{\it Helv.\ Phys.\ Acta }{\bf #1} (#2) #3}}
\newcommand\jhep[3]
{\href{http://jhep.sissa.it/stdsearch?paper=#1%28#2%29#3}
                {{\it J. High Energy Phys.\ }{\bf #1} (#2) #3}}
\newcommand\lmp[3]   {\@spires{LMPHD
                {{\it Lett.\ Math.\ Phys.\ }{\bf #1} (#2) #3}}
\newcommand\npa[3]   {\@spires{NUPHA
                {{\it Nucl.\ Phys.\ }{\bf A #1} (#2) #3}}
\newcommand\npb[3]    {\@spires{NUPHA
                {{\it Nucl.\ Phys.\ }{\bf B #1} (#2) #3}}
\newcommand\npps[3]  {\@spires{NUPHZ
                {{\it Nucl.\ Phys.\ }{\bf #1} {\it(Proc.\ Suppl.)} (#2)
#3}}
\newcommand\pla[3]   {\@spires{PHLTA
                {{\it Phys.\ Lett.\ }{\bf A #1} (#2) #3}}
\newcommand\plb[3]   {\@spires{PHLTA
                {{\it Phys.\ Lett.\ }{\bf B #1} (#2) #3}}
\newcommand\ppnp[3]  {\@spires{PPNPD
                {{\it Prog.\ Part.\ Nucl.\ Phys.\ }{\bf #1} (#2) #3}}
\newcommand\pr[3]    {\@spires{PHRVA
                {{\it Phys.\ Rev.\ }{\bf #1} (#2) #3}}
\newcommand\pra[3]   {\@spires{PHRVA
                {{\it Phys.\ Rev.\ }{\bf A #1} (#2) #3}}
\newcommand\prb[3]   {\@spires{PHRVA
                {{\it Phys.\ Rev.\ }{\bf B #1} (#2) #3}}
\newcommand\prc[3]   {\@spires{PHRVA
                {{\it Phys.\ Rev.\ }{\bf C #1} (#2) #3}}
\newcommand\prd[3]   {\@spires{PHRVA
                {{\it Phys.\ Rev.\ }{\bf D #1} (#2) #3}}
\newcommand\pre[3]   {\@spires{PHRVA
                {{\it Phys.\ Rev.\ }{\bf E #1} (#2) #3}}
\newcommand\prep[3]  {\@spires{PRPLC
                {{\it Phys.\ Rep.\ }{\bf #1} (#2) #3}}
\newcommand\prl[3]   {\@spires{PRLTA
                {{\it Phys.\ Rev.\ Lett.\ }{\bf #1} (#2) #3}}
\newcommand\ptp[3]   {\@spires{PTPKA
                {{\it Prog.\ Theor.\ Phys.\ }{\bf #1} (#2) #3}}
\newcommand\rmp[3]   {\@spires{RMPHA
                {{\it Rev.\ Mod.\ Phys.\ }{\bf #1} (#2) #3}}
\newcommand\zpc[3]   {\@spires{ZEPYA
                {{\it Z.\ Physik }{\bf C #1} (#2) #3}}
\newcommand\mpla[3]  {\@spires{MPLAE
                {{\it Mod.\ Phys.\ Lett.\ }{\bf A #1} (#2) #3}}
\newcommand\mplb[3]  {\@spires{MPLAE
                {{\it Mod.\ Phys.\ Lett.\ }{\bf B #1} (#2) #3}}
\newcommand\sjnp[3]  {\@spires{SJNCA
                {{\it Sov.\ J.\ Nucl.\ Phys.\ }{\bf #1} (#2) #3}}
\newcommand\jetp[3]  {\@spires{SPHJA
                {{\it Sov.\ Phys.\ JETP\/ }{\bf #1} (#2) #3}}
\newcommand\jetpl[3]  {\@spires{JTPLA
                {{\it Sov.\ Phys.\ JETP Lett.\ }{\bf #1} (#2) #3}}
\newcommand\zetf[3]  {\@spires{ZETFA
                {{\it Zh.\ Eksp.\ Teor.\ Fiz.\ }{\bf #1} (#2) #3}}
\newcommand\yf[3]    {\@spires{YAFIA
                {{\it Yad.\ Fiz.\ }{\bf #1} (#2) #3}}
\newcommand\nc[3]    {\@spires{NUCIA
                {{\it Nuovo Cim.\ }{\bf #1} (#2) #3}}
\newcommand\joth[3]  {\@spires{JOTHE
                {{\it J.\ Operator Theory }{\bf #1} (#2) #3}}
\newcommand\ibid[3]{{\it ibid.\ }{\bf #1} (#2) #3}
\newcommand\ijmpa[3] {\@spires{IMPAE
                {{\it Int.\ J.\ Mod.\ Phys.\ }{\bf A #1} (#2) #3}}
\newcommand\ijmpb[3] {\@spires{IMPAE
                {{\it Int.\ J.\ Mod.\ Phys.\ }{\bf B #1} (#2) #3}}
\catcode`\%=14
\catcode`\|=12
\newcommand{\hepth}[1]{\href{http://xxx.lanl.gov/abs/hep-th/#1}{\tt hep-th/#1}}
\newcommand{\hepph}[1]{\href{http://xxx.lanl.gov/abs/hep-ph/#1}{\tt hep-ph/#1}}
\newcommand{\heplat}[1]{\href{http://xxx.lanl.gov/abs/hep-lat/#1}{\tt hep-lat/#1}}
\newcommand{\hepex}[1]{\href{http://xxx.lanl.gov/abs/hep-ex/#1}{\tt hep-ex/#1}} 
\newcommand{\nuclth}[1]{\href{http://xxx.lanl.gov/abs/nucl-th/#1}{\tt nucl-th/#1}}
\newcommand{\nuclex}[1]{\href{http://xxx.lanl.gov/abs/nucl-ex/#1}{\tt nucl-ex/#1}}
\newcommand{\grqc}[1]{\href{http://xxx.lanl.gov/abs/gr-qc/#1}{\tt gr-qc/#1}}

\def\IZ{{\mathbbm Z}}
\def\IR{{\mathbbm R}}
\def\IC{{\mathbbm C}}
\def\IQ{{\mathbbm Q}}

\def\ignorethis#1{}
\def\be{\begin{equation}}
\def\ee{\end{equation}}
\def\bear{\begin{eqnarray}}
\def\eear{\end{eqnarray}}
\def\nn{\nonumber}

\def\tilde{\widetilde}
\def \tb#1{\left(\begin{array}#1\end{array}\right)}
\def \abs#1{\left|#1\right|}
\def \bra#1{\left\langle #1\right\vert}
\def \ket#1{\left\vert #1\right\rangle}
\def \dslash{{\not\!\partial}}
\def \exp{\mbox{exp}}
\def \tp{\mbox{tp\,}}
\def \ignoruj#1{}
\def \eqn#1#2{\begin{equation}#2\label{#1}\end{equation}}
\def \eq#1{\begin{equation}#1\end{equation}}
\def \rut{2/5 transformation}
\def \Rut{{\mathbbm G}}             
\def \lpl{L_{planck}}
\def \lst{L_{string}}
\def \ha{{1\over 2}}
\def \asd{anti-self-dual }
\def \asdW{anti-self-dual}
\def \Tr{\mbox{Tr\,}}
\def \PHI{\Phi}
\def \THETA{\Theta}
\def \ham{\mathcal H}
\def \Lag{\mathcal L}
\def \action{\mathcal A}

\def\lm#1{{\bf (!!! #1 !!!)}}  
\def\og#1{{\bf (((({#1}))))}}  
\def\wt{\widetilde}
\def\CHECK{{\em -WE HAVE TO CHECK THIS!!! \vskip .5cm}}

\newcommand\px[1]{{\partial_{#1}}}
\newcommand\qx[1]{{\partial^{#1}}}
\newcommand\ppx[1]{{{\partial\over{\partial {#1}}}}}
\newcommand\pypx[2]{{{{\partial {#1}}\over{\partial {#2}}}}}
\newcommand\lylx[2]{{{\delta {#1}} \over {\delta {#2}}}} 
\newcommand\llylxx[3]{{{\delta^2 {#1}} \over {\delta {#2} \delta {#3}}}}

\newcommand\inv[1]{{1\over{#1}}}

\newcommand\rep[1]{{\underline{\bf {#1}}}}      
\newcommand\tr[1]{{\mbox{tr}\{{#1}\}}}          
\newcommand\trr[2]{{{\mbox{tr}}_{#1}\{{#2}\}}}  
\newcommand\ttr[1]{{\mbox{Tr}\{{#1}\}}}         
\newcommand\ev[1]{{\bra {#1} \ket}}             
\newcommand\evtr[1]{{\bra \tr{{#1}} \ket}}      
\newcommand\com[2]{{\lbrack {#1},{#2}\rbrack}}  
\newcommand\acom[2]{{\{ {#1},{#2} \}}}          
\newcommand\cov[1]{{\nabla_{#1}}}               

\def\lbr{{\lbrack}}
\def\rbr{{\rbrack}}
\def\dg{{\dagger}}

\def\wdg{{\wedge}}                              

\newcommand\CD[1]{{\lbr{\cal D} {#1}\rbr}}      

\def\PO{{\cal P}}                               
\def\Id{{\bf I}}                                

\def\suv{{\sigma_{\mu\nu}}}                     

\newcommand\sqov[2]{{\sqrt{{{#1}\over {#2}}} }}
\def\sqh{{\inv{\sqrt{2}}}}                      

\def\Imx{{\mbox{Im}}}                           
\def\Rex{{\mbox{Re}}}                           
\def\Kah{{K\"ahler}}                            


\def\diag{\mathop{\rm diag}}
\def\mth{M-theory}
\def\calr{${\cal R\,}$}
\def\U{U}
\def\SU{SU}

\def\hacek{\accent20}                           
\def\a{{\alpha}}
\def\b{{\beta}}
\def\g{{\gamma}}
\def\d{{\delta}}
\def\del{{\delta}}
\def\e{{\epsilon}} 
\def\lam{{\lambda}}
\def\u{{\mu}}
\def\r{{\rho}}
\def\s{{\sigma}}
\def\t{{\tau}}
\def\h{{\eta}}
\def\x{{\xi}}
\def\z{{\zeta}}

\def\TH{\Theta}
\def\PH{\Phi}
\def\Ga{{\Gamma}}
\def\GPR{{\widetilde{\Gamma}}}

\def\eb{{\bar{\epsilon}}}
\def\bpsi{{\bar{\psi}}}
\def\bTH{{\bar{\Theta}}}


\newcommand\epd[1]{{\epsilon_{#1}}}              
\newcommand\epu[1]{{\epsilon^{#1}}}              
\newcommand\ept[2]{{{\epsilon_{#1}}^{#2}}}       


\def\btheta{{\bar{\theta}}}
\def\bD{{\bar{D}}}                   

\def\sc{{\widetilde{\varphi}}}                  
\def\bsc{{\overline{\widetilde{\varphi}}}}      

\def\CHP{{\Phi}}                     
\def\bCHP{{\overline{\Phi}}}         

\def\vph{{\varphi}}                  

\def\VEV{{\Phi_0}}                   
\def\scVEV{{\varphi_0}}              
\def\Ar{{A}}                         

\def\sig{{{\sigma}}}                 
\def\bsig{{\bar{\sigma}}}            

\def\MR#1{{{\IR}^{#1}}}               
\def\MC#1{{{\IC}^{#1}}}               
\def\MS#1{{{\bf S}^{#1}}}             
\def\MB#1{{{\bf B}^{#1}}}             
\def\MT#1{{{\bf T}^{#1}}}             
\def\MHT#1{{\hat{{\bf T}}^{#1}}}      

\def\Imx{{\mbox{Im}}}                           
\def\Rex{{\mbox{Re}}}                           

\def\Modsp{{{\cal M}}}                

\newcommand\Ox[1]{{ O({#1}) }}        

\def\da{{\dot{\a}}}
\def\db{{\dot{\b}}}

\def\pslash{{\not\! p}} 
\def\kslash{{\not\! k}}


\def\CW{{\cal W}}                    
\def\WPH{{\widetilde{\Phi}}}         
\def\WW{{\widetilde{W}}}             
\def\WF{{\widetilde{F}}}             

\def\bCW{{\overline{\CW}}}                
\def\bWPH{{\overline{\WPH}}}               
\def\bWW{{\bar{\WW}}}                
\def\bWF{{\bar{\WF}}}                

\def\HYP{{\varphi}}                  
\def\bHYP{{\bar{\HYP}}}              

\def\HPHI{{\varphi}}            
\def\HPSI{{\psi}}               
\def\HF{{F}}                    

\def \lh{L_{het}}
\def \gh{g_{het}}
\def \li{L_{type\,I}}
\def \lf{{L_{planck}^{four}}}
\def \lt{{L_{planck}^{ten}}}
\def \gi{g_{type\,I}}
\def \lple{{L_{planck}^{eleven}}}

%% file: abstract.tex
After a short introduction to Matrix theory, we explain how can one
generalize matrix models to describe toroidal compactifications of
M-theory and the heterotic vacua with 16 supercharges. This allows us, for
the first time in history, to derive the conventional perturbative type
IIA string theory known in the 80s within a complete and consistent
nonperturbative framework, using the language of orbifold conformal
field theory and conformal perturbation methods. A separate chapter is
dedicated to the vacua with Ho\v{r}ava-Witten domain walls that carry
$E_8$ gauge supermultiplets. Those reduce the gauge symmetry of the matrix
model from $U(N)$ to $O(N)$. We also explain why these models contain open
membranes. The compactification of M-theory on $T^4$ involves the 
so-called $(2,0)$ superconformal field theory in six dimensions,
compactified on $T^5$. A separate chapter describes an interesting
topological contribution to the low energy equations of motion on the
Coulomb branch of the $(2,0)$ theory that admits a skyrmionic solution
that we call ``knitting fivebranes''. Then we return to the orbifolds of
Matrix theory and construct a formal classical matrix model of the
Scherk-Schwarz compactification of M-theory and type IIA string theory as
well as type 0 theories. We show some disastrous consequences of the
broken supersymmetry. Last two chapters describe a hyperbolic
structure of the moduli spaces of one-dimensional M-theory.

%% file: thanks.tex
I am grateful
to the many people who helped me along the way, explained many things to
me and to those that have shared their friendship and love. For example, I
am grateful to Micha Berkooz, Michael Dine, Rami Entin, Michal Fabinger,
Tomeu Fiol, Willy Fischler, Ori Ganor, David Gross,
Petr Ho\v{r}ava, Marcos Mari{\~{n}}o, Greg Moore, Arvind
Rajaraman, Christian R\"omelsberger, Moshe Rozali,  Steve Shenker,
Andy Strominger, Lenny Susskind, Herman Verlinde,
Edward Witten and many others.

\vspace{5mm}

A lot of thanks goes to my advisor Tom Banks who not only taught me a
lot of things about physics but who was also a significant source of my
support in the USA.

\vspace{5mm}

My work was supported in part by the Department Of Energy under grant
number DE-FG02-96ER40559. I would like to thank the ITP in Santa
Barbara for hospitality during the final stages of this work.

%% file: chap1-intro.tex
\chapter{Introduction to M(atrix) theory}

Superstring theory has been known to lead to a consistent perturbative
S-matrix for a decade or so, since the discoveries of Green and Schwarz in
the middle 80s. Superstring theory is only consistent in ten spacetime
dimensions and it can be approximated by supergravity theories
at low energies (or supergravities coupled to supersymmetric Yang-Mills
theories).
However the most symmetric supergravity theory required eleven spacetime
dimensions and seemed to be disconnected from superstring theory
completely. 
Green, Schwarz and Witten wrote \cite{gswitten} in 1987:
``Eleven-dimensional supergravity remains an enigma. It is
hard to believe that its existence is just an accident, but it is
difficult at the present time to state a compelling conjecture for what
its role may be in the scheme of things.'' 

The state of affairs changed drastically in 1995 when Witten ignited the
second superstring revolution \cite{witten}, building upon previous works
by Duff \cite{duffrevo} and Hull and Townsend \cite{hulltownsend}.
It turned out that M-theory,
a consistent completion of 11-dimensional supergravity at all energy
scales, can be defined as the strong coupling limit of type IIA
superstring theory: the radius of the new (circular) eleventh dimension
increases with the coupling and D0-branes and their bound states become
light degrees of freedom at strong coupling and
play the
role of the Kaluza-Klein modes in this new direction. Type IIA fundamental
strings themselves are membranes (that couple to the three-form potential
$A_{MNP}$ of M-theory) wrapped on the circle. M-theory also contains
five-branes, magnetic duals of membranes. 

This line of reasoning led physicists to find an overwhelming evidence for
dualities that imply that what has been thought of as several independent
string theories are really just perturbative expansions of the same theory
around different points in the moduli space. A full nonperturbative
and background-independent
definition of the theory was still missing.
However in 1996,
Banks, Fischler, Shenker
and Susskind \cite{bfss} found the first nonperturbative definition of
M-theory. A recent review can be found in \cite{taylorreview}.

\section{Basics of M(atrix) theory}

While the nature of M-theory in 11 dimensions was a complete enigma in
1995, exactly this eleven-dimensional vacuum became the first one to be
described in so-called {\it Discrete Light Cone Quantization} (DLCQ)
\cite{motlb,newcon} (the authors of \cite{bfss} originally formulated
their model in terms of the {\it Infinite Momentum Frame}). The theory is
defined as a large $N$ limit of a quantum mechanical model with a $U(N)$
symmetry that can be derived as the dimensional reduction of ${\cal N}=1$
ten-dimensional supersymmetric Yang-Mills theory into $0+1$ dimensions.
This model has been known for some years \cite{dhn} as the regularization
of supermembrane theory in the light cone gauge.

Using a slightly more modern logic explained in \cite{SeiWHY} (see also 
\cite{SenWHY}) we can derive the M(atrix) theory as the description of
M-theory in the light cone gauge as follows.

\vspace{3mm}

We want to compactify M-theory on a lightlike circle i.e. identify points
$(x^+,x^-,x^i)$ and $(x^++2\pi kR,x^-,x^i)$ where $k\in\IZ$, $i=1...9$ are
the transverse directions and $x^\pm=(x^0\pm x^{10})/\sqrt 2$ are the
lightlike directions; $x^+$ plays the role of time in the light cone
gauge. The component of the momentum $p^+$ must then be quantized,
\eqn{quati}{p^+=\frac NR,\qquad N\in\IZ}
and furthermore this quantity must be positive (except for the vacuum with
$p^+=0$). Because $p^+$ is conserved, we can study subspaces of the full
Hilbert space with different values of $N$ separately. In order to find a
description of this sector of the Hilbert space, we identify the points in
an {\it almost} lightlike (but spacelike) direction:
\eqn{almostli}{(x^+,x^-,x^i)\approx (x^++2\pi kR,x^--k\epsilon,x^i),
\qquad \epsilon\to 0^+.}
If we chose $\epsilon$ negative, closed timelike curves would appear and
the theory would not be causal. However for a very small positive
$\epsilon$ we can make a large boost to convert the almost lightlike
vector $(2\pi R,-\epsilon,0)$ into a spacelike vector $(R',-R',0)$ with
a very small $R'$. Therefore the physics is equivalent to the physics of
M-theory compactified on a small circle i.e. type IIA superstring theory,
and the original longitudinal momentum $p^+=N/R$ is translated to $N$
D0-branes of type IIA superstring theory. Because of the large boost, the
D0-branes have very small velocities. This limit of their dynamics was
studied by DKPS \cite{dkps}. We obtain a quantum matrix Hamiltonian
\eqn{qmh}{H=p^-=\frac{p^2+m^2}{2p^+}=
R\Tr\left(
\frac 12 \Pi_i\Pi_i-
\frac{M_{11}^6}{16\pi^2}[X_i,X_j]^2-
\frac{M_{11}^3}{4\pi}\lambda\Gamma^0\Gamma^i[X^i,\lambda]
\right).}
Here $X_i$, $i=1\dots 9$, are $N\times N$ Hermitean matrices, $\Pi_i$ are
their canonical duals and $\lambda$ are Hermitean fermionic matrices that
have 16 components forming a spinor of $spin(9)$. The model is
supersymmetric, i.e. the Hamiltonian equals an anticommutator of
supersymmetries, and physical states are required to be $U(N)$ invariant.

BFSS explained \cite{bfss} that this model describes graviton multiplets
with any value of $p^+=N/R$: it is generally believed that there is a
unique bound state of the $SU(N)$ quantum mechanics for any value of $N$
(we will present some evidence in the next chapter) and that the traces
(the
$U(1)$ part) are responsible for the transverse momentum ($\Tr\Pi_i$) and
for the
128+128 polarizations of the supergraviton
($\Tr\lambda$). A state
containing several well-separated subsystems is described by a set of
block-diagonal matrices, so that each subsystem (cluster) lives in one of
the blocks. The existence of the exactly flat directions of the potential
is guaranteed by supersymmetry. The off-diagonal degrees of freedom
(``W-bosons'') stay in their ground states if the distances are large
enough, but generally, they are responsible for the interactions between
the
two blocks. Many nontrivial tests of the interactions between the
gravitons have been made in the framework of M(atrix) theory and an 
agreement with supergravity has been found. We should emphasize that the
``unphysical'' lightlike compactification becomes harmless and goes away 
in the limit $R\to\infty$ which implies (for a fixed $p^+$) large $N$, and
also that
some quantities are required to agree with supergravity for large $N$
only.

\subsection{Membranes}

The matrix model contains also two-dimensional objects, membranes. The
simplest
example is to show that the model contains membranes of toroidal topology.
We start with $n\times n$ matrices $U,V$ that satisfy
\eqn{uvksc}{U^n=V^n={\bf 1},\qquad
UV=VU\alpha,\qquad
\alpha=\exp(2\pi i/n).}
In a convenient basis they can be represented as the clock and the shift
operator:
\eqn{clockshit}{U=
\left(
\begin{array}{ccccc}
1&\circ&\circ&\dots&\circ\\
\circ&\alpha&\circ&\dots&\circ\\
\circ&\circ&\alpha^2\!&\dots&\circ\\
\vdots&\vdots&\vdots&\ddots&\circ\\
\circ&\circ&\circ&\dots&\alpha^{n-1}\!
\end{array}
\right),
\qquad
V=
\left(
\begin{array}{ccccc}
\circ&\circ&\circ&\dots&1\\
1&\circ&\circ&\dots&\circ\\
\circ&1&\circ&\dots&\circ\\    
\vdots&\vdots&\vdots&\ddots&\circ\\
\circ&\circ&\circ&\dots&\circ 
\end{array}
\right).
}
Then a general $n\times n$ matrix $X^i$ can be decomposed as
\eqn{dekom}{X^i=\sum_{r,s=1-[n/2]}^{n-[n/2]}X^{i}_{rs}U^r V^s.}
The coefficients $X^i_{rs}$ can be represented as the modes in the Fourier
expansion and we can define a periodic function of $p,q$ (with periods
$2\pi$) by
\eqn{peripq}{X^i(p,q)=\sum_{r,s=1-[n/2]}^{n-[n/2]}X^{i}_{rs}
\exp(ipr+iqs).}
We assume that the function $X^i(p,q)$ is smooth enough, or equivalently,
that the coefficients $X^i_{rs}$ decrease sufficiently with
$\abs{r},\abs{s}$.
In this map between matrices and functions, trace is identified with the
integral (the expressions below are the ``average'' values of a function
or diagonal entries of a matrix)
\eqn{traceint}{\frac{\Tr}{n}=\frac{\int{dp\,dq}}{(2\pi)^2}}
and the commutator becomes the Poisson bracket (to understand it, compute 
the commutators of expressions like $U^r V^s$):
\eqn{comupo}{[X^i,X^j]\to \frac{2\pi i}{n}
(\partial_p X^i\partial_q X^j-
\partial_q X^i\partial_p X^j)+O(n^2)}
In this new language, the BFSS Hamiltonian becomes 
\eqn{bfssmem}{R\int \frac{dp \,dq}{(2\pi)^2}
\left(
\frac n2\Pi_i\Pi_i
+\frac{M_{11}^6}{4n}
\{X_i,X_j\}_{PB}^2
-i\frac{M_{11}^3}{2}\lambda\Gamma^0\Gamma^i\{X^i,\lambda\}_{PB}
\right)...}
This is
the Hamiltonian for a supermembrane in the light cone gauge; it reduces
to a Hamiltonian for a superstring in ten dimensions
after the dimensional reduction. It is
useful to realize that the matrix model contains closed oriented membranes
of arbitrary topology;  the existence of flat directions is related to the
fact that we can always join two membranes by a very thin throat without
modifying the Hamiltonian.

%% file: chap2-screw.tex
\chapter{Matrix string theory}

In this chapter I show why the maximally supersymmetric Yang-Mills
theory in $1+1$ dimensions captures the nonperturbative physics of type
IIA superstrings in the light cone gauge. The relevant approximation of
the SYM theory is an orbifold conformal field theory. The original $U(N)$
gauge symmetry is broken to a semidirect product of $U(1)^N$ and the
permutation group $S_N$. The boundary conditions of the fields involve a
general permutation; each cycle of this permutation is uniquely associated
with a ``long'' string whose longitudinal momentum $p^+$ is determined by
the length of the cycle. The multi-string states are therefore represented
by block-diagonal matrices as is usual in M(atrix) theory.
The condition $L_0=\tilde L_0$ arises in the
large $N$ limit from the requirement of the invariance of the physical
Hilbert space under a $\IZ_N$ subgroup of the local group. The
interactions can be understood via the so-called DVV vertex operator, and
the correct scaling law between the radius of the $11^{th}$ circle and the
type IIA coupling constant can be derived.

\section{Reflections on the second quantization and its matrix competitor}

For a long time, the second quantization has been the only
way to accomodate multi-particle states that expressed naturally
the fact that they were identical. The first candidate
to replace this machinery turned out to have the form of matrix
models \cite{bfss}
that we sketched in the first chapter. Different particles ``live'' in
different blocks
of a block diagonal matrix and their permutation symmetry
(either Bose or Fermi) is contained in the gauge group of the
theory which is $U(N)$ in \cite{bfss}. The
Hamiltonian contains
squares of all the possible commutators, and thus for large distances
the physical states (of not too high energies) can be described
by almost commuting matrices which can be simultaneously
diagonalized. Only when the distances are small, the commutators
are not so dominant and the classical positions of particles
make no sense.
This mechanism offers a natural and very specific 
realization of the old paradigm that on distances smaller than 
the Planck
length the usual concepts of geometry do not work and are replaced by
something more
general that can perhaps be called ``noncommutative geometry'' in the
sense that coordinates become (non-commuting) matrices.

\vspace{3mm}

The phrase ``the second quantization'' is usually used in
an inaccurate meaning -- as the
canonical (first) quantization of a classical
field which has an infinite number of degrees of freedom,
for example the electromagnetic or the Dirac field.
The machinery of quantizing is however the same as
the machinery for quantizing one classical particle.
In a sense, a classical field carries
the same number of degrees of freedom
as a wave function of one particle.

But the procedure of the second quantization
can be also presented in a more orthodox fashion. The first
quantization makes classical observables such as
$x,p$ become operators 
and leads to a state vector (wave function). The second
quantization is based on the next step: we lift the values
of the wave function evaluated
at different points 
to operators. The whole procedure
can be written independently of basis: the first quantized state
vector $\ket\psi$ is replaced by an operator vector $\ket\Psi$
and the relation of orthonormality becomes
the (anti)commutation relation
\eq{\left[\left\langle u\vert\Psi\right\rangle,
\left\langle
\Psi\vert v\right\rangle\right]_{grad}=\left\langle u
\vert v\right\rangle.}
The graded commutator is an anticommutator 
if both 
states $\ket u$; $\ket v$ are Grassmann-odd, otherwise it is just a
commutator.
The operator-ket-vector $\ket\Psi$ contains annihilation
operators $a_i=\left\langle i\vert\Psi\right\rangle$, while
the conjugate bra-vector $\bra\Psi$ contains the Hermitean
conjugate operators i.e. creation operators $a^\dagger_i=\left\langle
\Psi\vert i\right\rangle$. Then we usually postulate the existence of
a ground
state $\ket 0$ annihilated by the whole $\ket\Psi$, and the excited
states are built by application of $\left\langle\Psi\vert
u\right\rangle$ creation operators.

Then we often write the free part of the second quantized
Hamiltonian
as the ``push-forward'' of the first quantized Hamiltonian to
the multiparticle Hilbert space.
For example, the first quantized operator $f$ is lifted to the
second quantized $F$:
\eq{F=\bra \Psi f \ket\Psi}
Namely the identity operator $1$ is replaced by the number operator $N$
that counts the particles. 
It is easy to show that graded commutators of the second quantized
operators arise from the corresponding first quantized ones. 
\eq{[\left\langle\Psi\vert f\vert\Psi\right\rangle,
\left\langle\Psi\vert g\vert\Psi\right\rangle
]_{grad}=\left\langle\Psi\vert[f,g]_{grad}\vert\Psi\right\rangle.}

Exactly this second quantization was used for strings in the light cone
gauge in papers by Green, Schwarz and others
as well as in the cubic covariant string field theory where the
classical spacetime fields are formally elements of the first quantized
Fock space of a string, too.
We build the
canonical
second quantized Hamiltonian from the one-string 
(first-quantized) Hamiltonian and
then we are adding interaction terms reponsible for
splitting and joining the strings (as well as the crossing-over
interactions and others required by a stringy version of locality).
These terms are proportional to the coupling constant
(and in the light-cone gauge we also need ``contact terms'' proportional
to its higher powers in order to cover the whole moduli space of Riemann
surfaces), which
itself is a modulus of the theory. For a long time only perturbative
calculations were possible.

An important physical property of this second-quantized scheme is the 
{\it cluster property}:
the states formed by two (or more) sufficiently distant
(and thus almost non-interacting) subsystems
are represented by the tensor product of these
subsystems. In this sense, there {\it is}
a qualitative difference between a one-string state and
multistring states.

On the contrary, from an aesthetical and intuitive point of view
there {\it should be no} qualitative difference between
e.g. a state with two touching closed strings and a state
with one closed string going along the same curve in space: just like the
number of particles in quantum field theory depends on the energy scales
at which we measure the system,
there should be no general definition of the number of strings in
string theory; 
we probably cannot write the number of strings as some
integral of a local quantity.

Therefore our desire is a formulation making no qualitative
distinction between one- and multi-string states. Of course, we must
preserve the cluster property. Matrix models obey both of
these requirements: the difference between states with
different numbers of gravitons
(or strings) really becomes well-defined only in the
limit of large distances
(or in the free string limit) and the cluster property is realized
in terms of a block decomposition of matrices. The 
permutation group of particles $S_N$ is naturally contained in the
$U(N)$ gauge group of these models.

\section{M(atrix) theory in $0+0$ and $0+1$ dimensions}

Let us begin almost at the beginning of the world -- in $0+0$
dimensions. We write the action as the dimensional reduction
of the $9+1$-dimensional Super-Yang-Mills theory to $0+0$
dimensions (this action, formally describing a collection of D-instantons
in type IIB theory in a certain energetic regime,
is often called the IKKT model -- or the ``Japanese matrix model'' -- and
has been claimed to be related to type IIB strings \cite{ikkt}):
\eq{S=\mbox{Tr}\left(\frac 14{[X_\mu,X_\nu][X^\mu,X^\nu]}+
\theta^T\gamma_{0}\gamma_\mu[X^\mu,\theta]+
\beta\cdot 1\right)}
Here the $X$'s and $\theta$'s are Hermitean matrices, $\theta$
is a real spinor of $spin(9,1)$ constrained to contain
sixteen components of one chirality only.
The auxiliary constant $\beta$ term corresponds in some sense to $P^+$ (or
$P_{11}$ in the infinite momentum language):
its trace is the total $P^+$, proportional to the size of matrices.
Physical systems must be invariant under the group fixing
a quantity like
\eq{v^\dagger\cdot P^+\cdot v=
v^\dagger e^{iX^\dagger} P^+ e^{-iX}v.}
The $X$'s must be Hermitean, because here, $P^+$ is a multiple of the unit
matrix. Other choices of $P^+$ would be inequivalent only if $P^+$ had a
different
signature. Such systems, naively describing branes and antibranes, have
been studied as well. The symmetry group is then noncompact and the
analytic continuation from $U(N)$ to $U(N-k,k)$ might be related to the
crossing symmetry \cite{pericross}.

Because the appearance
of $\gamma_{0}$, which must be included in the $32\times 32$
language (while in the $16\times 16$ language it can be
replaced by the unit matrix), may look unfamiliar, let us say a
few words about it: the invariants made out of spinors are formed
using the Dirac conjugate spinor $\bar\psi=\psi^\dagger\gamma_0$, just
like in four dimensions.
The $32\times 32$ matrices $\gamma_\mu$ can be written using
the $16\times 16$ ones of $spin(9)$ denoted by $g_\mu$ as
\eq{\gamma_0=\tb{{rr}0&-1\\ 1&0},\quad
\gamma_{\mu=1..9}=\tb{{cc}0&g_\mu\\ g_\mu&0},\quad
\gamma_{chir}=\tb{{rr}1&0\\ 0&-1}}
All the matrices are real, $\gamma_0$ is antisymmetric
while the other nine are symmetric as is clear from the
following explicit form of those. These $16\times 16$
matrices can be written as the following tensor products of
Pauli (or unit) matrices. Note the even number of
$y\equiv \sigma^2$ in each of them (which implies reality)
and the fact that they anticommute with each other:
\eq{g_{1\dots 9}=z111,x111,yz1y,yx1y,yyz1,yyx1,y1yz,y1yx,yyyy.}

Now we are at the level of the action;
we do not have a Hilbert space yet. In spite of that we can perform
a formal variation of the machinery described by Taylor \cite{taylor} to
obtain the theory in $0+1$ dimensions of BFSS \cite{bfss}.

We wish to mod out by a continuous symmetry isomorphic to $\IR$
containing all the shifts of $X_0$:
\eq{X_0\mapsto X_0+\Delta X_0.}
According to \cite{taylor} we must represent this group by a subgroup
of the gauge group. So we add two continuous indices
$t_m,t_n\in \IR$. The matrices $X^\mu_{mn}$ are now upgraded
to $X^\mu_{mn}(t_m,t_n)$. Here $t_m$ plays a similar role
as $m$ and $t_n$ as $n$. In other words, $t_m$ and $m$ together
form the left index, while $t_n$ and $n$ represent the right one.
In other words, the matrices are tensored with operators on the space
of complex functions of a real variable.
Thus, for instance, the hermiticity condition takes the form
\eq{X^\mu_{mn}(t_m,t_n)^\dagger=X^\mu_{nm}(t_n,t_m).}
These indices are sufficient to represent the $\Delta X_0$
shift as the operator $\exp (i\Delta X_0\cdot t)$, which
has matrix elements
\eq{Shift(t_m,t_n)=\delta(t_m-t_n)\exp(i\Delta X_0\cdot t_m).}
Note that if $t$ should be interpreted as a ``time'' then
the $\Delta X_0$ and thus also $X_0$ should be understood
as the dual quantity (``energy'').

Let us now write down
the conditions of \cite{taylor} that restrict the operators.
The shift of $X_0$ has no influence on the other matrices $Y$, so
they should obey (here $Y$ is understood as the operator
on the space of functions of $t$ here)
\eq{\exp(i\Delta X_0 t)Y\,\exp(-i\Delta X_0 t)=Y.}
It means that $Y$'s commute with all the Fourier modes and thus
with all the functions of $t$ -- therefore these matrices
are functions of $t$ i.e. they
have matrix elements proportional to $\delta(t_m-t_n)$.
Therefore we can omit one of the two indices $t$.
In the case of $X_0$ we must make one
modification only -- take the $X_0$ shift into account:
\eq{\exp(i\Delta X_0 t)X_0\,\exp(-i\Delta X_0 t)=X_0+\Delta X_0.}
Therefore $X_0$ has the form of a function of $t$ plus the derivative
with respect to $t$, creating the $\Delta X_0$ term.
\eq{X_0=x_0(t)+i\frac{\partial}{\partial t}.}
At this moment we derived the BFSS model \cite{bfss},
the $t$-derivative is generated properly and the function $x_0$
plays the role of the $A$ gauge potential in \cite{bfss}.
The trace in the $0+0$ model now includes also the
trace over the continuous indices $t_m,t_n$ and thus
the action will change to a $t$-integral. (We must be careful and
normalize the action properly; essentially we must divide it by
an infinite factor $\delta(0)$ in order for this procedure to work.)

We should perhaps emphasize that the physical
time $t$ arises as the {\it dual} variable
to $X_0$ of the $0+0$-dimensional model in our construction. In a similar
way, M(atrix) theory describing M-theory on a torus of dimension smaller
than four is described as the maximally supersymmetric Yang-Mills theory
on a {\it dual} torus.

\vspace{2mm}

As the IKKT model has no Hamiltonian, we can have serious doubts whether
it
has any physical meaning, and therefore we might consider the previous
paragraphs to be just a motivation. Nevertheless the final product of this
construction is the BFSS model that certainly {\it does} allow a
Hamiltonian treatment. The general idea of the
construction above -- that gauging a symmetry is guaranteed by requiring
the matrix to be gauge-equivalent to their transforms -- has been already
understood in \cite{bfss} and we can apply it to study type IIA string
theory, i.e. M-theory on a circle.

\subsection{An old-fashioned approach to the compactification of one
spatial coordinate}

Now we will compactify the coordinate $X^1$ on a circle
with period $R_1$. In other words, we will mod \cite{lmztwo} the theory
in 0+1 dimensions by the group (isomorphic to $\IZ$) of all
the shifts by multiples of $R_1$:
\eq{X^1\mapsto X^1+k\cdot R_1,\quad k\in \IZ.}

We will Fourier-transform the procedure of \cite{taylor} directly to the
continuous basis. The construction in \cite{taylor} involves
D0-branes copied to all the locations identified by the symmetry
and therefore two integer-valued indices
$m,n\in \IZ$ must be added to the matrices. We will present a direct
Fourier transformation of this procedure
and add two indices
$\sigma_1^m,\sigma_1^n$
parametrizing a circle in the same fashion as in the
previous section. The period of $\sigma_1^m,\sigma_1^n$
is taken to be $2\pi$.

In other words, we tensor the matrices with the space
of operators on the space of functions of an angular variable.

We represent the symmetry $X^1\mapsto X^1+k\cdot R_1$ by
the operator (gauge transformation) $\exp (ik\sigma_1)$. Note again that
the
$\sigma_1$ is a momentum-like variable. (Its period should
be $\approx 1/R_1$ if we want to make the dual torus manifest.)

Now we can repeat the arguments of the previous section and
show that all the operator-matrices except for $X^1$
are ordinary functions of $\sigma_1$. They contain $\delta(\sigma_1^m
-\sigma_1^n)$, so again, we can erase one $\sigma_1$ ``index''.
And we can also show that the $X^1$ operator contains
a sum of a function of $\sigma_1$ and the derivative
with respect to $\sigma_1$ (understood to be multiplied
by the unit matrix):
\eq{X^1=x^1(\sigma_1)+iR_1\frac{\partial}{\partial \sigma_1}.}
This derivative acts on the other matrices, for instance
\eq{[X^1,X^2]=[x^1,x^2]+iR_1\frac{\partial x^2}{\partial
\sigma_1}.}
The Hamiltonian was written as a trace and now the trace must
include also the trace (the integral) over the $\sigma_1$
variable. Therefore the Hamiltonian has the form 
\eq{H\approx \int_0^{2\pi}\!\!d\sigma_1\mbox{Tr}
\left(
\frac{\Pi_i(\sigma_1)\Pi_i(\sigma_1)}2
-\frac {[X^i(\sigma),X^j(\sigma)]^2}4+
\theta^T(\sigma_1)\gamma_i[X^i(\sigma_1),\theta(\sigma_1)]
\right).}
Since $X^1$ is understood to contain also $iR_1\partial/
\partial\sigma_1$, a commutator with such a
``matrix'' is equal to ($iR_1$ times)
the covariant derivative.

We had to divide the trace by an infinite factor, namely the
trace of the unit matrix over the continuous $\sigma_1$.
(The same was true even in the first step which gave
us the $0+1$-dimensional theory from the $0+0$-dimensional theory.)
But this
factor should be compensated by the same factor by which
we must rescale the longitudinal momentum $P^+$ (or $P_{11}$
in the infinite momentum frame formulation): the $P^+$
should be also proportional to that trace and thus we can
scale $P^+$ to be only a product of the minimal quantum and the size $N$
of the matrices at each point.

This procedure has been essentially known to the authors of \cite{bfss}
although it was not clear from the beginning that a different background
requires a {\it different} matrix model.

\subsection{Limits of the resulting type IIA theory}

Let us now have a look at the Hamiltonian we got
in the case of a single D-brane with the gauge group $U(1)$. 
The model is always a description of D1-branes of type IIB string theory
on a circle (we start with type IIA string theory and D0-branes on a
subplanckian circle and therefore we must T-dualize the circle to get
D1-branes in type IIB string theory) at low energies. For a single
D1-brane, we get a free theory in which the S-duality of type IIB string
theory is manifest: the Hamiltonian is nothing else than the light cone
gauge Green-Schwarz Hamiltonian of a type IIA string. 

The 16-component spinor $\theta$
of $spin(9)$ includes 8-component spinors of
$spin(8)$ of both chiralities $\gamma_1$; one chirality becomes
left-moving, the other chirality becomes right-moving, just like for
a type IIA string.\footnote{You might be puzzled why a description of
type IIB D1-branes, which are S-dual to type IIB fundamental strings,
contains left-moving and right-moving spinors of the 
{\it opposite} $spin(8)$ chirality. But 
do not forget that while the type IIA matrix strings are
described in the light cone gauge, the type IIB D1-branes are 
in the static gauge.}
In the $N=1$ case all the commutators are zero and the Hamiltonian equals
\eq{H\approx \int_0^{2\pi}\!\!d\sigma_1
\left(
\frac{\Pi_i(\sigma_1)\Pi_i(\sigma_1)}2
+\frac{R_1^2}4\sum_{j=2}^9
\left(\frac{\partial X^j(\sigma)}{\partial\sigma_1}\right)^2+
iR_1\cdot\theta^T(\sigma_1)\gamma_1
\frac{\partial\theta(\sigma_1)}{\partial \sigma_1}
\right).}

The limits
$R_1\to 0$ and $R_1\to \infty$ behave as expected.
In the case $R_1\to\infty$,
from the $[X^1+iR_1\cdot d/d\sigma,X^i]$
the most important term is the $\partial/ \partial
\sigma X^i$ term
which causes the $X$'s
to be essentially
independent
of $\sigma_1$. We can replace $X^i(\sigma_1)$ by
$X^i$ and we are back to the original D0-branes model
of \cite{bfss}. This is the decompactification, strongly coupled limit of
type IIA string theory.

\vspace{2mm}

In the case $R_1\to 0$ the $\sigma_1$ derivative terms are negligible.
Then the typical configuration should have the commutators sufficiently
small because otherwise they dominate the energy. Therefore for every
$\sigma_1$, $X^i(\sigma)$, $i=2\dots 9$ can be approximated by commuting
matrices. But the basis in which they are simultaneously diagonalizable
can differ as we change $\sigma_1$ and the derivative of the basis is
stored in the gauge field $X^1(\sigma_1)$.  This change can be fixed
together with the gauge but there is a very important global effect which
we will discuss in the following section. 

\section{Representation of the ``long'' strings,
screwing strings to matrices}

In  Czech  there
is a single word both for a {\it matrix} and a {\it nut}: ``matice''.
This provides the motivation for the phrase ``screwing strings
to matrices'' (in Czech ``\v sroubov\'an\'\i{} strun do matic'').
Some people prefer to call this model
{\it matrix strings} \cite{dvv}.
But what is the idea?

\vspace{2mm}

If we ask how the multistring states are represented, we find the
usual answer of \cite{bfss} with a natural modification: 
in an appropriate gauge,
the matrices
$X^i(\sigma_1)$ whose matrix elements are the functions
$X^i_{mn}(\sigma_1)$ have a (block) diagonal form where
each block corresponds to one string. (The real physical
state is obtained from such an idealized one by the symmetrizing
over all the gauge group and other procedures.)

But now we face a new question: the longitudinal momentum $P^+$
in the light cone gauge
(or $P_{11}$ in the infinite momentum frame ideology)
is naturally given again by the size $N$ of the matrices.
Note that in the light cone gauge superstring field theories
the total $p^+$ was always proportional to the length of
strings. Now the length is a multiple of the minimal quantum. 

So the question is: how can we represent a string with $P^+$
greater than the quantum of $P^+$ carried by the $N=1$ string?
We have already said that in the $R_1\ll 1$ case the matrices
$X^i(\sigma)$, $i=2\dots 9$ should be simultaneously
diagonalizable but the basis in which all of them have the
diagonal form can change with $\sigma_1$. This evolution
with $\sigma_1$ is stored in $X^1(\sigma)$ which plays the role
of a component of the vector gauge field.
Typically, because of the local $U(N)$ symmetry of the model,
the basis can be locally chosen to be independent of $\sigma_1$,
but there can also be certain global effects.

We have the condition that after
adding a period $2\pi$ to $\sigma_1$, the basis 
transforms again into a basis
where $X$'s are diagonal. But does it mean that
the monodromy must be the identity? Are there other transformations
keeping the diagonal form?

Of course, there are: these transformations are the
{\it permutations of the eigenvectors}: if we conjugate a diagonal matrix 
by a permutation of its eigenvectors, we obtain another diagonal matrix.
Every
permutation can be decomposed into a product of cycles. And what does
a cycle permuting $k$ eigenvectors represent? It
simply represents a string of length ($P^+$) equal to $k$ (times the
length of the minimal string).

It is perhaps useful to write down some explicit formulae for the
classical configuration describing long strings.
Let $X^i$, $i=2\dots 9$
denote the functions with period $2\pi k$ in which the embedding of a
string
of length $k$ into spacetime is encoded. 
How does the matrix model store the information about these functions in
the simultaneously diagonalizable $k\times k$
matrix variables $X^i_{k\times k}(\sigma)$ of period $1\cdot 2\pi$?
The answer is
\eqn{mast}{X^i_{k\times k}(\sigma)=U(\sigma)
\left(\begin{array}{cccc}
\!\!X^i(\sigma+2\pi)\!\!&\circ&\dots&\circ\\
\circ&\!\!X^i(\sigma+4\pi)\!\!&\dots&\circ\\
\circ&\circ&\ddots&\vdots\\
\circ&\circ&\dots&\!\!X^i(\sigma+2k\pi)\!\!
\end{array}\right)
U^{-1}(\sigma)}
where the unitary matrix-valued function $U(\sigma)$ must obey the
following relation in order to make $X^i_{k\times k}(\sigma)$ periodic
with the period $2\pi$ (written for the $k=4$ case for the sake of
simplicity) 
\eqn{mastobej}{U(\sigma+2\pi)=U(\sigma)
\tb{{cccc}
\circ&1&\circ&\circ\\
\circ&\circ&1&\circ\\
\circ&\circ&\circ&1\\
1&\circ&\circ&\circ\\}}
Here $\circ$ denotes $0$.
In the last matrix we could write arbitrary phases $e^{i\alpha_j}$
instead of $1$'s. Such a matrix function $U(\sigma)$
can be explicitly found for finite $k$ because $U(N)$
is connected. Note that the connection $X^1$ is pure gauge and
carries
the information about the dependence of $U(\sigma)$
on $\sigma$ 
\eq{X^1\sim i(\frac{\partial U(\sigma)}{\partial
\sigma})U^{-1}(\sigma)}
In other words, generic diagonal expectation values of the matrices
$X^i(\sigma)$ break the group $U(N)$ into a semidirect product of
$U(1)^N$ and the permutation group $S_N$. Configurations involving a
nontrivial bundle with the permutations of the eigenvalues are part of the
theory and are responsible for the existence of the long strings. States
of a long string can be also understood as a bound state of $k$ D-branes.
Namely the type IIA string of a general length $k$
has a massless supergraviton multiplet as its
ground state and this state is continuously connected to the bound state
of $k$ D0-branes in the $SU(k)$
BFSS model: if we accept that the number of the
ground states (or at least the index) is independent of the coupling
constant and that the 1+1-dimensional matrix model becomes the free type
IIA string theory or the BFSS model in the two respective limits, we can
view the existence of long strings as a proof of the existence of the
D0-brane bound states.

\section{The origin of the level matching conditions and the interactions}

The correct explanation of the condition $L_0=\tilde L_0$ that a closed
type IIA string must satisfy
has been clarified by Dijkgraaf, Verlinde and Verlinde \cite{dvv} and it
holds for very long strings only.

The cyclic permutation matrix in (\ref{mastobej}) is a global element of
the gauge group. The matrices $X$ in the adjoint of $U(N)$ transform under
this permutation matrix and we can see in (\ref{mast}) that the
transformation has the effect of shift of $\sigma$ by $2\pi$ which is
$1/k$ of the circumference of the long string. Because the stringy wave
functional must be invariant under this gauge transformation
and this shift can be written as $\exp(2\pi i k(L_0-\tilde L_0))$,
$L_0-\tilde L_0$ must be a
multiple of $k$ (in the usual normalization) for all physical states
and we see that for large $k$
$L_0-\tilde L_0$ must vanish, otherwise the state would have an infinite
energy.

\vspace{3mm}

The fact that the $1+1$-dimensional supersymmetric Yang-Mills theory
reproduces type IIA string theory holds beyond the free 
theory approximation.
At weak type IIA string coupling, the Hilbert space contains states with
an arbitrary number of free type IIA strings of length $k$ (whose sum
equals
$N$); these are described by a conformal field theory. How do the strings
interact? Two strings of lengths $k_1,k_2$ can join into a string of
length $k_1+k_2$ (or vice versa, a longer string can split into two).
At the level of the permutations, we see that two cycles should merge into
one: this happens if we compose the permutation, containing two cycles,
with a transposition. Physically such an interaction can occur on the
``worldsheet'' (on which the gauge theory is defined) if two strings
touch: then two vectors of the eigenvalues $X^i(\sigma)$ are equal and a
$U(2)$ symmetry is restored. The approximation of free strings breaks
down, the basis in which the matrices are diagonal is not well defined 
near the interaction point and the strong dynamics can change the
permutation monodromy associated with a given state. This is the origin of
the interactions.

Can we derive the correct scaling law between the radius of the 11-th
circle of M-theory and the string coupling constant? The circumference of
the circle in the gauge theory scales like $(R_{11}/l_{11})^{-3/2}$ in the
gauge-theoretical natural units of $1/g_{YM}^2$ as we have seen before. In
a revised version of \cite{motlb} I argued that the amplitude for an
interaction to occur is inversely proportional to this length: a
collection of long type IIA strings can be embedded in the theories with
different values of the coupling constant (i.e. of the circumference of
the
circle in the gauge theory
in units of $1/g_{YM}^2$) and a different value of $N$, so that the total 
length of the string in units of $1/g_{YM}^2$ is identical in both cases
(this corresponds to fixing $R_{11}$ in string units): but the shorter the
circumference of
this circle is, the more the long string must be ``wound'' around it and
consequently the greater is number of the points which a given point
on the worldsheet can interact with (in the large $N$ limit, the
interaction can effectively occur between {\it any} two points
and the density makes the only difference). Therefore
the coupling constant is inversely proportional to the circumference of
the circle in the gauge theory (if we set $g_{YM}^2=1$ so that locally the
interaction, based on the restoration of the $U(2)$ symmetry, looks the
same regardless of what the coupling constant is), i.e. $g_s\sim
R_{11}^{3/2}$ as
expected. This argument holds for the $E_8\times E_8$ heterotic
strings, too.

\subsection{The DVV interaction vertex}

DVV offered a derivation that allows one to check more technical
details \cite{dvv}.  They used the language of conformal perturbation
theory. The free theory is the conformal field theory describing free type
IIA strings in various sectors determined by the permutation. 
Let us scale $\sigma$ again to have a circumference $2\pi$, independent of
$R_{11}$; the dimensionful coupling $g_{YM}^2$ has dimension of mass and
scales like $(R_{11}/l_{11})^{-3/2}$. The theory
is perturbed by some corrections because a gauge group $U(2)$ or more
generally $U(M)$ can be restored. 
\eqn{dvvexp}{S=S_{CFT}+\lambda\int d^2\sigma\, V_{int}+\dots}
These corrections are irelevant in the
infrared and we should look for the ``least irelevant'' one. This
correction is nothing else than the DVV vertex.

To explain its nature, consider first the case of the ``bosonic matrix
string'', i.e. the 1+1-dimensional purely bosonic $U(N)$
gauge theory with 24
transverse scalars $X^2\dots X^{25}$. (The lack of supersymmetry makes a
spacetime
interpretation of such a matrix model impossible, but let us not worry
about it.)
Two vectors of eigenvalues $X^i_a(\sigma)$ and $X^i_b(\sigma)$,
corresponding to strings $a,b$, can suddenly permute for
$\sigma>\sigma_I$,
if an interaction occurs at $\sigma_I$. 
It means that the sum $X^i_a+X^i_b$ remains constant while
the $X^i_a-X^i_b$ flips the overall sign.
Such a twist is generated by the
twist field $\tau$ associated with the $X^i_a-X^i_b$ degrees of freedom
-- the vertex operator of the ground state of the
sector where $X^i_a-X^i_b$ are antiperiodic. This $\tau$ is the product
\eqn{prodtau}{\tau=\prod_{i=1}^{24}\tau_i}
of the twist fields associated with each scalar. Each twist field has
dimension $1/16$, and there are 24 of them, so the total weight of such a
field
is $(24/16,24/16)=(3/2,3/2)$, if we take the product of the twist fields
in both the left-moving and the right-moving sectors of the CFT. The total
scaling dimension is therefore $3$.
The vertex operator
is integrated over $\int d^2\sigma$
so the coefficient still must have a dimension of $length^1$. The only
local length scale in the problem is $1/g_{YM}^2$
(the masses of the W-bosons that we integrate out when we use
the low energy description of CFT are inversely proportional 
to $1/g_{YM}^2$),
which scales like
$(R_{11}/l_{11})^{3/2}$, and therefore also the coefficient of the
interaction will scale in this way.

And we know that this is the correct scaling of the string coupling $g_s$
in the superstring case, both in type IIA and heterotic $E_8\times E_8$. 
Actually the dimension of the DVV vertex is always $(3/2,3/2)$. In the
heterotic case we must replace the right-moving bosonic twist field by its
supersymmetric counterpart. In the type IIA superstring case we must do it
for the left-movers, too. In both cases the supersymmetric partner
$\tau_S$ has again weight equal to $3/2$. It can be written as
\eqn{taus}{\tau_S=\tau_{exc}^i\Sigma^i,}
where the sum is over the vector indices $i=1\dots 8$. 
The symbol $\tau_{exc}^i$
denotes the vertex operator for the first excited state in the
antiperiodic sector of $X$'s (an excited twist field), namely the state
$\alpha_{-1/2}^i\ket{0}_{antiper}$. The $\alpha^i_{-1/2}$ excitation
contributes
$1/2$ to the weight, the ground state itself carries $1/2=8\times 1/16$,
and the remaining $1/2$ is the weight of the spin field
$\Sigma^i$ which is the vertex operator for the 
degenerate ground state of the
periodic sector of the spinorial fermions. The ground state for
$\theta_\alpha$ transforming in $8_s$ transforms as $8_c+8_v$. The
familiar fact that the Ramond-Ramond ground state of $\psi^\mu$
transforms as $8_s+8_c$ is related to the fact we need here by triality.
Anyway, half of the spin fields associated with 
the spinorial fermion transform as a
vector $8_v$. This is the reason why we must contract them with the {\it
excited} $\tau_{exc}^i$. Once again, the total dimension is
$(3/2,3/2)$ and we reproduce the correct relation between the string
coupling and the size of the eleventh dimension. The total action in the
conformal perturbative expansion is
\eqn{conper}{S=S_{CFT}+g_s\sum_{i<j=1}^N\int d^2\sigma
\,\tau_{exc, i,j}^k(\sigma)\Sigma^k_{i,j}(\sigma)\,\otimes
\,\tilde\tau_{exc, i,j}^l(\sigma)\tilde\Sigma^l_{i,j}(\sigma)+\dots,}
where the sum goes over all pairs of the eigenvalues $i,j$ and the tilde
denotes the right-moving sector of the CFT. The interaction operator,
essentially known to Mandelstam decades ago, is known to respect the
$SO(9,1)$ Lorentzian symmetry. This might be viewed as a hint that also
the original BFSS model restores this symmetry in the large $N$ limit.

\subsection{Compactification to nine dimensions}

It is possible to compactify M(atrix) theory on $T^2$ and $T^3$, too.
For instance M-theory on a subplanckian $T^2$ should be described by type
IIB strings. In M(atrix) theory we can see this as follows. The matrix
model for M-theory on $T^2$ is a $2+1$-dimensional maximally
supersymmetric $U(N)$ gauge theory on a dual torus. If the spacetime torus
shrinks, the worldvolume torus of the matrix model becomes infinite.
Therefore the physics is dominated by the infrared fixed point of the
gauge theory which is a $2+1$-dimensional superconformal field theory
where the $spin(7)$ R-symmetry enhances to $spin(8)$. 
The M-theoretical reason is that $2+1$-dimensional supersymmetric
Yang-Mills theory describes D2-branes (for $\alpha'\to 0$) and the
D2-branes of type IIA string theory become M2-branes of M-theory at strong
coupling.
The gauge field can be electromagnetically dualized in $2+1$ dimensions to
a scalar
\eqn{dvebe}{F^{\mu\nu}=\epsilon^{\mu\nu\lambda}\partial_\lambda \phi_{8}}
and this $\phi_8$ is nothing else than the extra dimension appearing in
the type IIB description, corresponding to M2-branes wrapped on the small
two-torus. Matrix theory offers quite rigorous explanations of many
conjectures associated with string duality. This should not be surprising
because M(atrix) theory {\it is} a nonperturbative definition of M-theory
in certain backgrounds, in fact the first definition that people found.
But now we want to move on to the question of heterotic strings. Do they
admit a description in terms of matrix models?

%% file: chap3-hetm.tex
\chapter{Heterotic matrix models}

In this chapter I propose a nonperturbative definition
of heterotic string theory on multidimensional tori and discuss the
question of their matrix string limits and the origin of open membranes.

\section{Introduction}

The matrix model of uncompactified M-theory \cite{dhn},
\cite{towns}, \cite{bfss} has
been generalized to arbitrary toroidal compactifications of type IIA and
IIB string theory. For M-theory on $T^k$, $k\leq 3$ (i.e. type II
strings on $T^{k-1}$), one describes
the physics by a $k+1$-dimensional supersymmetric Yang-Mills theory on a
dual $T^k$. For $k=4$, the ``4+1-dimensional SYM'' is replaced by a
5+1-dimensional $(2,0)$ theory on $T^5$
(the momentum in the new, sixth dimension represents
the winding number of the longitudinal fivebranes
on the four-torus in spacetime, $Z_{-ijkl}$, this is
the first time where the total
fivebrane charge can be nonzero). In the case
$k=5$
one needs a
little string theory on $T^5$ and for $k>5$ the matrix models do not
decouple from gravity anymore.
All these models can be understood as particular large $M$
limits of the original matrix model, in the sense that they may be viewed
as the dynamics of a restricted class of large $M$ matrices, with the
original matrix model Lagrangian.

A separate line of reasoning has led to a description of the Ho\v rava-Witten 
domain wall in terms of matrix quantum mechanics
\cite{orbi}-\cite{lmztwo}.
Here, extra degrees of freedom have to be added to the original matrix
model. As we will review below, if these new variables, which transform in
the vector representation of the gauge group, are not added, then 
a potential causes that the
model does not live in an eleven dimensional spacetime, but only on its
boundary.  Although it is, by construction, a unitary quantum mechanics,
it probably does not recover ten dimensional Lorentz invariance in the
large $M$ limit.  Its nominal massless particle content is the ten
dimensional $N = 1$ supergravity (SUGRA) multiplet, which is anomalous.

With the proper number of vector variables added, the theory does have an
eleven dimensional interpretation.  It is possible to speak of states far
from the domain wall and to show that they behave exactly like the model
of \cite{bfss}.  Our purpose in the present paper is to compactify this
model on spaces of the general form $S^1 / \IZ_2 \times T^d$.  We begin
by
reviewing the argument for the single domain wall quantum mechanics, and
generalize it to an $S^1 / \IZ_2$ compactification.  The infinite momentum
frame Hamiltonian for this system is practically identical to the static
gauge $O(M)$ Super Yang Mills (SYM) Hamiltonian for $M$ heterotic D
strings in type I string theory.  They differ only in the boundary
conditions imposed on the fermions which transform in the vector of
$O(M)$.  These fermions are required for $O(M)$ anomaly cancellation in
both models, but the local anomaly does not fix their boundary conditions.
Along the moduli space of the $O(M)$ theory, the model exactly reproduces
the string field theory Fock space of the $E_8\times E_8$
heterotic string theory.  The
inclusion of both Ramond and Neveu-Schwarz boundary conditions for the
matter fermions, and the GSO projection, are simple consequences of the
$O(M)$ gauge invariance of the model.

Generalizing to higher dimensions, we find that the heterotic matrix model
on $S^1 / \IZ_2 \times T^d$ is represented by a $U(M)$ gauge theory on
$S^1
\times T^d / \IZ_2$.  On the orbifold circles, the gauge invariance
reduces
to $O(M)$.  We are able to construct both the heterotic and open string
sectors of the model, which dominate in different limits of the space of
compactifications.

In the conclusions, we discuss the question of whether the heterotic
models which we have constructed are continuously connected to the
original uncompactified eleven dimensional matrix model.  The answer to
this question leads to rather surprising conclusions, which inspire us to
propose a conjecture about the way in which the matrix model solves the
cosmological constant problem.  It also suggests that string vacua with
different numbers of supersymmetries are really states of {\it different}
underlying theories.  They can only be continuously connected in limiting
situations where the degrees of freedom which differentiate them decouple.

\section{Heterotic matrix models in ten and eleven dimensions}

In \cite{evashamit} an $O(M)$ gauged supersymmetric matrix model for a
single Ho\v rava-Witten domain wall embedded in eleven dimensions was
proposed.  It was based on an extrapolation of the quantum mechanics
describing D0-branes near an orientifold plane in type IA string theory
\cite{orbi}.  The model was presented as an orbifold of the original
\cite{bfss} matrix model in \cite{lmztwo}. 

\subsection{Heterotic string theory as an orientifold of M-theory}

There exists a formal way how to derive a matrix description of an
orbifold theory. If we start with M-theory in 11 dimensions and divide it
by a group $G$, physical states should be invariant under $G$. Therefore
the elements of $G$ must be identified with some elements of the gauge
group: the matrices $X^i$ and $\theta^i$ are constrained to
satisfy this identification. In the
previous chapter we saw that the procedure for the case of the
compactification on a circle, $G=\IZ$, has the effect of the dimensional
``oxidation'' of the matrix model: a new circular spatial dimension
appears and the matrix $X^i$ where $i$ points along the circle is lifted
to a component of the covariant derivative with respect to the new
dimension of the matrix model.

Can we apply the same logic to construct the $\IR / \IZ_2$
(and then $S^1/ \IZ_2$) compactification
of M-theory? Yes, we can. First we must realize how does the symmetry
$\IZ_2$ act. It flips the sign of $X^1$, the direction perpendicular to
the domain wall at $X^1=0$, and correspondingly multiplies the spinors
$\theta$ by $\pm\gamma^1$ (the sign in front of $\gamma^1$ determines the
chirality of the $E_8$ multiplet as well as gravitino on the boundary,
let
us consider the plus sign). What we have said so far is not a symmetry:
in M-theory we know that we must also change the sign of the 3-potential
(to keep $\int C_3\wedge F_4\wedge F_4$ invariant, for example): the
Ho\v{r}ava-Witten domain wall is really an orientifold plane since it
flips the orientation of membranes. In the matrix model we can see this
because the term of the form $\Tr\theta[X,\theta]$ in the matrix model
Hamiltonian is {\it odd} under the $\IZ_2$ generator described so far. It
is easy to fix this problem. The true symmetry must also {\it transpose}
all the matrices.  Matrices must be invariant under this combined
operation.  As a result, $X^1$ (and $\Pi^1$) as well as the spinors with
$\gamma^1\theta=-\theta$ become {\it antisymmetric} Hermitean matrices and
$X^2\dots X^9$, their conjugate momenta and the rest of the spinors become
{\it symmetric} Hermitean matrices. Clearly, $X^1$ and its superpartners
transform in adjoint of a restricted gauge group $O(N)$ while the
remaining fields transform in the symmetric tensor representation of
$O(N)$. We can see that the number of bosonic degrees of freedom exceeds
the number of the fermionic degrees of freedom (since most of the bosons
are symmetric and the symmetric tensor has more components than the
antisymmetric tensor, by $N$): 
this can be viewed as the first hint that we need to add
some more fermions transforming in the fundamental representation and we
will explain details in one of the subsections below. 

\subsection{Open membranes in M(atrix) theory}

What is the interpretation of the transposition of matrices explained in
the previous subsection in terms of membranes? We see that while the
matrix $U$ from the Chapter 1 is symmetric, the transposition of the
matrix $V$ is $V^{-1}$. Because $V$ represents $\exp(iq)$ where $q$ is a
spatial coordinate of the membrane, the transposition corresponds to a
reflection of the membrane (of its coordinate $q$, more precisely). And
because the matrices $X^2\dots X^9$ must be invariant under this
reflection (they are symmetric), the corresponding functions $X^i(p,q)$
defined on the toroidal
membrane also satisfy $X^i(p,-q)=X^i(p,q)$ and the toroidal membrane is
effectively reduced to a cylinder. The matrix $X^1$ is antisymmetric,
therefore it is an odd function of $q$ and $X^1(q=0)=X^1(q=\pi)=0$: the
open membrane must terminate on the end of the world.

A more detailed analysis was subsequently made by Kim and Rey \cite{orbi}.
The
orthogonal and symplectic group was shown to be capable to describe the
$\IZ_2$ orbifolds of the sphere $S^2$ and the torus $T^2$ summarized in
the table below.

$$\begin{array}{|c|c|c|}
\hline \mbox{Gauge group}&\mbox{Matrix generators}&\mbox{Membrane
topology}\\
\hline
SO(2N) \mbox{ and } SO(2N+1)&Y_{lm}-(-1)^{l+m}Y_{lm}&\mbox{Disk}\\
\hline
USp(2N)&Y_{lm}-(-1)^{l}Y_{lm}&\mbox{Projective plane}\\
\hline
SO(2N) \mbox{ and } SO(2N+1)&J_{r,s}-J_{r,-s}&\mbox{Cylinder}\\
\hline
SO(2N)&J_{r,s}-(-1)^sJ_{r,-s}&\mbox{Cylinder}\\
\hline
SO(2N) \mbox{ and } USp(2N)&J_{r,s}-(-1)^rJ_{r,-s}&\mbox{Klein bottle}\\
\hline
SO(2N) \mbox{ and } USp(2N)&J_{r,s}-(-1)^{r+s}J_{r,-s}&\mbox{Klein
bottle}\\
\hline
SO(2N) \mbox{ and } SO(2N+1)&J_{r,s}-J_{s,r}&\mbox{M\"obius strip}\\
\hline
SO(2N)&J_{r,s}-(-1)^{r+s}J_{s,r}&\mbox{M\"obius strip}\\ \hline
\end{array}$$
\begin{table}[tp]
\caption{Matrix representations of $\IZ_2$ orbifolds
of the sphere $S^2$ and the torus $T^2$: here $Y_{lm}$ are the spherical
harmonics and $J_{r,s}$ are the Fourier modes on the torus. Both are
represented by some matrix generators in the regularized case.}
\end{table}

\subsection{Additional degrees of freedom}

The orbifold derivation presented above has a formal character and one
should rather follow Seiberg's more rigorous derivation of the matrix
model. In order to describe M-theory near a Ho\v{r}ava-Witten domain wall,
we need to consider type IIA string theory near the corresponding
orientifold plane i.e. type IA string theory. Since the dilaton is
constant, there must be eight D8-branes stuck at this orientifold plane.
In the type IA context it is
necessary to add to the matrix model the low-energy
degrees of freedom transforming in the vector of $O(M)$ 
and corresponding to the existence of D8-branes and the $08$-strings
connecting them to the D0-branes. Since D8-branes are movable in type
IA theory, there are consistent theories both with and without these extra
degrees of freedom.  That is, we can consistently give them masses, which
represent the distances between the D8-branes and the orientifold.
However, as first pointed out by \cite{orbi}, unless the number of
D8-branes sitting near the orientifold is exactly $8$, the D0-branes feel a
linear potential (resulting from the dilaton gradient)
which either attracts them to or repels them from the
orientifold. This is the expression in the quantum mechanical
approximation, of the linearly varying dilaton first found by Polchinski
and Witten \cite{jopo}. This system was studied further by Kim and Rey 
and by Banks, Seiberg and Silverstein in
\cite{orbi}-\cite{bss}. In the latter work the supersymmetry and gauge
structure of
model were clarified, and the linear potential was shown to correspond to
the fact that the ``supersymmetric ground state'' of the model along
classical flat directions representing excursions away from the
orientifold was not gauge invariant.

From this discussion it is clear that the only way to obtain a model with
an eleven dimensional interpretation is to add sixteen massless real
fermions transforming in the vector of $O(M)$, which is the model proposed
in \cite{evashamit}.  In this case, D0-branes can move freely away from
the wall, and far away from it the theory reduces to the $U([{M\over 2}])$
model of \cite{bfss} \footnote{Actually there is a highly nontrivial
question which must be answered in order to prove that the effects of the
wall are localized. In \cite{bss} it was shown that supersymmetry allowed
an arbitary metric for the coordinate representing excursions away from
the wall. In finite orders of perturbation theory the metric falls off
with distance but, as in the discussion of the graviton scattering
amplitude in \cite{bfss}, one might worry that at large $M$ these could
sum up to something with different distance dependence.  In \cite{bfss} a
nonrenormalization theorem was conjectured to protect the relevant term in
the effective action.  This cannot be the case here.}.

Our task now is to construct a model representing two Ho\v rava-Witten end
of the world $9$-branes separated by an interval of ten dimensional space.
As in \cite{lmztwo} we can approach this task by attempting to mod out the
$1+1$ dimensional field theory \cite{bfss}, \cite{taylor}, \cite{motlb},
\cite{bs}, \cite{dvv} which describes M-theory compactified on a circle.  
Following the logic of \cite{lmztwo}, this leads to an $O(M)$ gauge
theory.
The $9$-branes are stretched around the longitudinal direction of the
infinite momentum frame (IMF) and the $2-9$ hyperplane of the transverse
space.  $X^1$ is the differential operator $${R_1 \over i}{\partial \over
\partial\sigma} - A_1$$ where $\sigma$ is in $[0,2\pi ]$, and $A_1$ is an
$O(M)$ vector potential.  The other $X^i$ transform in the ${\bf {M(M+1)
\over 2}}$ of $O(M)$.  There are two kinds of fermion multiplet. $\theta$
is an ${\bf 8_c}$ of the spacetime $SO(8)$, a symmetric tensor of $O(M)$
and is the superpartner of $X^i$ under the eight dynamical and eight
kinematical SUSYs which survive the projection.  $\lambda$ is in the
adjoint of $O(M)$, the ${\bf 8_s}$ of $SO(8)$, and is the superpartner of
the gauge potential.  We will call it the gaugino.

As pointed out by Kim and Rey \cite{orbi} and \cite{bss}, this model is
anomalous. One
must add $32$ Majorana-Weyl fermions $\chi$ in the ${\bf M}$ of $O(M)$.
For sufficiently large $M$, this is the only fermion content which can
cancel the anomaly.  The continuous $SO(M)$ anomaly does not fix the
boundary conditions of the $\chi$ fields. There are various consistency
conditions which help to fix them, but in part we must make a choice which
reflects the physics of the situation which we are trying to model.

The first condition follows from the fact that our gauge group is $O(M)$
rather than $SO(M)$.  That is, it should consist of the subgroup of $U(M)$
which survives the orbifold projection.  The additional $\IZ_2$ acts only
on
the $\chi$ fields, by reflection.  As a consequence, the general
principles of gauge theory tell us that each $\chi$ field might appear
with either periodic or antiperiodic boundary conditions, corresponding to
a choice of $O(M)$ bundle.  We must also make a projection by the discrete
transformation which reflects all the $\chi$'s. What is left undetermined
by these principles is choice of relative boundary conditions among the
$32$ $\chi$'s.

The Lagrangian for the $\chi$ fields is \eqn{chilag}{\chi (\partial_t +
2\pi R_1\partial_{\sigma} - i A_0 - i A_1) \chi.} In the large $R_1$
limit, the volume of the space on which the gauge theory is compactified
is small, and its coupling is weak, so we can treat it by semiclassical
methods.  In particular, the Wilson lines become classical variables.  We
will refer to classical values of the Wilson lines as expectation values
of the gauge potential $A_1$. (We use the term expectation value loosely,
for we are dealing with a quantum system in finite volume.  What we mean
is that these ``expectation values'' are the slow variables in a system
which is being treated by the Born-Oppenheimer approximation.) An
excitation of the system at some position in the direction tranverse to
the walls is represented by a wave function of $n \times n$ block matrices
in which $A_1$ has an expectation value breaking $O(n)$ to $U(1) \times
U([n/2])$. In the presence of a generic expectation value, in $A_0 = 0$
gauge, the $\chi$ fields will not have any zero frequency modes. The
exceptional positions where zero frequency modes exist are $A_1 = 0$ (for
periodic fermions) and $A_1 = \pi R_1$ (for antiperiodic fermions). These
define the positions of the end of the world $9$-branes, which we call the
walls.  When $R_1 \gg l_{11}$, all of the finite wavelength modes of all
of the fields have very high frequencies and can be integrated out. In
this limit, an excitation far removed from the walls has precisely the
degrees of freedom of a $U([{n\over2}])$ gauge quantum mechanics.  The
entire content of the theory far from the walls is $U([{M\over 2}])$ gauge
quantum mechanics.  It has no excitations carrying the quantum numbers
created by the $\chi$ fields, and according to the conjecture of
\cite{bfss} it reduces to eleven dimensional M-theory in the large $M$
limit.  This reduction assumes that there is no longe range interaction
between the walls and the rest of the system.

In order to fulfill this latter condition it must be true that at $A_1 =
0$, and in the large $R_1$ limit, the field theory reproduces
the $O(M)$ quantum mechanics described at the beginning of this section
(and a similar condition near the other boundary).
We should find $16$ $\chi$ zero modes near each wall.  {\it Thus,
the theory must contain a sector in which
the $32$ $1+1$ dimensional $\chi$ fields are grouped
in groups of $16$ with opposite periodicity}. Half of the fields will
supply the required zero modes near each of the walls.  Of course, the
question of which fields have periodic and which antiperiodic boundary
conditions is a choice of $O(M)$ gauge.  However, in any gauge
only half of the $\chi$ fields will have zero modes located at any
given wall.   We could of course consider sectors of the
fundamental $O(M)$ gauge theory in which there is a different
correlation between boundary conditions of the $\chi$ fields.
However, these would not have an eleven dimensional interpretation at
large $R_1$.  The different sectors are not mixed by the Hamiltonian so
we may as well ignore them.

To summarize, we propose that M-theory compactified on $S^1 / \IZ_2$ is
described by a
$1+1$ dimensional $O(M)$ gauge theory with $(0,8)$ SUSY.  Apart from the
$(A_{\mu}, \lambda )$ gauge multiplet, it contains
a right moving $X^i, \theta$ supermultiplet in the symmetric tensor
of $O(M)$ and 32 left moving fermions, $\chi$, in the vector.  The
allowed gauge bundles for $\chi$ (which transforms under the discrete
reflection which leaves all other multiplets invariant), are those in
which two groups of $16$ fields have opposite periodicities.
In the next section we will generalize this construction to
compactifications on general tori.

First let us see how heterotic strings emerge from this formalism in the
limit of small $R_1$.  It is obvious that in this limit, the string
tension term in the SYM Lagrangian becomes very small.  Let us rescale
our $X^i$ and time variables so that the quadratic part of the
Lagrangian is independent of $R_1$.  Then, as in \cite{bs},
\cite{motlb}, \cite{dvv},
the commutator term involving the $X^i$ gets a coefficient
$R^{-3}$ so that we are forced onto the moduli space in that limit.  In
this $O(M)$ system, this means that the $X^i$ matrices are diagonal,
and the gauge group is completely broken
to a semidirect product of $\IZ_2$ (or
$O(1)$) subgroups which reflect the individual components of the
vector representation, and an $S_M$ which permutes the eigenvalues of
the $X^i$.  The moduli space of low
energy
fields\footnote{We use the term moduli space to refer to the space of
low energy fields whose effective theory describes the small $R_1$
limit (or to the target space of this effective theory).
These fields are in a Kosterlitz-Thouless phase and do not have
expectation values, but the term moduli space is a convenient shorthand
for this subspace of the full space of degrees of freedom.}
consists of diagonal $X^i$ fields, their superpartners $\theta_a$ (also
diagonal matrices), and the $32$ massless left moving $\chi$ fields.
The gauge bosons and their superpartners $\lambda^{\dot{\alpha}}$
decouple in the small $R_1$ limit.  All of the $\chi$ fields are light
in this limit.

\subsection{Screwing heterotic strings}

As first explained in \cite{motlb} and elaborated in
\cite{bs}, and \cite{dvv},
twisted sectors under $S_N$ lead to strings of arbitrary
length\footnote{These observations are mathematically identical to
considerations that arose in the counting of BPS-states in black hole
physics \cite{BPS}.}.
The
strings of conventional string theory, carrying continuous values of the
longitudinal momentum, are obtained by taking $N$ to infinity
and concentrating on
cycles whose length is a finite fraction of $N$.
The new feature which arises in the heterotic string is that the
boundary conditions of the $\chi$ fields can be twisted by the discrete
group of reflections.

A string configuration of length $2\pi k$, $X_S^i (s)$, $0 \leq s \leq
2\pi k$, is represented by a diagonal matrix:

 \eqn{screw}{
X^i(\sigma)=\tb{{cccc}
X_S^i(\sigma)&&&\\
&X_S^i(\sigma+2\pi)&&\\
&&\ddots&\\
&&&X_S^i(\sigma+2\pi(N-1))}.}
This satisfies the twisted boundary condition
$X^i(\sigma+2\pi)=E_O^{-1}X^i(\sigma) E_O$ with
\eqn{bcx}{
E_O=\tb{{ccccc}
&&&&\epsilon_k\\
\epsilon_1&&&&\\
&\epsilon_2&&&\\
&&\ddots&&\\
&&&\epsilon_{N-1}&},}
and $\epsilon_i = \pm 1$.
The latter represent the $O(1)^k$ transformations,
which of course do not effect $X^i$ at all.

To describe the possible twisted sectors of the matter fermions we
introduce the matrix $r^a_b = diag (1\ldots 1 ,-1 \ldots -1)$, which
acts on the $32$ valued index of the $\chi$ fields.
The sectors are then defined by
\eqn{chibc}{\chi^a (\sigma + 2\pi ) = r^a_b E_O^{-1} \chi^b
(\sigma )}

As usual, inequivalent sectors
correspond to conjugacy classes of the gauge group.
In this case, the classes can be described by a permutation
with a given set of cycle
lengths, corresponding to a collection of
strings with fixed longitudinal momentum fractions,
and the determinants of the $O(1)^k$ matrices inside each cycle.
In order to understand the various gauge bundles,
it is convenient to write the ``screwing
formulae'' which express the
components of the vectors $\chi^a$ in terms of string
fields $\chi_s^a$ defined on the interval $[0, 2\pi k]$.
The defining boundary conditions are
\eqn{bcchi}{\chi_i^a (\sigma + 2\pi ) =
\epsilon_i r^a_b \chi_{i + 1}^b (\sigma )}
where we choose the gauge in which $\epsilon_{i < k} =1$
and $\epsilon_k = \pm 1$ depending
on the sign of the determinant.  The vector index $i$ is counted modulo $k$.
This condition is solved by
\eqn{soln}{\chi_i^a (\sigma ) = (r^{i - 1})^a_b \chi_S^b
(\sigma + 2\pi (i - 1))}
where $\chi_S$ satisfies
\eqn{hetbc}{\chi_S^a (\sigma + 2\pi k)
= (r^k)^a_b \epsilon_k \chi_S^b (\sigma )}
For $k$ even, this gives the PP and AA sectors
of the heterotic string, according to the
sign of the determinant.
Similarly, for $k$ odd, we obtain the AP and PA sectors. As we explained
in a paper with Susskind \cite{sussmotl}, the $E_8$ symmetry is broken to
$SO(16)$ by a longitudinal Wilson line around the light-like direction
that is equivalent to a $2\pi$
rotation in $SO(16)$: therefore the states containing an odd number of
spinors of $SO(16)$ appear in the models with odd $k$. Recall that
${\bf 248}$
of $E_8$ decomposes as the adjoint ${\bf 120}$ plus a chiral spinor 
${\bf 128}$ under $SO(16)$.

As usual in a gauge theory,
we must project on gauge invariant states.  It turns out that
there are only two independent kinds of
conditions which must be imposed.  In a sector characterized by a
permutation $S$, one can be chosen to be
the overall multiplication of $\chi$ fields associated with a given
cycle of the permutation (a given string)
by $-1$.
This GSO operator anticommuting with all the 32 $\chi$ fields
is represented by the ${\bf -1}$ matrix from the gauge group $O(N)$.
The other is the projection associated with the cyclic
permutations themselves.
It is easy to verify that under the latter transformation
the $\chi_S$ fields transform as
\eqn{chitransf}{\chi_S^a (\sigma )
\rightarrow r^a_b \chi_S^b (\sigma + 2\pi)}
Here $\sigma \in [0,2\pi k]$ and we are
taking the limit $M\rightarrow \infty$,
$k/M$ fixed.  In this limit the $2\pi$
shift in argument on the righthand side of
(\ref{chitransf}) is negligible,
and we obtain the second GSO projection of the heterotic string.

Thus, $1+1$ dimensional $O(M)$
SYM theory with $(0,8)$ SUSY, a left moving
supermultiplet in the symmetric
tensor representation and 32 right moving fermion
multiplets in the vector (half
with P and half with A boundary conditions)
reduces in the weak coupling, small (dual) circle limit to two copies of
the Ho\v rava-Witten domain wall
quantum mechanics, and in the strong coupling large (dual)
circle limit, to the string field theory of the $E_8 \times E_8$
heterotic string.

\section{Multidimensional cylinders}

The new feature of heterotic compactification on $S^1 /\IZ_2 \times T^d$
is that the coordinates in the toroidal dimensions are represented by
covariant derivative operators with respect to new world volume
coordinates.  We will reserve $\sigma$ for the periodic coordinate dual
to the interval $S^1 / \IZ_2$ and denote the other coordinates by
$\sigma^A$.   Then,
\eqn{covder}{X^A = {2\pi R_A \over i}{\partial
\over \partial \sigma^A} - A_A
(\sigma ); \quad A = 2\ldots k+1.}
Derivative operators are antisymmetric, so in order to implement the
orbifold projection, we have to include the transformation $\sigma^A
\rightarrow  - \sigma^A$, for $A = 2\ldots d + 1$,
 in the definition of the orbifold symmetry.
Thus, the space on which SYM is compactified is $S^1 \times (T^d /
\IZ_2)$.   There are $2^d$ {\it orbifold circles} in this space, which are
the fixed manifolds of the reflection.
Away from these singular loci, the gauge group
is $U(M)$ but it will be restricted to $O(M)$ at the singularities.
We will argue that there must be
a number of $1 + 1$ dimensional fermions living only on these circles.
When $d = 1$ these orbifold lines can be thought of as the boundaries of
a {\it dual cylinder}.
Note that if we take $d = 1$ and rescale the $\sigma^A$ coordinates
so that their lengths are $1/R_A$ then a long thin cylinder in spacetime
maps into a long thin cylinder on the world volume, and a short fat
cylinder maps into a short fat cylinder.  As we will see, this
geometrical fact is responsible for the emergence of Type IA
and heterotic strings in the appropriate limits.

The boundary conditions on the world volume fields are
\eqn{bca}{X^i (\sigma ,\sigma^A ) = \bar X^i (\sigma ,  - \sigma^A ),\quad
A_a (\sigma ,\sigma^A ) = \bar A_a (\sigma ,  - \sigma^A ),}
\eqn{bcc}{A_1 (\sigma ,\sigma^A ) = -\bar A_1 (\sigma ,  - \sigma^A)}
\eqn{bcd}{\theta (\sigma ,\sigma^A ) = \bar\theta(\sigma ,-\sigma^A),\quad
\lambda (\sigma ,\sigma^A ) = -\bar\lambda (\sigma ,-\sigma^A )}
All matrices are Hermitean,
so transposition is equivalent to complex conjugation.
The right hand side of the boundary condition (\ref{bcc}) can also be
shifted
by $2\pi R_1$, reflecting the fact that $A_1$ is an angle variable.

Let us concentrate on the cylinder case, $d=1$.
In the limits $R_1 \ll l_{11} \ll R_2$ and
 $R_2 \ll l_{11} \ll R_1$, we will find that
the low energy dynamics is completely
described in terms of the moduli space,
which consists of commuting $X^i$ fields.
In the first of these limits,
low energy fields have no $\sigma^2$ dependence, and
the boundary conditions restrict
the gauge group to be $O(M)$, and force
$X^i$ and $\theta$ to be real symmetric matrices.
Anomaly arguments
then inform us of the existence of $32$
fermions living on the boundary circles.
The model reduces to the $E_8 \times E_8$
heterotic matrix model described in the previous
section, which, in the indicated limit,
was shown to be the free string field theory of
heterotic strings.

\subsection{Type IA strings}

The alternate limit produces something novel.
Now, low energy fields are restricted
to be functions only of $\sigma^2$.
Let us begin with a description of closed strings.
We will exhibit a solution of the boundary conditions for each
closed string field $X_S (\sigma )$ with periodicity $2\pi k$.  Multiple
closed strings are constructed via the usual block diagonal procedure.

\eqn{uopen}{
X^i(\sigma^2)=U(\sigma^2)
D
U^{-1}(\sigma^2),}
\eqn{diagstring}{\begin{array}{c}D=\mbox{diag}(X_s^i(\sigma^2),\epsilon
X_s^i(2\pi-\sigma^2),X_s^i(2\pi+\sigma^2),
\epsilon X_s^i(4\pi-\sigma^2),\\
\dots,
X_s^i(2\pi(N-1)+\sigma^2),\epsilon X_s^i(2\pi N-\sigma^2)).
\end{array}}
where $\epsilon$ is $+1$ for $X^{2\dots 9}$ and $\theta$'s,
$-1$ for $A^1$ and $\lambda$'s.
{}From this form it is clear that the matrices will commute
with each other for any value of $\sigma^2$.
We must obey Neumann boundary conditions
for the real part of matrices and Dirichlet conditions
for the imaginary parts (or for $\epsilon=-1$ vice versa),
so we must use specific values of the unitary matrix
$U(\sigma^2)$ at the points $\sigma^2=0,\pi$.
Let us choose
\eq{U'(0+)=U'(\pi-)=0}
(for instance, put $U$ constant on a neighbourhood
of the points $\sigma^2=0,\pi$)
and for a closed string,
\eq{U(\pi)=\tb{{cccc}
m&&&\\
&m&&\\
&&\ddots &\\
&&&m},\qquad
U(0)=C\cdot
U(\pi)\cdot C^{-1},}
where $C$ is a cyclic permutation matrix
\eq{C=\tb{{ccccc}
&1&&&\\
&&1&&\\
&&&\ddots&\\
&&&&1\\
1&&&&}}
where $m$ are $2\times 2$ blocks (there are $N$ of them)
(while in the second matrix the $1$'s are $1\times 1$ matrices so that
we have a shift of the $U(\pi)$ along the diagonal
by half the size of the Pauli matrices.
The form of these blocks guarantees the conversion of $\tau_3$
to $\tau_2$:
\eq{m=\frac{\tau_2+\tau_3}{\sqrt 2}.}
This $2\times 2$ matrix causes two ends to be connected
on the boundary. It is easy to check that the right
boundary conditions will be obeyed.

To obtain open strings, we just change the
$U(0)$ and $U(\pi )$.
An open string of odd length is obtained by
throwing out the last element in (\ref{diagstring}) and taking
\eq{
U(0)=\tb{{ccccc}
1_{1\times 1}&&&&\\
&m&&&\\
&&m&&\\
&&&\dots &\\
&&&&m
},\quad\,
U(\pi)=\tb{{ccccc}
m&&&\\
&m&&\\
&&m&\\
&&&\dots&\\
&&&&1_{1\times 1}}}
Similarly, an open string of even length  will have one of the matrices
$U(0),U(\pi)$ equal to what it was in the closed string case $m\otimes
1$ while
the other will be equal to
\eq{
U(0)={\small\tb{{cccccc}
1_{1\times 1}&&&&&\\
&m&&&&\\
&&m&&&\\
&&&\dots &&\\
&&&&m&\\
&&&&&1_{1\times 1}
}}}

Similar constructions for the fermionic coordinates are straightforward
to obtain.  We also note that we have worked above with the original
boundary conditions and thus obtain only open strings whose ends are
attached to the wall at $R_1 = 0$.  Shifting the boundary condition
(\ref{bcc}) by $2\pi R_1$ (either at $\sigma^2 =0$ or $\sigma^2 = \pi$ or
both) we obtain strings attached to the other wall, or with one end on
each wall.  Finally, we note that we can perform the gauge
transformation $M \rightarrow \tau_3 M \tau_3$ on our construction.
This has the effect of reversing the orientation of  the string fields,
$X_S (\sigma^2 ) \rightarrow X_S (- \sigma^2 )$.  Thus we obtain
unoriented strings.

We will end this section with a brief comment about moving D8-branes
away from the orientifold wall.  This is achieved by adding explicit
$SO(16) \times SO(16)$ Wilson lines to the Lagrangian of the $\chi^a$
fields.  We are working in the regime $R_2 \ll l_{11} \ll R_1$, and we
take these to be constant gauge potentials of the form $\chi^a {\cal
A}_{ab} \chi^b$, with ${\cal A}$ of order $R_1$.  In the presence of
such terms $\chi^a$ will not have any low frequency modes, unless we
also shift the $O(M)$ gauge potential $A_1$ to give a compensating shift
of the $\chi$ frequency.  In this way we can construct open strings
whose ends lie on D8-branes which are not sitting on the orientifold.

In this construction, it is natural to imagine that $16$ of the $\chi$
fields live on each of the boundaries of the dual cylinder.  Similarly,
for larger values of $d$ it is natural to put ${32 \over 2^d}$ fermions
on each orbifold circle, a prescription which clearly runs into problems
when $d > 4$. This is reminiscent of other orbifold constructions in M
theory in which the most symmetrical treatment of fixed points is not
possible (but here our orbifold is in the dual world volume).  It is
clear that our understanding of the heterotic matrix model for general
$d$ is as yet quite incomplete.  We hope to return to it in a future
paper.

\section{Conclusions}

We have described a class of matrix field theories which incorporate the
Fock spaces of the the $E_8 \times E_8$
heterotic/Type IA string field theories
into a unified quantum theory.  The underlying gauge dynamics provides a
prescription for string interactions.  It is natural to ask what the
connection is between this nonperturbatively defined system and previous
descriptions of the nonperturbative dynamics of string theories with
twice as much supersymmetry.  Can these be viewed as two classes of
vacua of a single theory?  Can all of these be obtained as different
large $N$ limits of a quantum system with a finite number of degrees of
freedom?

The necessity of introducing the $\chi$ fields into our model suggests
that the original eleven dimensional system does not have all the
necessary ingredients to be the underlying theory.  Yet we are used to
thinking of obtaining lower dimensional compactifications by restricting
the degrees of freedom of a higher dimensional theory in various ways.
Insight into this puzzle can be gained by considering the limit of
heterotic string theory which, according to string duality, is supposed
to reproduce M-theory on $K3$.  The latter theory surely reduces
to eleven dimensional M-theory in a continuous manner as the radius of
$K3$ is taken to infinity.
Although we have not yet worked out the details of
heterotic matrix theory on higher dimensional tori, we think that it
is clear that the infinite $K3$
limit will be one in which the $\chi$ degrees
decouple from low energy dynamics.

The lesson we learn from this example is that {\it decompactification of
space time dimensions leads to a reduction in degrees of freedom}.
Indeed, this principle is clearly evident in the prescription for
compactification of M-theory on tori in terms of SYM theory.  The more
dimensions we compactify, the higher the dimension of the field theory we
need to describe the compactification.  There has been some discussion
of whether this really corresponds to adding degrees of freedom since
the requisite fields arise as limits of finite matrices.  However there
is a way of stating the principle which is independent of how one
chooses to view these constructions. Consider, for example, a graviton
state in M-theory compactified on a circle.  Choose a reference energy
$E$ and ask how the number of degrees of freedom with energy less than
$E$ which are necessary to describe this state, changes with the radius
of compactification.  As the radius is increased, the radius of the dual
torus decreases.  This decreases the number of states in the theory with
energy less than $E$, precisely the opposite of what occurs when we
increase the radius of compactification of a local field theory

\subsection{Cosmological constant problem}

It seems natural to speculate that this property, so counterintuitive
from the point of view of local field theory, has something to do with
the cosmological constant problem.  In \cite{tbcosmo}
one of the authors
suggested that any theory which satisfied the 't Hooft-Susskind
holographic principle
would suffer a thinning out of degrees
of freedom as the universe expanded, and that this would lead to an
explanation of the cosmological constant problem.  Although the
speculations there did not quite hit the mark, the present ideas suggest
a similar mechanism.  Consider a hypothetical state of the matrix model
corresponding to a universe with some number of Planck size dimensions
and some other dimensions of a much larger size, $R$.  Suppose also that
SUSY is broken at scale $B$, much less than
the (eleven dimensional) Planck scale.
The degrees of freedom
associated with the compactified dimensions all have energies
much higher than the
SUSY breaking scale. Their zero point fluctuations will lead
to a finite, small (relative to the Planck mass) $R$ independent,
contribution to the total vacuum
energy.
As $R$ increases, the number of degrees of freedom at scales less than
or equal to $B$ will decrease.
Thus, we expect a corresponding decrease in the
total vacuum energy.
The total vacuum energy in the large $R$ limit is thus bounded
by a constant, and is dominated by the
contribution of degrees of freedom associated with
the small, compactified dimensions.
Assuming only the minimal supersymmetric cancellation
in the computation of the vacuum energy,
we expect it to be of order $B^2 l_{11}$.  This
implies a vacuum energy density of order
$B^2 l_{11} / R^3$, which is too small to be of
observational interest for any plausible values of the parameters.
If a calculation of this nature turns
out to be correct, it would constitute a prediction
that the cosmological constant is essentially zero in the matrix model.

It should not be necessary to emphasize how premature
it is to indulge in speculations
of this sort (but we couldn't resist the temptation).
We do not understand supersymmetry
breaking in the matrix model and we are even
further from understanding its cosmology.
Indeed, at the moment we do not even
have a matrix model derivation of the fact\footnote{Indeed
this ``fact'' is derived by rather indirect arguments in perturbative
string theory.} that
parameters like the radius of compactification are dynamical variables.
Perhaps the
most important lacuna in our understanding
is related to the nature of the large $N$
limit.  We know that many states of the system
wander off to infinite energy as $N$
is increased.
Our discussion above was based on extrapolating results of the finite $N$
models, without carefully verifying that the
degrees of freedom involved survive
the limit.  Another disturbing thing about our discussion is the absence
of a connection to Bekenstein's area law for the number of states.  The
Bekenstein law
seems to be an integral part of the physical picture of the matrix model.
Despite these obvious problems, we feel
that it was worthwhile to present this preliminary discussion of
the cosmological constant problem
because it makes clear that the spacetime picture
which will eventually emerge from the matrix model is certain to be
very different from the one implicit in local field theory.

%% file: chap4-knit.tex
\chapter{Knitted fivebranes in the (2,0) theory}

In this chapter which results from a collaboration with Ori Ganor
\cite{GANORMOTL}
we study non-linear corrections to the low-energy
description of the (2,0) theory. We argue for the existence of a
topological correction term similar to the $C_3\wedge X_8(R)$ in M-theory.
This term can be traced to a classical effect in supergravity and to a
one-loop diagram of the effective 4+1D Super Yang-Mills. We study other
terms which are related to it by supersymmetry and discuss the
requirements on the subleading correction terms from M(atrix) theory. We
also speculate on a possible fundamental formulation of the theory.

\section{Introduction}
During the last four years a lot of attention has been devoted to the
newly discovered 5+1D theories \cite{somecom}.
The version of these theories with (2,0) supersymmetry
arises as a low-energy description of type-II B on an $A_{N-1}$
singularity \cite{somecom} or as the dual low-energy
description of $N$ coincident 5-branes in M-theory \cite{openp}.
Part of the attention
\cite{rozali,berozaliseiberg,berozali}
is due to the r\^ole they play in
compactified M(atrix) theory \cite{bfss}, part is because they provide
testing grounds to M(atrix) theory ideas
\cite{prem,WitQHB,Lowe,KSAB,ABS},
and another part is because they
shed light on non-perturbative phenomena in 3+1D gauge theories
\cite{somecom}.
These theories are also very exciting on their own right. They
lack any parameter which will allow a classical perturbative expansion
(like the coupling constant of SYM).
Thus, these theories have no classical limit (for finite $N$).
The only possible classical expansion is a derivative expansion
where the energy is the small parameter.

One of our goals will be to explore the low-energy description
of the (2,0) theory.
At low energies, and a generic point in moduli space the zeroth order
approximation is $N$ free tensor multiplets which contain the
chiral \asd 2-forms.
Since the theory contains chiral 2-forms it is more convenient
to write down the low-energy equations of motion rather
than the non-manifestly covariant Lagrangian (there is the
other option of using the manifestly covariant formulation
of \cite{APS,APPS}, but using the equations of motion
will be sufficient for our purposes). These equations
are to be interpreted \`a la Wilson, i.e. as quantum equations
for operators but with a certain unspecified UV cutoff.
The leading terms in the Wilsonian low-energy description
are the linear equations of motion for the $N$ free tensor
multiplets. We will be looking for the first sub-leading corrections.
Those corrections will be non-linear and are a consequence of the
interacting nature of the full (2,0) theory.
In general at high enough order in the derivative expansion
the terms in the Wilsonian action are cutoff dependent.
However, we will see that the first order corrections
are independent of the cutoff.
We will argue that the low-energy equations contain a topological
term somewhat analogous to the subleading $C_3\wdg X_8(R)$ term of
M-theory \cite{VWSCT,DLM}
and which describes a topological correction term
to the \asd string current.
We will then study the implications of supersymmetry.

The chapter is organized as follows.
Section \ref{s2} is a review of the (2,0) theory.
In section \ref{s3} we derive the topological term from the supergravity
limit of $N$ 5-branes of M-theory. Our discussion will be an
implementation
of results described in \cite{WTHR}.
In section \ref{s4}
we discuss the implied correction
terms after compactification to 3+1D, and
we find related terms which are implied by supersymmetry.
In section \ref{s5} we discuss the currents in 5+1D.
Finally, in sections \ref{s6}, \ref{s7} we speculate on  a possible
``deeper''
meaning of these correction terms.



\section{Review of the (2,0) theory}
\label{s2}

This section is a short review of some facts we will need
about the (2,0) theory.

\subsection{Realization}

The $(2,0)_N$ theory is realized either as the low-energy decoupled
degrees of freedom from an $A_{N-1}$ singularity (for $N\ge 2$)
of type IIB \cite{somecom}
or from the low-energy decoupled degrees of freedom of $N$ 5-branes
of M-theory \cite{openp}.
This is a conformal 5+1D theory which  is interacting for $N>1$.
It has a chiral (2,0) supersymmetry with 16 generators.  
One can deform the theory away from the conformal point.
This corresponds to separating the $N$ 5-branes (or blowing
up the $A_{N-1}$ singularity). If the separation scale $x$
is much smaller than the 11D Planck length $M_p^{-1}$ then at energies
$E\sim M_p^{3/2} x^{1/2}$ one finds a massive decoupled theory
whose low-energy description is given by $N$ free tensor multiplets.  

Each free tensor multiplet in 5+1D comprises of 5 scalar fields
$\PH^A$ with $A=1\dots 5$, one tensor field $B_{\u\nu}^{(-)}$ where
the $(-)$ indicates that its equations of motion force it to be
\asdW, and 4 multiplets of chiral fermions $\TH$.
The (2,0) supersymmetry in 5+1D has $Sp(2) = Spin(5)_R$ R-symmetry.
The scalars $\PH^A$ are in the $\rep{5}$ whereas the fermions are
in the $(\rep{4},\rep{4})$ of $SO(5,1)\times Sp(2)$ but with
a reality condition. Thus there are 16 real fields in $\TH$.

For the low-energy of the $(2,0)_N$ theory there are $N$ such tensor
multiplets. The moduli space, however, is not just $(\MR{5})^N$
because there are discrete identifications given by the permutation
group. It is in fact $(\MR{5})^N/S_N$.
Let us discuss what happens for $N=2$. The moduli space can be written
as $\MR{5}\times (\MR{5}/\IZ_2)$. The first $\MR{5}$ is the
sum of the two tensor multiplets. In 5+1D this sum is described
by a free tensor multiplet which decouples from the rest of the theory
(although after compactification, it has some global effects which
do not decouple).
The remaining $\MR{5}/\IZ_2$ is the difference of the two tensor
multiplets. This moduli space has a singularity at the origin
where the low-energy description is no longer two free tensor
multiplets but is the full conformal theory.

\subsection{Equations of motion for a free tensor multiplet}
To write down the lowest order equations of motion for a free
tensor multiplet we use the field strength
$$
H_{\a\b\g} = 3\px{[\a}B_{\b\g]}^{(-)}\,.
$$
This equation does not imply that $H$ is \asd but
does imply that $H$ is a closed form. It is possible to modify
this equation such that $H$ will be manifestly \asdW.
We will define $H$ to be \asd part of $dB$ according to,
\eqn{hfromb}{H_{\a\b\g}=
\frac 32(\partial_{[\a}B_{\b\g]})
-\frac 14\epsilon_{\a\b\g}{}^{\a'\b'\g'}(\partial_{\a'}B_{\b'\g'})\,.}
This definition is the same as the previous one for \asd
$dB$, it trivially implies that $H$ is \asd and it
does not lead to the equation $dH=0$ which we will find useful later on.
In any case, we will use the equations of motion for $H$ only
and $B$ will therefore not appear.
For the fermions it is convenient to use 11D Dirac matrices
$$
\Ga^\u,\, \u=0\dots 5\,,\qquad
\Ga^A,\, A=6\dots 10\,,
$$
with commutation relations
$$
\acom{\Ga^\u}{\Ga^\nu} = 2 \eta^{\u\nu}\,,\qquad
\acom{\Ga^A}{\Ga^B} = 2\delta^{AB}\,,\qquad
\acom{\Ga^A}{\Ga^\u} = 0\,.
$$
We define
$$
\GPR = \Ga^{012345} = \Ga^0\Ga^1\cdots\Ga^5 = \Ga^6\Ga^7\cdots\Ga^{10}\,.
$$
The spinors have positive chirality and satisfy
$$
\TH = \GPR\TH\,.
$$
The $Spin(5)_R$ acts on $\Ga^A$ while $SO(5,1)$ acts on $\Ga^{\u}$.
The free equations of motion are given by
\begin{eqnarray}
H^{\u\nu\s} &=& {1\over 6}{\epsilon_{\t\r\g}}^{\u\nu\s} H^{\t\r\g}
\equiv - \frac16\epsilon^{\u\nu\s}{}_{\t\r\g}H^{\t\r\g}\,,\\
\partial_{\lbr\t}{H_{\u\nu\s\rbr}} &=& 0,\\
\Box \PH^A &=& 0\,,\\
\dslash \TH &=& 0\,.
\end{eqnarray}
The supersymmetry variation is given by
\begin{eqnarray}
\delta H_{\a\b\g} &=& -\frac{i}{2} \bar\epsilon
                       \Ga_\d\Ga_{\a\b\g}\partial^\d\TH \label{strengvarr}
\\
\delta \PH_A      &=& -i \bar\epsilon\Ga_A\TH           \label{phivarr}\\
\delta \TH        &=& (\frac{1}{12} H_{\a\b\g}
                 \Ga^{\a\b\g}+ \Ga^\a\partial_\a
                 \PH_A\Ga^A)\epsilon     \,.            \label{fermivarr}
\end{eqnarray}

The quantization of the theory is slightly tricky.
There is no problem with the  fermions $\TH$ and bosons $\PH^A$,
but the tensor field is self-dual and thus has to be quantized
similarly to a chiral boson in 1+1D.
This means that we second-quantize a free tensor field  
without any self-duality constraints and then set to zero
all the oscillators with self-dual polarizations.
The action that we use in 5+1D is:
$$
\action = -{1\over {4\pi}}\int \left\{
 \px{\u}\Phi^A \partial^\u\Phi_A + {3\over 2}\px{[\u}B_{\s\t]}
\partial^{[\u}B^{\s\t]}
 + i
\bTH
\dslash\TH
 \right\}d^6\s\,.
$$
Here we have defined $\bTH=\TH^T\Ga^0$.
The normalization is such that integrals of $B_{\s\t}$ over closed
2-cycles
live on circles of circumference $2\pi$.
In appendix A we list some more useful formulae.

\section{Low-energy correction terms -- derivation from SUGRA}
\label{s3}

In this section we will derive a correction term to the zeroth order
low-energy terms.

Let us consider two 5-branes in M-theory.
Let their center of mass be fixed.
The fluctuations of the center
of mass are described by a free tensor multiplet.
Let us assume that the distance between the 5-branes at infinity
$|M_p^{-2}\PH_0|$ is much larger than the 10+1D
Planck length $M_p^{-1}$ and let us consider
the low-energy description of the system for energies
$E\ll |\PH_0|$. The description at lowest order is given by
supergravity in the 10+1D bulk and by a 5+1D tensor multiplet
with moduli space $\MR{5}/\IZ_2$ (we neglect the free tensor
multiplet coming from the overall center of mass).
The lowest order equations of motion for the tensor multiplet are
the same linear equations as described in the previous section.
We would like to ask what are the
leading nonlinear corrections to the linear equations.

We will now argue that according to the arguments given in \cite{WTHR}
there is a topological contribution to the $dH$ equation of
motion (here $\PH^{(ij)} \equiv \PH^{(i)} - \PH^{(j)}$)
\begin{equation}\label{eqh}
\px{[\a}H_{\b\g\d]}^{(i)} =
\sum_{j=1\dots N}^{(j\ne i)}
{3\epu{ABCDE}\over {16\pi |\PH^{(ij)}|^5}}
\PH^{E,(ij)}
\px{[\a}\PH^{A,(ij)}\px{\b}\PH^{B,(ij)}\px{\g}\PH^{C,(ij)}\px{\d]}
\PH^{D,(ij)}\,.
\end{equation}
Here $A\dots E = 1\dots 5$. $\PH^A$ are the scalars of
the tensor multiplet and $H_{\a\b\g}$ is the \asd field
strength.
Note that the RHS can be written as a pullback $\pi^{*}\omega_4$
of a closed form on the
moduli space which is
$$
\Modsp \equiv \MR{5}/\IZ_2 - \{0\}\,.
$$
Here 
$$
\pi: \IR^{5,1}\longrightarrow \Modsp = \MR{5}/\IZ_2 - \{0\}
$$
is the map $\PH^A$ from space-time to the moduli space
and
$$
\omega_4 = {3\over {8\pi^2 |\PH|^5}}
\epu{ABCDE}\PH^E  d\PH^A\wdg d\PH^B\wdg d\PH^C\wdg d\PH^D
$$
is half an integral form in $H_4(\frac 12\IZ)$, i.e.
$$
\int_{S^4/Z_2} \omega_4 = {1\over 2}\,.
$$

Let us explain how (\ref{eqh}) arises.
When $\PH^A$ changes smoothly and slowly, the supergravity picture
is that each 5-brane ``wraps'' the other one.
Each 5-brane is a source for the (dual of the) $F_4 = dC_3$ 4-form
field-strength of 10+1D supergravity. When integrated on a sphere $S^4$
surrounding the 5-brane we get $\int_{S^4} F_4 = 2\pi$.
The other 5-brane now feels an effective $C_3$ flux on its world-volume.
This, in turn, is a source for the 3-form \asd low-energy  
field-strength $dH = dC_3$. It follows that the total string charge
measured
at infinity of the $\IR^{5,1}$ world-volume of one 5-brane is,
$$
\int dH = \int dC_3 = \int F_4\,.
$$
The integrals here are on $\IR^4$ which is a subspace of $\IR^{5,1}$
and they measure how much effective string charge passes through
that $\IR^4$. The integral on the RHS can now be calculated.
It is the 4D-angle subtended by the $\IR^4$ relative to the second
5-brane which was the source of the $F_4$. But this angle can be expressed
solely in terms of $\PH^A$ and the result is the integral over $\omega_4$.


These equations can easily be generalized to $N$ 5-branes.
We have to supplement each field with an index $i=1\dots N$.
We can also argue that there is a correction
\begin{equation}\label{eqphi}
\Box\PH^{D,(i)} =
-\sum_{j=1\dots N}^{(j\ne i)}
{{\epu{ABCDE}}\over {32\pi |\PH^{(ij)}|^5}}
\PH^{E,(ij)}\px{\a}\PH^{A,(ij)}\px{\b}\PH^{B,(ij)}\px{\g}\PH^{C,(ij)}
H^{\a\b\g,(ij)} + \cdots
\end{equation}
Here $\PH^{(ij)} \equiv \PH^{(i)} - \PH^{(j)}$ and similarly
$H^{(ij)} = H^{(i)} - H^{(j)}$.
The term $(\cdots)$ contains fermions and other contributions.
  
The equation
(\ref{eqphi}) for $\Box \PH$ can be understood as the equation for force
between a tilted fivebrane and another fivebrane which carries
an $H_{\alpha\beta\gamma}$ flux.
As far as BPS charges go, the $H$ flux inside a 5-brane
is identified in M-theory with a membrane flux.
This means that (after compactification)
as a result of a scattering of a membrane on a 5-brane   
an $H$-flux can be created and the membrane can be annihilated.
The identification of the $H$-flux with the membrane charge
is also what allows a membrane to end on a 5-brane \cite{openp}.
Consistency implies that a 5-brane with an $H$ flux should
exert the same force on other objects as a 5-brane and a membrane.
This is indeed the case, as follows from the $C_3\wdg H$ interaction
on the 5-brane world-volume \cite{openp}.

The Lorentz force acting on a point like particle equals (in its rest
frame)
  \eqn{lorfor}{m\frac{d^2}{dt^2} x^i=e\cdot F^{0i}\,.}
As a generalization for a force acting on the fivebrane because of the
flux $H$ in the other 5-brane, we can replace $d^2/{dt^2}$ by
$\Box$ and write
  \eqn{lorfiv}{\Box \PH_A = F_{A\a\b\g}H^{\a\b\g\,}.}
But we must calculate the four-form supergravity field strength at the
given point. Only components with one Latin index and three Greek indices
are important. We note that the electric field strength in the real
physical 3+1-dimensional electrostatics is proportional to
  \eqn{elstatf}{F_{0A}\propto \frac{r_A}{r^3}\propto \frac 1{r^2}\,.}
The power 3 denotes 3 transverse directions, $F$ contains all the indices
in which the ``worldvolume'' of the particle is stretched. As an
analogue for fivebrane stretched exactly in $012345$ directions,  
\eqn{statf}{*F_{012345A}\propto \frac{\PH^{(ij)}_A}{|\PH_{(ij)}|^5}
\propto \frac 1{|\PH_{(ij)}|^4}\,.}
We wrote star because we interpret the fivebrane as the ``magnetic''
source. $F$ in (\ref{lorfiv}) has one Latin index and three Greek
indices, so its Hodge dual has four Latin indices and three Greek indices.
$*F$ in (\ref{statf}) contains only one Latin index but   
when the 5-branes are tilted by infinitesimal angles
$\partial_\g\PH_C$ we get also a contribution
to the desired component of $F$:
  \eqn{desired}{*F_{\a\b\g ABCD}=
*F_{\a\b\g\d\s\t D}
\partial^\d \PH^{(ij)}_A
\partial^\s \PH^{(ij)}_B
\partial^\t \PH^{(ij)}_C\,.}
Now if we substitute (\ref{statf}) to (\ref{desired}) and the result
insert to (\ref{lorfiv}), we get the desired form of the $\Box \PH$
equations.


Similarly, there is an equation for $\TH$,
\bear
\dslash
\TH^i &\propto&
\sum_{j=1\dots N}^{(j\neq i)}   
\frac{\epsilon^{ABCDE}}{\abs{\PH^{(ij)}}^5}
(\PH^{(ij})_E
\partial_\a(\PH^{(ij)})_A
\partial_\b(\PH^{(ij)})_B
\partial_{\g}(\PH^{(ij)})_C
\Ga^{\a\b\g}\Ga_D\TH^{(ij)}\,.
\label{eqtheta}
\eear


Our goal in this chapter is to deduce the corrections in the derivative
expansion in the low-energy of the (2,0) theory.
We cannot automatically deduce that (\ref{eqh}), (\ref{eqphi})
and (\ref{eqtheta})
can be extrapolated to the (2,0) theory because this description
is valid only in the opposite limit, when $|\PH|\ll M_p$, and supergravity
is not a good approximation.
However, the RHS of (\ref{eqh}) is a closed 4-form on
the moduli space $\Modsp = \MR{5}/\IZ_2 - \{0\}$
which is also half integral, i.e. in
$H_4(\Modsp,{1\over 2}\IZ)$.
It must remain half-integral as we make $|\PH|$
smaller. Otherwise, Dirac quantization will be violated.
(Note that the wrapping number is always even.)
Eqn. (\ref{eqphi}) follows from the same term in the action
as (\ref{eqh}). As for other correction terms,
if we can show that they are implied by (\ref{eqh}) and supersymmetry,
then we can trust them as well.
This will be the subject of the next section.

We would like to point out that this reasoning is somewhat
similar to that of \cite{GG,GGV} who related the $R^4$ terms in 11D
M-theory to the $C\wdg X_8(R)$ term of \cite{VWSCT,DLM}.

\section{Compactification}
\label{s4}

In this section we will study the reduction of the terms to 3+1D by 
compactifying on $\MT{2}$.
Let $\Ar$ be the area of $\MT{2}$ and $\tau$ be its complex structure.
At low-energy in 3+1D we obtain a free vector multiplet of $N=4$ with
coupling constant $\tau$. We are interested in the subleading corrections  
to the Wilsonian action. We will study these corrections as a function of
$\Ar$. Let us first note a few facts (see \cite{SeiSXN} for a detailed
discussion).

When one reduces classically a free tensor multiplet from 5+1D down
to 3+1D one obtains a free vector-multiplet with one photon and
6 scalars. Out of the 6 scalars one is compact. This is the
scalar that was obtained from $B_{45}$. We denote it by $\sigma$.
$$
\sigma = (\Imx\tau)^{-1/2}\Ar^{-1/2}\int_\MT{2} B_{45}\,.
$$
We have normalized its kinetic energy so as to have an $\Imx\tau$
 in front, like  3+1D SYM.
The radius of $\sigma$ is given by
\begin{equation}
\label{sigper}
\sigma\sim \sigma + 2\pi (\Imx\tau)^{-1/2}\Ar^{-1/2}\,.
\end{equation}
In 5+1D there was a $Spin(5)_R$ global symmetry.
$N=4$ SYM has $Spin(6)_R$ global symmetry but the dimensional 
reduction of the (2,0)-theory has only $Spin(5)_R$.
Let us also denote by $\VEV$ the square root of sum of squares  
of the VEV of the 5 scalars other than $\sigma$.

Now let us discuss the interacting theory.
When $\VEV\Ar\ll 1$ we can approximate the 3+1D theory at
energy  scales $E\ll \Ar^{-1}$ by 3+1D SYM. In this case the
$Spin(5)_R$ is enhanced, at low-energy, to $Spin(6)_R$.
 For $\VEV\Ar\gg 1$ the ``dynamics'' of the theory occurs at length
scales well below the area of the $\MT{2}$ where the theory is
effectively (5+1)-dimensional.
The 3+1D low-energy is therefore the classical dimensional 
reduction of the 5+1D low-energy. Thus, from our 3+1D results
below we will be able to read off the 5+1D effective low-energy
in this regime.

\subsection{Dimensional reduction of the correction term}
Let us see what term we expect to see at low-energy in 3+1D.
We take the term,
$$
\px{[\a}H_{\b\g\d]} =
{3\over {16\pi |\PH|^5}}\epd{ABCDE}\PH^E
\px{\a}\PH^A\px{\b}\PH^B\px{\g}\PH^C\px{\d}\PH^D\,,
$$
and substitute $0123$ for $\a\b\g\d$.
The field $H_{\b\g\d}$ is,
$$
H_{\b\g\d} = -\ept{\b\g\d}{\a}H_{\a 45}
= -(\Imx\tau)^{1/2}\Ar^{-1/2}\epd{\b\g\d\a}\qx{\a}\sigma\,.
$$
The equation becomes
$$
\qx{\u}\px{\u}\sigma = -{1\over {32\pi}}
(\Imx\tau)^{-1/2}\Ar^{1/2}
{1\over {|\PH|^5}}\epd{ABCDE}\PH^E
\px{\a}\PH^A\px{\b}\PH^B\px{\g}\PH^C\px{\d}\PH^D
\epsilon^{\a\b\g\d}\,.
$$
Here $\PH^A\dots\PH^E$ are the six-dimensional fields.
The 4-dimensional fields are defined by
\begin{equation}
\label{sixfour}
\PH^A = (\Imx\tau)^{1/2}\Ar^{-1/2}\vph^A\,.
\end{equation}
Thus, the action should contain a piece of the form
\bear
\lefteqn{
{1\over {32\pi}}(\Imx\tau)\int d^4 x\,\,\px{\u}\sigma\qx{\u}\sigma
-}\qquad\nn\\&&
-{1\over {32\pi}}(\Imx\tau)^{1/2}\Ar^{1/2}\epd{ABCDE}
\int d^4 x\,\, {{\sigma}\over {|\vph|^5}} \epu{\a\b\g\d}
\vph^E\px{\a}\vph^A\px{\b}\vph^B\px{\g}\vph^C\px{\d}\vph^D\,. 
\label{sigfd}
\eear
Note that this is the behavior we expect when   
$\VEV\Ar\gg 1$. When $\VEV\Ar\sim 1$ the approximation
of reducing the 5+1D effective action is no longer valid as
explained above.

Let us first see how to write such a term in an $N=1$ superfield
notation.
Let us take three chiral superfields, $\CHP$ and $\CHP^I$
($I=1,2$). We assume that
$$
\CHP = \scVEV + \delta\vph + i\sigma\,.
$$
$\sigma$ is the imaginary part of $\CHP$ and
$\scVEV$ is the VEV of the real part.
Below, the index $I$ of $\CHP^I$ is lowered and raised with the
anti-symmetric $\epd{IJ}$.

\subsection{Interpolation between 3+1D and 5+1D}
In the previous section we assumed that we are in the region
$\VEV A \gg 1$. This was the region where classical
dimensional reduction from 5+1D to 3+1D is a good approximation.
However, the question that we are asking about the low-energy effective
action makes sense for any $A$. For $\VEV A \sim 1$ quantum effects
are strong.
Let us concentrate on another possible 
term which appears in the 5+1D effective
action and behaves like,
\begin{equation}
\label{vfrt}
\int d^6 x{{(\partial\PH)^4}\over {|\PH|^3}}\,.
\end{equation}
This term is of the same order of magnitude as (\ref{sigfd})
and its existence in the 5+1D effective action is suggested
by M(atrix) theory. It would give the correct $v^4/r^3$ behavior
for the potential between far away gravitons in M-theory
compactified on $\MT{4}$. We will also see below how terms
similar in structure to  (\ref{vfrt}) are related to (\ref{sigfd})
by supersymmetry.

After dimensional reduction to 3+1D we obtain a term which
behaves like
\begin{equation}
\label{vfrtdr}
(\Imx\tau)^{1/2}\Ar^{1/2}
\int d^4 x{{(\partial\vph)^4}\over {|\vph|^3}}\,.
\end{equation}
This is valid when $\VEV\Ar \gg 1$. On the other hand, when   
$\VEV\Ar \ll 1$, $N=4$ SYM with a coupling constant
given by the combination $\tau$ is a good approximation, at low enough
energies (around the scale of $\VEV \Ar^{1/2}$).
In SYM, 1-loop effects can produce a term that behaves like (see
\cite{DS}),
\begin{equation}
\label{vsymonlp}
\int d^4 x {{(\partial\vph)^4}\over {|\vph|^4}}\,.
\end{equation}
Note that this term contains no $\tau$, and no $\Ar$.

How can we interpolate between (\ref{sigfd}) and (\ref{vsymonlp})?
  
The answer lies in the periodicity of $\sigma$.
For any value of $\VEV\Ar$ the formula must be periodic
in the 6th scalar $\sigma$, according to (\ref{sigper}).
Thus, we propose to write
\begin{equation}
\int d^4 x{(\partial\vph)^4}
\sum_{k\in\IZ} {1\over
 {[\sum_{A=1}^5|\vph_A|^2
   + (\sigma + 2 k \pi (\Imx\tau)^{-1/2}\Ar^{-1/2})^2]^2}}\,.
\end{equation}
For small $\Ar$ we can keep only the term with $k=0$ and recover
(\ref{vsymonlp}). For large $\Ar$ we have to approximate
the sum by an integral and we obtain
\begin{eqnarray}
\lefteqn{
\sum_{k\in\IZ} {1\over
 {[\sum_{A=1}^5|\vph_A|^2
    + (\sigma + 2 k \pi (\Imx\tau)^{-1/2}\Ar^{-1/2})^2]^2}}
}\qquad\qquad \nn\\
&\sim&
\int_{-\infty}^\infty {{dk}\over
 {[\sum_{A=1}^5|\vph_A|^2
     + (\sigma + 2 k \pi (\Imx\tau)^{-1/2}\Ar^{-1/2})^2]^2}}\nn\\
 &=& {1\over 4}(\Imx\tau)^{1/2}\Ar^{1/2}
{1\over {(\sum_{A=1}^5|\vph_A|^2)^{3/2}}}\,.
\nn
\end{eqnarray}
Thus we recover roughly (\ref{vfrtdr}).
One can make a similar conjecture for the generalization of
(\ref{sigfd}) by changing the power of the denominator in the denominator
from $2$ to $5/2$ and modifying the numerator according to
(\ref{sigfd}).
It is also easy to see, by Poisson resummation, that the corrections
to the integral fall off exponentially like (using (\ref{sixfour})),
$$
\exp\left\{-(\Imx\tau)^{1/2}\Ar^{1/2}(\sum_{A=1}^5|\vph_A|^2)^{1/2}\}
\right\}
 =
e^{-\VEV \Ar}\,,
$$
and so are related to instantons made by strings wrapping the $\MT{2}$.
There are no corrections which behave like Yang-Mills instantons,
i.e. $e^{2\pi i\tau}$. The reason for this was explained
in \cite{DS},  in the SYM limit.

\subsection{A derivation from 4+1D SYM}
When we compactify the $(2,0)_{(N=2)}$ theory
on $\MS{1}$ of radius $L$, we find a low-energy
description of $U(1)^2$ SYM. When $\VEV L^2 \ll 1$ and when the
energies are much smaller than $L^{-1}$, the effective 4+1D
SYM Lagrangian with $U(2)$ gauge group is a good approximation.

The moduli space is $\MR{5}/\IZ_2$ and the term (\ref{eqh})
implies that there is a term in the Lagrangian which is 
proportional to (we have switched to physical units),
$$
g\epd{ABCDE}
\int d^5 x\,\, {{1}\over {|\vph|^5}} \epu{\a\b\g\d\u} A_{\u}
\vph^E\px{\a}\vph^A\px{\b}\vph^B\px{\g}\vph^C\px{\d}\vph^D\,.
$$
This term can actually be seen as a 1-loop effect!
Let us consider a loop of a charged gluino with
4 external legs of scalars and 1 external leg of a photon.
Let the external momenta be
$$
k_1,k_2,\cdots, k_5
$$
The loop behaves as,
\begin{equation}
\label{fivloop}
g^{5} \tr{t^1  t^2  t^3  t^4}
\int d^5 p\,\, \tr{\gamma_\u
{1\over {\pslash - m}}   
\cdot
{1\over {\pslash + \kslash_1 -  m}}
\cdots
{1\over {\pslash + \kslash_1 +\cdots \kslash_4 -  m}}}\,.
\end{equation}
Here $m$ is the mass of the gluino and is proportional
to $g\vph_0$.
The coupling constant $g$ is proportional to $\sqrt{L}$ (see appendix).  
The term with $\epu{\a\b\g\d\u}$ comes from expanding (\ref{fivloop})
in the $\kslash_i$.
We find
$$
g^{5}
\tr{t^1 t^2 t^3 t^4}
 \tr{\gamma_\u\kslash_1\kslash_2\kslash_3\kslash_4}
m\int {{d^5 p}\over {(p^2 + m^2)^5}}
\sim
g^5 m^{-4}
\epu{ABCDE}
\epd{\a\b\g\d\u} k_1^\a k_2^\b k_3^\g k_4^\d\,.
$$
This is the behavior that we want.
It would be interesting to check
if a similar term appears in the low energy description of
the M-theory on $T^6$ as a matrix model
\cite{whycorrect}-\cite{HL}.
In a certain regime we can approximate
by 6+1D Yang-Mills. For the $SU(2)$ case the moduli space is   
$\IR^3/ \IZ_2$. A similar effect could generate
term of the form below.
$$
\int A\wdg F\wdg F\wdg \epsilon_{ABC}
{{\phi^A d\phi^B\wdg d\phi^C}\over {|\phi|^3}}\,.
$$
After completion of this work, we have found out that
such terms were indeed calculated in \cite{thomp}.
We are grateful to G. Thompson for pointing this out to us.

\subsection{Component form}
Let us see how to write the term (\ref{sigfd})
in an $N=1$ superfield notation.
Let us take three chiral superfields, $\CHP$ and $\CHP^I$
($I=1,2$). We assume that
$$
\CHP = \scVEV + \delta\vph + i\sigma\,.
$$
$\sigma$ is the imaginary part of $\CHP$ and
$\scVEV$ is the VEV of the real part.
Below, the index $I$ of $\CHP^I$ is lowered and raised with the
anti-symmetric $\epd{IJ}$.

Let us examine the following term: 
\bear 
I_1 &=&
{1\over {32\pi}}(\Imx\tau)^{1/2}\Ar^{1/2}
\int d^4 x d^4\theta
  {1\over {(\CHP\bCHP + \CHP^I\bCHP_I)^{3/2}}}
  \bD^\da\bCHP^I D^\a\CHP_I
  \sig^{\u}_{\a\da}\bCHP^J\px{\u}\CHP_J + c.c.\hspace{12pt}
\label{ione}
\eear  
We can expand
\bear
\lefteqn{
\int d^4 x d^4\theta
  {1\over {(\CHP\bCHP + \CHP^I\bCHP_I)^{3/2}}}
  \bD^\da\bCHP^I D^\a\CHP_I
  \sig^{\u}_{\a\da}\bCHP^J\px{\u}\CHP_J
} \qquad\qquad \nn\\
&=&
{1\over {\scVEV^3}}  \int d^4 x d^4\theta \bD^\da\bCHP^I D^\a\CHP_I
  \sig^{\u}_{\a\da}\bCHP^J\px{\u}\CHP_J
\nn
\\&-& {3\over {2\scVEV^4}}  \int d^4 x d^4\theta
  (\CHP + \bCHP - 2\scVEV)
  \bD^\da\bCHP^I D^\a\CHP_I 
  \sig^{\u}_{\a\da}\bCHP^J\px{\u}\CHP_J
+\Ox{{1\over {\scVEV^5}}}\,.\hspace{12pt}
\eear
Let us denote
\bear
I_2 &=& {i\over {8\scVEV^3}}  \int d^4 x d^4\theta \bD^\da\bCHP^I
D^\a\CHP_I
  \sig^{\u}_{\a\da}\bCHP^J\px{\u}\CHP_J + c.c.\,,
\nn\\
I_3 &=& {i\over {8\scVEV^4}}  \int d^4 x d^4\theta\,\,
  (\CHP + \bCHP - 2\scVEV)
  \bD^\da\bCHP^I D^\a\CHP_I
  \sig^{\u}_{\a\da}\bCHP^J\px{\u}\CHP_J + c.c.\,,
\eear
Let us check the bosonic part of $I_1$.
We use $\sc$ and $\sc^I$ for the scalar
components of $\CHP$ and $\CHP^I$.
We will expand in inverse powers of $\scVEV$ and
keep only leading terms.

It is easy to see that $I_3$ contains the term
\be
 {1\over {\scVEV^4}} \int d^4 x\,\,
  \sigma \epu{\a\b\g\d}\px{\a}\sc^I\px{\b}\sc_I\px{\g}\bsc^J\px{\d}\bsc_J
\ee
At the order of $1/{\scVEV^3}$ there are a few more terms that 
do not include $\CHP$. They are listed below.
\bear
J_1 &=&  {1\over {\scVEV^3}} \int d^4\theta
    \px{\u}\CHP^I\qx{\u}\CHP^J\bCHP_I\bCHP_J
     +   {1\over {\scVEV^3}} \int d^4\theta
    \px{\u}\bCHP^I\qx{\u}\bCHP^J\CHP_I\CHP_J\,,
\nn\\
J_2 &=& {1\over {\scVEV^3}} \int d^4\theta
    \px{\u}\CHP^I\qx{\u}\bCHP_I\CHP^J\bCHP_J\,,
\nn\\
J_3 &=& {1\over {\scVEV^3}} \int d^4\theta
    \px{\u}\CHP^I\qx{\u}\bCHP^J\CHP_{I}\bCHP_{J}\,,
\nn\\
J_4 &=& {1\over {\scVEV^3}} \int d^4\theta
    \sig^{\u}_{\a\da} D^\a\CHP^I\bD^\da\bCHP_I \px{\u}\CHP^J\bCHP_J\,,
\eear
We now write down the bosonic terms of the above
\bear
J_3
&=& {1\over {\scVEV^3}}\int d^4x\,\{
 \sc_I\qx{\u}\sc^I\qx{\nu}\bsc_J\px{\u}\px{\nu}\bsc^J
-\sc_I\px{\u}\px{\nu}\sc^I\bsc_J\qx{\u}\qx{\nu}\bsc^J
\}\nn\\
J_2
&=& {1\over {2\scVEV^3}}\int d^4x\,\{
    -4\sc_I\px{\u}\px{\nu}\sc_J\bsc^I\qx{\u}\qx{\nu}\bsc^J
    +2\sc_I\px{\u}\px{\nu}\sc_J\bsc^J\qx{\u}\qx{\nu}\bsc^I
\nn\\&&
    -2\sc_I\px{\u}\px{\nu}\sc_J\qx{\nu}\bsc^I\qx{\u}\bsc^J
    -3\sc_I\sc_J\px{\u}\px{\nu}\bsc^I\qx{\u}\qx{\nu}\bsc^J
\nn\\&&
     +\sc_I\px{\u}\sc^I\px{\nu}\bsc_J\qx{\u}\qx{\nu}\bsc^J
\}\nn\\
J_1
&=& {1\over {2\scVEV^3}}\int d^4x\,\{
    6\sc_I\px{\u}\px{\nu}\sc_J\qx{\u}\bsc^I\qx{\nu}\bsc^J
   +3\sc_I\px{\u}\px{\nu}\sc_J\bsc^J\qx{\u}\qx{\nu}\bsc^I
\nn\\&&+3\sc_I\px{\u}\px{\nu}\sc_J\bsc^I\qx{\u}\qx{\nu}\bsc^J
    +\sc_I\sc_J\px{\u}\px{\nu}\bsc^I\qx{\u}\qx{\nu}\bsc^J
\}\nn\\
J_4
&=& {8\over {\scVEV^3}}\int d^4x\,\{   
\px{\u}\bsc_I\px{\nu}\bsc^I\qx{\u}\sc_J\qx{\nu}\sc^J
+\px{\nu}\bsc_J\qx{\nu}\bsc^I\qx{\u}\sc_I\px{\u}\sc^J
\nn\\&&-2\bsc_J\qx{\u}\bsc^I\qx{\nu}\sc_I\px{\u}\px{\nu}\sc^J
\}\,.
\eear
We will now check which combination has
the following symmetry which is part
of $SO(5)$ and doesn't involve $\sc$ and $\bsc$,
\be
\delta \sc^I = \bsc^I\,.
\ee  
We find
\bear
\delta J_1
&=& {1\over {\scVEV^3}} \int d^4x\,\{
    4\sc_I\bsc_J\px{\u}\qx{\nu}\bsc^I\px{\nu}\qx{\u}\bsc^J
   -4\sc_I\px{\u}\sc^I\px{\nu}\sc_J\qx{\u}\qx{\nu}\bsc^J
\}\nn\\
\delta J_2
&=& {1\over {2\scVEV^3}}\int d^4x\,\{
     5\sc_I\px{\u}\bsc^I\px{\nu}\bsc_J\qx{\u}\qx{\nu}\bsc^J
    -3\sc_I\bsc_J\px{\u}\px{\nu}\bsc^I\qx{\u}\qx{\nu}\bsc^J\nn\\
   &&+8\sc_I\px{\nu}\sc^I\px{\u}\sc_J\qx{\u}\qx{\nu}\bsc^J
\}\nn\\
\delta J_3
&=& {1\over {\scVEV^3}}\int d^4x\,\{
    2\sc_I\px{\u}\bsc^I\px{\nu}\bsc_J\qx{\u}\qx{\nu}\bsc^J
   +2\sc_I\px{\u}\sc^I\px{\nu}\sc_J\qx{\u}\qx{\nu}\bsc^J
\}\nn\\
\delta J_4
&=& {8\over {\scVEV^3}}\int d^4x\,\{
2\sc_I\px{\nu}\bsc_J\px{\u}\bsc^I\qx{\u}\qx{\nu}\bsc^J
+2\sc_I\bsc_J\px{\u}\px{\nu}\bsc^I\qx{\u}\qx{\nu}\bsc^J
\}\,.
\eear
This puts some restrictions on the possible
${1 / {\VEV^3}}$ term
\bear
\lefteqn{\scVEV^3 \delta (C_1 J_1 + C_2 J_2 + C_3 J_3 + C_4 J_4)
}\qquad\qquad\nn\\&=&
    (4C_1 - \frac 32 C_2 + 16 C_4)
\sc_I\bsc_J\px{\u}\qx{\nu}\bsc^I\px{\nu}\qx{\u}\bsc^J
  \nn\\&& +(\frac 52 C_2 + 2C_3 + 16 C_4)
\sc_I\px{\u}\bsc^I\px{\nu}\bsc_J\qx{\u}\qx{\nu}\bsc^J
\nn\\&&   +(-4C_1 + 4C_2 + 2C_3)
       \sc_I\px{\nu}\sc^I\px{\u}\sc_J\qx{\u}\qx{\nu}\bsc^J\,.
\eear
We see that
$$\frac 52 C_2 + 2C_3 + 16C_4 = 0\,,\qquad
4C_1 - \frac 32 C_2 + 16C_4 = 0\,.
$$
Thus, we need to take the following $SO(4)$ invariant combination:
$$C(3J_1+8J_2-10J_3)+C'(4J_1+8J_3-J_4)\,,
$$
where $C,C'$ are undetermined. We have not checked
if one can extend it to a
supersymmetric and $SO(5)$ invariant
combination by including interactions with
$\CHP$.
We thank Savdeep Sethi for discussions on this point.

\section{Conserved quantities}
\label{s5}

We can check that the overall ``center of mass'' decouples.
We can write it as a conservation equation for the
total dissolved membrane charge ($j_Z$), total transverse momentum
($j_{\PH}$) and kinematical supersymmetry ($j_\TH$):
\eqn{kincur}{
j_Z^{\alpha,\beta\gamma}=\frac1{2\pi}
\sum_{i=1}^N H_i^{\alpha\beta\gamma}\,,\quad
j^{\alpha,A}_{\PH}=\frac1{2\pi}
\sum_{i=1}^N\partial^\alpha\PH_i^A\,,\quad
j^\alpha_\TH=\frac1{2\pi}\sum_{i=1}^N \Ga^\alpha\TH_i\,.
}
They are conserved simply because $\partial_\alpha j^\alpha$ gives
the sum over $i,j$ of the right hand sides of
 (\ref{eqh}, \ref{eqphi}, \ref{eqtheta})
but the summand is $ij$ antisymmetric. The charges are defined as the
integrals of the $\alpha=0$ (lowered index) components
\eqn{charin}{
Z^{IJ}=\int \frac{d^5\s}{2\pi} \sum_{i=1}^N H_0^{i,IJ},\quad P^A=\int
\frac{d^5\s}{2\pi}\sum_{i=1}^N\partial_0\PH_i^A\,,\quad Q^{KIN}=\int
\frac{d^5\s}{2\pi}\sum_{i=1}^N\Ga_0\TH_i\,.
}
 We use the terms ``dissolved membranes'' and
``thin membranes'' for membranes of M-theory with 0 or 1 directions
transverse to the fivebranes, respectively. The thin membrane charge
appears as a central charge in the supersymmetry algebra \cite{SWSIXD}.
The
reason is that $\{Q,\bar Q\}$ in M-theory contains momenta, twobrane and
fivebrane charges. But in (2,0)  theory, only the generators with $\GPR
Q=Q$ i.e. $\bar Q \GPR=-\bar Q$ survive. So we see that $\{Q,\bar Q\}$ is
a matrix anticommuting with $\GPR$ (i.e. containing an odd number of Greek
indices). For momenta it means that only momenta inside the fivebrane
worldvolume appear on RHS of supersymmetry algebra because $\Ga_\u$
anticommutes with $\GPR$ while the transverse $\Ga_A$ commutes with
$\GPR$.

Only membrane charges contain $\Ga_\u\Ga_A$ which anticommutes
with $\GPR$ while $\Ga_{\u\nu}$ and $\Ga_{AB}$ commute with $\GPR$. This
is
an explanation why the thin membranes (looking like strings) with one
direction transverse to the fivebrane occur on the RHS of the
supersymmetry algebra.
There are also 3-form central charges which appear with
$\Ga_{\u\nu\s}\Ga_A$ in the SUSY algebra. These correspond to
tensor fluxes of the 3-form $H$ (analogous to electric and magnetic fluxes
in Yang-Mills theories).
But let us return to the thin membranes. We should be able to find the
corresponding current. The answer is (up to an overall normalization)
\begin{equation}
M_{\a,\b}^A=-\frac 1{12\pi}
\epsilon_{\a\b\g\d\e\zeta}\sum_{i=1}^N
\partial^\g(\PH^A_i H_i^{\d\e\zeta})=
\frac{1}{2\pi}
\sum_{i=1}^N \partial^\g(H^i_{\a\b\g}\PH^A_i)\,.
\label{thincur}
\end{equation}
The conservation law $\partial^\a M_{\a\b}^A$ is a simple consequence
of $\alpha\gamma$ anti-symmetry of $\epsilon_{\a\b\g\d\e\zeta}$. It is
also easy to see that for a configuration containing a membrane, the total
integral $\int d^5\s M_{0I}^A=W_I\cdot \Delta\PH^A$ measures the membrane
charge. Here $W_I$ is the winding vector of the induced string and $\Delta
\PH^A$ is the asymptotic separation of the two fivebranes.

There must be also a current corresponding to the $SO(5)$ R-symmetry.
It is given by
\eqn{Rcur}{
R_\a^{AB}= {1\over {2\pi}}
\sum_{i=1}^N
\left(
2\PH_i^{[A}\partial_\a\PH_i^{B]}
-\frac i2\bTH_i\Ga_\a\Ga^{AB}\TH_i
\right) + \mbox{corrections}.}

It is also quite remarkable that the corrected equations
conserve the stress energy tensor known from free theory.
For the initial considerations, let us restrict our attention to the
bosonic part of the stress tensor and choose the sign so that
$T_{00}>0$ i.e. $T^{0}_{\,\,\,0}<0$.
Ignoring the requirement of the vanishing trace (i.e. without
the second derivatives that we discuss below),
the bosonic part
of our stress tensor is given by
\eqn{strtensor}{T_{\alpha\beta}^{try}= {1\over {2\pi}}
\sum_{i=1}^N
\left(\frac 14 H^i_{\alpha\gamma\delta}H_\beta^{i,\gamma\delta}+
\partial_\alpha \PH_A^i\partial_\beta\PH_A^i-
\frac 12\eta_{\alpha\beta}\partial_\gamma\PH_A^i
\partial^\gamma\PH_A^i
\right).}
%
Note that the $\PH$ part has nonzero trace. The divergence
of this symmetric tensor can be written as
\eqn{diverg}{
\partial^\alpha T^{try}_{\alpha\delta} = {1\over {2\pi}}
\sum_{i=1}^N\left(
\frac 12 H^i_{\delta\gamma\delta'}
(\partial_\alpha H^{i,\alpha\gamma\delta'})
+(\Box \PH_D^i)\partial_\delta \PH_D^i
\right).
}
If we substitute $\Box\PH_D$ from (\ref{eqphi}) and
$\px{\a}H^{i,\a\g\d'}$ from (\ref{eqh}) we obtain
$\partial^\alpha T^{try}_{\alpha\delta} = 0$.   

We should note one thing that could be confusing. In the M-theory
containing $N$ fivebranes, the stress tensor is not equal to zero
but rather to\footnote{The minus sign in (\ref{cosmostr}) is
because our choice of the spacelike metric and $T_{00}>0$.}
  \eqn{cosmostr}{T^{M}_{\alpha\beta}=-N\tau^{(5)}\eta_{\alpha\beta}
+ T_{\alpha\beta}\,,}
where our $T_{\alpha\beta}$ is just a small correction to the infinite
first term given by the tension of the fivebrane $\tau^{(5)}$. The
first term is in the limit of (2,0) theory infinite because
$\tau^{(5)}$ is of order $l_{Planck}^{-6}$ and $l_{Planck}$ is much
smaller than a typical distance inside fivebranes studied by (2,0) theory.
Nevertheless, gravity in this limit decouples and thus the
``cosmological'' term in (\ref{cosmostr}) plays no role.

\subsection{Traceless stress tensor and supercurrent}

In this subsection, we exhibit a traceless version of the stress tensor
and the supercurrent. We will use the adjective ``traceless'' both for the
supercurrent $J_\alpha$ and the stress tensor $T_{\alpha\beta}$
which means that
  \eqn{tlesuper}{\Ga^\alpha J_\alpha=0,\quad
T_\alpha^\alpha=0\,.}
The supercurrent has positive chirality $(\GPR-1) J_\alpha=0$ -- it
means
that the total supercharges have positive chirality as well.
We will also require continuity for stress tensor and supercurrent.
  \eqn{cont}{\partial^\alpha J_\alpha=0,\quad
\partial^\alpha T_{\alpha\beta}=0\,.}
Our definition of the stress tensor will be finally
\begin{eqnarray}
T_{\alpha\beta} &=& {1\over {2\pi}} \left\{
\sum_{i=1}^N
\frac 14 
H^i_{\alpha\gamma\delta}H_\beta^{i,\gamma\delta}+\label{st}\right.\nn\\
&+&\sum_{i=1}^N
\frac 3{5}\left(\partial_\alpha \PH_A^i\partial_\beta\PH_A^i-
\frac 16\eta_{\alpha\beta}\partial_\gamma\PH_A^i
\partial^\gamma\PH_A^i\right)-
\frac 2{5}   
\PH_A^i\left(\partial_\alpha\partial_\beta-\frac 16
\eta_{\alpha\beta}\Box\right)\PH_A^i+\nonumber\\
&+& \left.\sum_{i=1}^N
\frac{(-i)}2
\bar\TH^i\left(
\Ga_{(\alpha}\partial_{\beta)}-\frac 16\eta_{\alpha\beta}
\dslash
\right)\TH^i
\right\}\,.
\label{stfermi}
\end{eqnarray}
We fixed a normalization for $H,\PH,\TH$ in this equation. The factors
$1/6$ inside the parentheses guarantee the tracelessness while the
relative factor $-3/2$ between the parentheses ensures vanishing of the
dangerous terms in $\partial^\alpha T_{\alpha\beta}$
which cannot be expressed from the equations of motion, namely
$\partial^{\alpha}\PH\partial_{\alpha\beta}\PH$. The $H^2$ part of the
stress tensor is traceless identically.
An explicit calculation shows that for the divergence of the stress   
tensor we get
\begin{eqnarray}
\partial^\alpha T_{\alpha\beta} &=& {1\over {2\pi}}
\sum_{i=1}^N
\left[
\frac 12(\partial^\alpha
H^i_{\alpha\gamma\delta})H_\beta^{i,\gamma\delta}
+\frac 23 (\partial_\beta\PH_A^i)(\Box\PH^A_i)-
\frac 13\PH^i_A(\partial_\beta\Box
\PH_i^A)+\right.\label{divst}
\nn\\&+&
\frac{7i}{12}(\partial_\beta\bar\TH^i)(
\dslash
\TH^i)-
\left.\frac i4\bar\TH^i\Ga_\beta\Box\TH^i-  
\frac i6\bar\TH^i\partial_\beta
\dslash
\TH^i\right].
  \end{eqnarray}

A similar approach can be used for the supercurrent as well. Here also
the $H\TH$ part is traceless identically
while for the other parts it
is ensured by the $1/6$ factors. The relative factor $-3/2$ between the
parentheses is again chosen to cancel the dangerous $\partial^\alpha \PH
\partial_\alpha \TH$ terms in $\partial^\alpha J_\alpha$.
Note that the structure of $J_\alpha$ mimics the form of $T_{\a\b}$.
\begin{eqnarray}
J_\alpha &=&  {1\over {24\pi}}\sum_{i=1}^N
H_i^{\beta\gamma\delta}\Ga_{\beta\gamma\delta}\Ga_\alpha\TH^i
+\label{suscur}\nonumber\\
&+&
{1\over {\pi}}
\sum_{i=1}^N \left[
\frac 35\left(\partial_\alpha \PH_A^i
-\frac 16 \Ga_\alpha  
\dslash
\PH_A^i\right)\Ga_A\TH^i
-\frac 25\PH_A^i\left(\partial_\alpha
-\frac 16 \Ga_\alpha
\dslash
\right)
\Ga^A\TH^i
\right].
\end{eqnarray}
We can compute also a similar continuity equation for the
supercurrent as we did for the stress tensor. The result is
\bear
\partial^\a J_\a &=&
{1\over {4\pi}}\partial_\a H^{\a\b\g}\Ga_{\b\g}\TH
+
{{1}\over {24\pi}} H^{\b\g\d}\Ga_{\b\g\d}\dslash\TH \nn\\
&+&
{1\over {2\pi}}
\left[(\Box\PH_A)\Ga^A\TH-
\frac 13\dslash\PH_A\Ga^A(\dslash\TH)
-\frac 23\PH_A\Ga^A(\Box\TH)\right].
 \eear
Using the equations of motion and the integration by parts, the
Hamiltonian and the total supercharge
defined as
\eqn{hasus}{
\ham=\int d^5\s T_{00}\,, \qquad
Q=\int d^5\s J_0
}
can be easily expressed as
  \eqn{hafr}{
\ham= {1\over {2\pi}}
\int d^5\sigma\left(
\frac 12 \left(\Pi^2+(\nabla \PH)^2\right)
+\frac 1{12} H_{KLM}H^{KLM}
+\frac{(-i)}2
\bar\TH \Ga^J\partial_J\TH \right).
}
(we use conventions with
$\com{\Pi(x)}{\PH(y)} = -2\pi i \delta^{(5)}(x-y)$ and $\{
\TH^s(x),\bTH_{s'}(y)
\}=\pi((1+\GPR)\Ga_0)^{s}{}_{s'}\delta^{(5)}(x-y)$)
and
\eqn{susr}{
Q={1\over {2\pi}} \int d^5\s
\left(
\frac{1}{6} H^{IJK}\Ga_{IJK}\Ga_0\TH
-\partial_\b\PH_A\Ga^\b\Ga^0\Ga^A\TH\right).
}
For convenience, we can also easily compute
\eqn{susrb}{
\bar Q={1\over {2\pi}} \int d^5\s
\left(
{1\over 6}
\bTH\Ga^0\Ga_{IJK}H^{IJK}+
\bTH \Ga^0\Ga^\b\Ga^A\partial_\b\PH_A\right)}
using a simple identity
  \eqn{barlo}{\overline{(\TH\Ga_{\u_1}\dots\Ga_{\u_N})}
=(-1)^N \bTH\Ga_{\u_N}\dots\Ga_{\u_1}\,.}
Now we can consider the supersymmetry transformation.
A variation of a field F will be written as
  \eqn{varf}{\d F = [\eb Q, F] = [\bar Q \e,F]\,.}
Note that $\eb Q$ is an antihermitean operator because the components of
$Q$ or $\e$ are hermitean anticommuting operators or numbers,
respectively. Using the canonical commutation relations
we can easily compute the variations of the fields
\begin{equation}
\label{vth}
\d \TH=-[\TH,\bar Q\e]=
\left(
\frac{1}{6}\Ga_{IJK}H^{IJK} +
\Ga^\b\Ga^A\partial_\b\PH_A\right)\e \,.
\end{equation}
This agrees with the transformations written before.
This, together with the normalization of $\acom{Q}{\bar{Q}}$ is
how we determined the relative coefficients.
Similarly
\eqn{vph}{
\d\PH_A=[\eb Q,\PH_A]=\eb[Q,\PH_A]
=\eb[{1\over {2\pi}}
\smallint d^5\s 
\Pi_B\Ga^B\TH,\PH_A]=
-i 
\eb\Ga_A\TH\,,
}
which agrees with previous definitions.
A similar but more tedious calculation gives us
\eqn{vha}{ 
\d H_{IJK}=\frac i2
\eb\cdot \epsilon_{0J'K'IJK}
\Ga^{I'J'K'}\Ga_0\partial_{I'}\TH\,,
}
which also agrees with the previous definition.

Let us summarize some formulae that are useful in understanding the
commutator of two supersymmetry transformations:
\bear
\d f=[\eb Q,f]=[\bar Q\e,f],& &
\partial_\a f = -i[P_\a,f],\qquad P^0=\ham>0\\
 \{Q_s,\bar Q^{s'}\}&=&-2P^\u ((1+\GPR)/2\cdot\Ga_\u)  
{}_s{}^{s'} + \mbox{thin}\\
\Rightarrow \delta Q&=&-[Q,\bar Q\e]=2P^\u\Ga_\u\e+ \mbox{thin}\\
\d J_\a &=&-2 T_{\a\b}\Ga^{\b}\e + \mbox{thin}\\
(\d_1\d_2-\d_2\d_1)f   
&=&[\eb_1 Q,[\eb_2 Q,f]]-[\eb_2 Q,[\eb_1 Q,f]]=[[\eb_1 Q,\eb_2 Q],f]\\
&=&[\eb_1^s\{Q_s,\bar Q^{s'}\}\e_{2,s'},f]=-2[P^\u\eb_1\Ga_\u\e_2,f]
+ \mbox{thin}\\
&=&-2i(\eb_1\Ga^\u\e_2)\partial_\u f+ \mbox{thin}
\eear

\section{Speculations about the fundamental formulation}
\label{s6}

In this section we would like to speculate on whether a fundamental
formulation of the (2,0) theory can be constructed from
the equations we discussed above. 
We warn the reader in advance that this section could cause
some gritting of teeth!    
Of course, the correction terms are not renormalizable if treated
as ``fundamental'' but let us go on, anyway. Perhaps some hidden
symmetry makes them renormalizable after all?
 
The model has the following virtues:   
\begin{itemize}
\item there are absolutely no new fields. We use only $N$ copies of the
field strength $H_{MNP}$, five scalars $\PH_A$ and the 16 component
fermion $\TH$. Because of that, restriction to $N$ copies for distant
fivebranes is almost manifest.

\item the string current automatically satisfies the quantization
condition as a right winding number. This is related to the fact
that our current is automatically conserved (obeys the continuity
equation) which is necessary to allow us to insert it to equation
$dH=J$ -- and it has the correct dimension $mass^4$.

\item the total charge (sum over $1...N$) vanishes. The string (membrane
   connecting fivebranes) brings correctly minus source to one fivebrane
   and plus source to the other which agrees with the fact that   
   the oriented membrane is outgoing from one fivebrane and incoming to
   another fivebrane -- and with $e_i-e_j$ roots of $U(N)$

\item the model is symmetric with respect to the correct
Ho\hacek rava-Witten symmetry \cite{howitten} that accompanies the
reflection
$\PH_A^i\to-\PH_A^i$ by changing sign of $C_{MNP}$ (i.e. of $H_{MNP}$).

\item string states are given by strange configuration
   of fivebranes so that the vector of direction between two $\PH$'s
   draws whole $S^4$ (surface of ball in $\IR^5$) if one moves in
   the 4    transverse directions of the string.
 
\item $U(N)$ is not manifest, it arises due to the string states --
  perhaps in analogy with  the way enhanced symmetries appear in string
  theory because of D-brane    bound states.
\end{itemize}

What does a string look like? It is a solution constant in the time
and in one spatial direction, with a given asymptotical value of
$\Delta\PH=\abs{\PH^i-\PH^j}$ in infinity. We can show that such É
solution will
have typical size of order $\Delta\PH^{-1/2}$ in order to
minimize the tension (energy per unit of length of the string).

The value of $\partial\PH$ is of order $\Delta\PH/s$, integral of its
square over the volume $s^4$ is of order $(s\Delta\PH)^2$. On the  
contrary, such a topological charge makes the field $H$ to behave
like $1/r^3$ where $r$ is the distance from the center of the solution.
Therefore $H$ inside the solution is of order $1/s^3$ which means that
the contribution of $H^2$ to the tension is of order $s^4/s^6=1/s^2$.
The total tension $(s\Delta\PH)^2+1/s^2$ is minimal for
$s=(\Delta\PH)^{-1/2}$ and the tension is therefore of order
$\Delta\PH$.
The field $\PH$ tries to shrink the solution while $H$ attempts to blow
it up.
In the next section we will describe the solution more concretely.


\section{String-like solution of the (2,0) theory}
\label{s7}
We will try to describe the string-like solution of the bosonic part
of the equations, considering only the topological term of $dH$ and the
corresponding term in $\Box\PH$ equation.
The following discussion
is somewhat reminiscent of a related discussion in \cite{CRRU}    
for the effect of
higher order derivative terms on monopole solutions in $N=2$ Yang-Mills
but our setting is different.


\subsection{A rough picture}
Our solution will be constant in $\s^0,\s^5$ coordinates but it will
depend on the four coordinates $\s^1,\s^2,\s^3,\s^4$.
We are looking for a solution that minimizes the energy.
If the size of the solution in these
four directions is of order $s$, then the ``electric'' field, going like
$1/r^3$, is of order $1/s^3$ inside the solution and therefore the
integral
$d^4\s (H^2)$, proportional to the tension, is of order $s^4/(s^3)^2$.

On the contrary, for the asymptotic separation $\Delta\PH$ quantities
$\partial \PH$ are of order $\Delta\PH/s$ inside the typical size of the
solution and therefore the contribution to the tension
$d^4\s (\partial\PH)^2$ is of order $s^4(\Delta\PH/s)^2$.

Minimizing the total tension $1/s^2+s^2 \Delta\PH^2$ we get the typical
size $s=(\Delta\PH)^{-1/2}$ and the tension of order $\Delta\PH$. In this
reasoning, we used the energy known from the free theory because the 
bosonic part of the interacting stress energy tensor equals the free
stress energy tensor. The fact that the solution corresponds to the
interacting theory (and not to the free theory) is related to the
different constraint for $(dH)_{IJKL}$.

\subsection{The Ansatz}

We will consider $N=2$ case of the (2,0) theory, describing two
fivebranes. Our solution will correspond to the membrane stretched between
these two fivebranes. Denoting by $(1)$ and $(2)$ the two fivebranes, we
will assume $\PH_{(1)}=-\PH_{(2)}$, $H_{(1)}=-H_{(2)}$ and we denote
$\PH_{(1)}$ and $H_{(1)}$ simply as $\PH$ and $H$.

Our solution will be
invariant under $SO(4)_D$ rotating spacetime
and the transverse directions together.
The variable
$$r=\sqrt{\s_1^2+\s_2^2+\s_3^2+\s_4^2}$$
measures the distance from the center of the solution.
We choose
the asymptotic
separation to be in the 10th direction and we denote it as
$$\PH^{10}(\infty)=\frac 12\Delta\PH\,.$$

Now there is an arbitrariness in the identification of the coordinates
$1,2,3,4$ and $6,7,8,9$. So there is in fact a moduli space of classical
solutions, corresponding to the chosen identification of these
coordinates. According to our Ansatz, the solution will be determined in
the terms of the three functions.
$$\PH^{I+5}=\s^If_1(r),\,\,I=1,2,3,4\quad
\PH^{10}=f_2(r)\,,\quad
B_{05}=f_3(r).$$
We set the other components of $B_{\u\nu}$ to zero and define $H$ as the
\asd part of $dB$,
$$H_{\a\b\g}=\frac 32 \partial_{[\a}B_{\b\g]} - \mbox{dual
expression}\,.$$ 
It means that $H_{05I}=1/2\cdot\partial_I f_3$ and the selfduality  
says
$$
H_{051}=-H_{234},\quad
H_{052}=H_{134},\quad
H_{053}=-H_{124},\quad
H_{054}=H_{123}.$$
Now we can go through the equations. $dH$ equations for $1,2,3,4$
determines $-4\partial_{[1}H_{234]}=\partial_IH_{05I}=\frac 12\Delta f_3$
where we used $\Delta=\Box$ because of the static character.
Therefore $dH$ equation says
$$\Delta f_3=-8c_1\frac{f_1^3}{(f_2^2+r^2f_1^2)^{5/2}}   
(-rf_1f'_2+f_1f_2+rf'_1f_2).$$
The three factors $f_1$ arose from $\partial_2\PH_7$,
$\partial_3\PH_8$,
$\partial_4\PH_9$,
we calculated everything at $\sigma^{1,2,3,4}=(r,0,0,0).$
At this point, only $EABCD=10,6789$ and $6,10,789$ from
$\epsilon$ symbol contributed.
Here $\Delta$ always denotes the spherically symmetric
part of the laplacian in 4 dimensions, i.e.
$$\Delta=\frac{\partial^2}{\partial r^2}+\frac 3r\frac{\partial}{\partial
r}\,.$$
Similarly, we get hopefully two equations from $\Box\PH$. For $\PH^{10}$
(in the direction of asymptotic separation), we seem to get
$$\Delta f_2=6\frac{c_2}2\partial_1(-B_{05})
\frac{\e^{7,8,9,10,6}}{(f_2^2+r^2f_1^2)^{5/2}}rf_1^4\,.$$
Similarly, for the four other components we have
$$\Delta(rf_1)=-3c_2f'_3\frac{f_2 f_1^3}{(f_2^2+r^2f_1^2)^{5/2}}\,.$$

\subsection{Numerical solution, tension and speculations}

The functions $f_1,f_2,f_3$ are all even, therefore their derivatives
are equal to zero for $r=0$. The value of $f_2(0)$ finally determines
$f_2(\infty)$ which we interpret as $\Delta\PH /2$. The value of $f_1(0)$
must be fixed to achieve a good behavior at infinity and
$f_3(0)$ has no physical meaning, because only derivatives of $f_3=B_{05}$
enter the equations.

We can calculate the tension and we can compare the result
with the BPS formula. If we understand our equations just as some low
energy
approximation, there should be no reasons to expect that the calculated
tension will be precise, because
the approximation breaks down at the core.
  
The tension expected from SYM theory is something like
$$M_W/L_5=\Delta\PH^{SYM}\cdot g/L_5=\sqrt{2\pi/L_5}\Delta\PH^{SYM}
=\sqrt{2\pi}\Delta\PH^{(2,0)}\,.$$
We just used simple formula for W boson masses, W bosons are string
wound around 5th direction and the $\PH$ fields of SYM and (2,0) are
related by $\sqrt{L_5}$ ratio as well.

The tension from our (2,0) theory is just twice (the same contribution
from two fivebranes) the integral
$$2\int d^4\s \frac 1{4\pi}\left(H_{05I}^2+(\partial_I\PH^A)^2\right).$$
Because of the spherical symmetry, we can replace $\int d^4\s$
by $\int_0^\infty dr\cdot 2\pi^2 r^3$. Work is in progress.

\section{Discussion}
\label{s8}

Recently, a prescription for answering questions about
the large $N$ limit of the (2,0) theory has been proposed
\cite{Juan}.  In particular, the low-energy effective
description for a single 5-brane separated from $N$
5-branes has been deduced \cite{Juan}.
The topological term that we have discussed is, of course,
manifestly there. This is because a 5-brane probe in an
$AdS_7\times S^4$ feels the 4-form flux on $S^4$ and
and this will induce the anomalous $dH$ term.

What does M(atrix) theory have to say about non-linear
corrections to the low-energy of the (2,0) theory?
This is a two-sided question as the (2,0) theory
is a M(atrix) model for M-theory on $\MT{4}$
\cite{rozali,berozaliseiberg} and has a M(atrix) model
of its own \cite{prem,WitQHB}.

In order to be able to apply our discussion of the uncompactified
5+1D (2,0) theory to the M(atrix) model for M-theory
on $\MT{4}$ we need to be in a regime such that the VEV of
the tensor multiplet is much larger than the size of
$\MHT{5}$. This means that for a scattering process of
two gravitons in M-theory on $\MT{4}$ the distance between
the gravitons must remain much larger than the compactification
scale which we assume is of the order of the 11D Planck scale.
In this regime we expect the potential to behave
as $v^4/r^3$ (in analogy with $v^4/r^7$ in 11D).
Thus, things would work nicely if there were a term
\begin{equation}
\label{phfour}
{{(\partial\PH)^4}\over {|\PH|^3}}
\end{equation}
in the effective low-energy description in 5+1D.
In the large $N$ limit, the existence of this term has been observed
in \cite{Juan}.
The term (\ref{phfour}) will also be the leading term
in the amplitude for a low-energy scattering of two
massless particles in the (2,0) theory.
It should thus be possible to calculate it from the M(atrix)
model of the (2,0) theory, with a VEV turned on.

It is also interesting to ask whether a term like (\ref{phfour})
is renormalized or not. An analysis which addresses such a question
in 0+1D will appear in \cite{PSS}. Perhaps a similar analysis
in 5+1D would settle this question.

\section{Appendix A: Formulae for SUSY transformations} 

In this text, we will use
the $SO(10,1)$ formalism for spinors, inherited from the M-theory
containing $N$ fivebranes, and
the space-like metric (in 5, 6 and 11 dimensions)
\bear
\eta_{\mu\nu} &=& \mbox{diag}(-++++++++++)\,, \quad
\mu,\nu=0,1,\dots 10.\nonumber\\
ds^2 &=& \eta_{\alpha\beta}dx^{\alpha}dx^{\beta} = -dx_0^2
+dx_1^2+dx_2^2+dx_3^2+dx_4^2+dx_5^2\,.
\label{spacelike}
\eear

\subsection{SUSY transformation}  
The SUSY transformations of the free tensor multiplet in 5+1D
is given by
\begin{eqnarray}
\delta H_{\a\b\g} &=& -\frac{i}{2} \bar\epsilon
                       \Ga_\d\Ga_{\a\b\g}\partial^\d\TH
                   =
                    -3 i \eb \Ga_{\lbr\a\b}\partial_{\g\rbr}\TH
                    + \frac{i}{2} \eb\Ga_{\a\b\g}\Ga_{\d}\partial^\d\TH
\nn\\
\delta \PH_A      &=& -i \bar\epsilon\Ga_A\TH
\nn\\
\delta \TH        &=& (\frac{1}{12} H_{\a\b\g}
                 \Ga^{\a\b\g}+ \Ga^\a\partial_\a
                 \PH_A\Ga^A)\epsilon\,.
\nn
\end{eqnarray}
Since we are dealing with corrections to the low-energy
equations of motion, it is important to keep terms which vanish
by the equations of motion.
The SUSY commutators are thus given by
\begin{eqnarray}
(\delta_1\delta_2 -\delta_2\delta_1) H_{\a\b\g} &=&
   - 2 i (\eb_1\Ga^{\u}\e_2) \partial_{\u} H_{\a\b\g}
\nn\\
&& - \frac{i}{2}
\ept{[\a\b}{\d\a'\b'\g'}\eb_1\Ga_{\g]}\e_2\partial_{[\d}H_{\a'\b'\g']}
+4 i (\eb_1\Ga^{\d}\e_2) \partial_{[\d}H_{\a\b\g]}
\nn\\
(\delta_1\delta_2-\delta_2\delta_1)\TH &=&
   - 2 i (\eb_1\Ga^\u\e_2)\partial_{\u}\TH
\nn\\
&& -\frac{i}{24} \left\{ 18(\eb_2\Ga_\u\e_1)\Ga^\u
  - 6(\eb_2\Ga_\u\Ga_{A}\e_1)\Ga^\u\Ga^{A}
\right\} \Ga_\b\partial^{\b}\TH
\nn\\
(\delta_1\delta_2 - \delta_2\delta_1) \PH_A &=&
   -2 i (\eb_1\Ga^{\u}\e_2)\partial_{\u}\PH_A\,. 
\nn  
\end{eqnarray}  
The equations of motion transform according to,
\begin{eqnarray}
\delta\left(\Ga_{\d}\partial^\d\TH\right) &=&
     \frac{1}{6} \Ga^{\d\a\b\g}\e\partial_{[\d}H_{\a\b\g]}
     + \Ga^A\e\partial_{\a}\partial^{\a}\PH_A\,,
\nn\\
\delta\left(\partial_{[\u}H_{\a\b\g]}\right) &=&
     \frac{i}{2}
     \eb\Ga_{[\a\b\g}\partial_{\u]}\left(\Ga_\d\partial^\d\TH\right),
\nn\\
\delta\left(\partial_\u\partial^\u\PH_A\right) &=&
     -i \eb\Ga_A\left(\Ga_{\d'}\partial^{\d'}\right)
     \Ga_\d\partial^\d\TH=-i\eb\Ga_A\Box\TH\,. 
\nn
\end{eqnarray}
                    
\section{Appendix B: Quantization}

The quantization of the free tensor multiplet was discussed
at length in \cite{WitFBE}.
There is no problem with the  fermions $\TH$ and bosons $\PH^A$,
but the tensor field is self-dual and thus has to be quantized
similarly to a chiral boson in 1+1D.
This means that we second-quantize a free tensor field 
without any self-duality constraints and then set to zero
all the oscillators with self-dual polarizations.

The analogy with chiral bosons is made more explicit if
we compactify on $\MT{4}$ and take the low-energy limit
we we can neglect Kaluza-Klein states. We obtain a 1+1D
conformal theory. This theory is described by compact chiral
bosons on a $(3,3)$ lattice. This is the lattice of
fluxes on $\MT{4}$. For $\MT{4}$ which is a product of
four circles with radii $L_i$ ($i=1\dots 4$),  we get
3 non-chiral compact bosons with radii
$$
{{L_1 L_2}\over {L_3 L_4}},\,\,
{{L_1 L_3}\over {L_2 L_4}},\,\,
{{L_1 L_4}\over {L_2 L_3}}\,.
$$
Of course, in 1+1D, T-duality can replace each radius $R$ with
$1/R$ and thus $SL(4,\IZ)$ invariance is preserved.

If we further compactify on $\MT{5}$ the zero modes
will be described by quantum mechanics on $\MT{10}$, where
$\MT{10}$ is the unit cell of the lattice of fluxes.

\subsection{Commutators}

Let us write down the commutation relations.

We want to reproduce the equations of motion by the Heisenberg equations
\eqn{heis}{\partial_0 (L) = i[\ham,L]
\mbox{\quad where\quad} \ham=\int d^5\sigma T_{00}\,.}
We should be allowed to substitute $H,\PH,\TH$ for the operator $L$.
In the following text we will use indices $I,J,K,\dots$ for the spatial
coordinates inside the
fivebrane. We will keep the spacelike metric and the convention
  \eqn{levicc}{\epsilon_{12345}=\epsilon^{12345}=1\,.}
     
We have the equations $H=-*H$ and $dH=0$. Among the fifteen equations
for the vanishing four-form $dH=0$ we find ten equations with
index 0. These will be satisfied as the Heisenberg equations
(\ref{heis}). Remaining five equations with space-like indices will only
play a role of some constraints that are necessary for consistent
quantization as we will see.
Let us take the example of equations of motion for $(dH)_{0345}$.
\eqn{dh}{
0=\partial_0H_{345}
-\partial_3 H_{450}+\partial_4 H_{503}-\partial_5H_{034}=
\partial_0 H_{345}+\partial_3H_{123}
+\partial_4H_{124}+\partial_5H_{125}\,.
}
It means that we should have the commutator
\eqn{ihcom}{
i[\ham,H_{345}(\sigma')]=
-\partial^{(\sigma')}_IH_{12I}(\sigma')\,,
}
where the important part of hamiltonian is
\eqn{hham}{
\ham_H= {1\over {8\pi}}\int d^5\sigma
H_{0IJ}H_0^{\,\,IJ}=\int d^5\sigma
\frac {1}{24\pi} H_{KLM}H^{KLM}\,.
}
But it is straightforward to see that the relation (\ref{ihcom})
will be satisfied if the commutator of $H$'s will be
\eqn{hcom}{
[H_{IJK}(\sigma),H_{LMN}(\sigma')]=
-6\pi i
\partial_{[I}^{(\sigma)}\delta^{(5)}(\sigma-\sigma')
\epsilon_{JK]LMN}\,.
}

What does all this mean for the particles of the $H$ field?
Let us study Fourier modes of $H$'s with $\pm p_I$ where  
$p_I=(0,0,0,0,p)$.
Then we can see that $H_{125}(p)=H_{125}(-p)^\dagger$ is
a dual variable to $H_{345}(-p)=H_{345}(p)^\dagger$ and similarly
for two other pairs which we get using cyclic permutations
$12,34\to 23,14\to 31,24$. So totally we have three physical polarizations  
of the tensor particle (which is of course the same number like
that of polarizations of photon in $4+1$ dimensional gauge theory).

We can also easily see from (\ref{hcom})
that the $p$-momentum modes of variables
that do not contain index ``5'', namely
$H_{123},H_{124},H_{134},H_{234}$ commute with everything.
They (more precisely their $\partial_5$ derivatives)
exactly correspond to the components of $dH$,
namely
  \eqn{dhcom}{(dH)_{1235},(dH)_{1245},(dH)_{1345},(dH)_{2345}}
that we keep to vanish as the constraint part of $dH=0$.
Let us just note that $(*_5dH)_I=0$ contains {\it four}
conditions only because $d(dH)=0$ is satisfied identically.  
Anyhow, there are no quantum mechanical variables
coming from the components of $(dH)_I$. The
variables $dH$ are the generators of the two-form gauge
invariance
  \eqn{twoforinv}{B_{IJ}\mapsto B_{IJ}+\partial_I\lambda_J
-\partial_J\lambda_I\,.}
Note that for $\lambda_I=\partial_I\phi$ we get a trivial transformation
of $B$'s which is the counterpart of the identity $d(dH)=0$.

But what about the zero modes, the integrals of $H_{IJK}$ over
the five-dimensional space? These are the ten fluxes that should
be quantized, i.e. they should belong to a lattice. In the 4+1 dimensional
SYM theory they appear as four electric and six magnetic fluxes.
In the matrix model
of M-theory on $T^4$ these ten variables are interpreted as four compact
momenta and six transverse membrane charges.

The fact that ``unpaired'' degrees of freedom
are restricted to a lattice
is an old story. For
instance, in the bosonic formulation of the heterotic string in 1+1
dimensions we have 16 left-moving
(hermitean)
bosons (``\asd field strengths'')
$\alpha^i$, $i=1,\dots,16$ with commutation relations
  \eqn{hetcom}{[\alpha^i(\sigma),\alpha^j(\sigma')]=i\delta'(\sigma
-\sigma')\delta^{ij}\,.}
After combining them to Fourier modes
  \eqn{hetfour}{\alpha^i(\sigma)=\sqrt{\frac 2\pi}\sum_{n\in\IZ}
\alpha^i_n e^{-2i\sigma n}
\quad\Leftrightarrow\quad
\alpha^i_n=\frac 1{\sqrt{2\pi}}\int_0^\pi \alpha^i(\sigma) 
e^{2i\sigma n}d\sigma\,,}
we get relations  
  \eqn{hettc}{[\alpha^i_m,\alpha^j_n]=m\delta_{m+n}\delta^{ij}\,,\quad
(\alpha^i_m)^\dagger=\alpha^i_{-m}}
and we can interpret $\alpha^i_n$ and $\alpha^i_{-n}$
for $n>0$ as annihilation
and creation operators respectively. The modes $\alpha^i_0$ are then
restricted to belong to a selfdual lattice. Roughly speaking,
$\alpha^i_0$ equals the total momentum and it equals to
the total winding vector due to selfduality -- but these two must
belong to mutually dual lattices. The lattice must be even
in order for the operator
  \eqn{lhet}{L=:\frac 12\sum_{n\in\IZ}
\alpha^i_{-m}\alpha^i_m: = :\frac 14\int_0^\pi \alpha(\sigma)^2d\sigma:}
to have integer eigenvalues. We see that the 480 ground level states
${\ket 0}_{\alpha^i_0}$
with $(\alpha_0^i)^2=2$ give the same value $L=1$ as the sixteen
lowest excited states $\alpha^i_{-1}{\ket 0}_{\alpha_0=0}$. These combine
to the perfect number 496 of the states.

\subsection{Correspondence to Super Yang Mills}

We will use the normalization of the gauge theory with Lagrangian and
covariant derivative as follows
\eqn{symconv}{
 \Lag=-\frac 1{4g^2}F^{\mu\nu}F_{\mu\nu}\,,\qquad
 D_\alpha=\partial_\alpha+iA_\alpha\,.
}
The hamiltonian for the $U(1)$ theory then can be written as  
($i,j=1,2,3,4$)
\eqn{symham}{
\ham_{SYM}=
  \frac 1{2g^2}\int d^4\sigma
  \left[\sum_i (E_i)^2 + \sum_{i<j} (F_{ij})^2\right].
}
Let us consider compactification on a rectangular $T^5$
(the generalization for other tori is straightforward)
of volume $V= L_1 L_2 L_3 L_4 L_5$.
We should get (\ref{symham}) from our hamiltonian. Let us
write $d^5\sigma$ as $L_5 d^4\sigma$ (we suppose that the fields are
constant in the extra fifth direction).
\eqn{osymham}{
  \ham_{SYM}^{(2,0)}=
  \frac {L_5}{4\pi}\int d^4\sigma\sum_{I<J<K} (H_{IJK})^2\,.
}
So it is obvious that we must identify (up to signs)
$F_{\alpha\beta}$ with $H_{\alpha\beta5}\cdot g\sqrt{L_5/(2\pi)}$
e.g.
\eqn{symid}{
  H_{234}=\frac{E_1}{g\sqrt{L_5/(2\pi)}}\,,\quad
  H_{125}=\frac{F_{12}}{g\sqrt{L_5/(2\pi)}}\,.
}
To change $A_i$ of the SYM theory by a constant, we must take the
phase $\phi$ of the gauge transformation to be a linear function
of coordinates. But it should
change by a multiple of $2\pi$ after we go around a circle. Thus
\eqn{phisym}{
  \phi=\frac{2\pi n_i}{L_i}\sigma_i\,,\quad
  A_i\to A_i+\frac{2\pi n_i}{L_i}\,.
}
The dual variable to the average value of $A_i$ is the integral  
of $E_i/g^2$. We just showed that the average
value of $A_i$ lives on a circle with radius
and therefore $L_1L_2L_3L_4\cdot E_i/g^2$ belongs to the lattice
with spacing $L_i$. Similarly, we can obtain a nonzero magnetic flux
from the configuration
($A_i$ can change only by a multiple of the quantum in (\ref{phisym}))
\eqn{magfl}{
  A_i=\frac{2\pi n_{ij}}{L_i}\cdot \frac{\sigma_j}{L_j}\,,
}
which gives the magnetic field
\eqn{magfi}{
  F_{ij}=\frac{2\pi n_{ij}}{L_i L_j\,}.
}
Therefore for the spacings of the average values of $E_i,F_{ij}$ we have
\eqn{fspac}{
\Delta E_i=\frac{g^2L_i}{L_1L_2L_3L_4},\quad
\Delta F_{ij}=\frac{2\pi}{L_i L_j}.
}
Looking at (\ref{symid}) we can write for the averages of $H$'s e.g.
\eqn{hspac}{
\Delta H_{234}=
\frac{gL_1}{L_1L_2L_3L_4\sqrt{L_5/(2\pi)}}\,,\quad
\Delta H_{125}=\frac{2\pi}{L_1L_2g\sqrt{L_5/(2\pi)}}\,,
}
which can be extended to a six-dimensionally covariant form only
using the following precise relation between the coupling
constant and the circumference $L_5$
\eqn{radcou}{
g = \sqrt{2\pi L_5}\,,
}
giving us the final answer for the spacing
\eqn{skok}{
\Delta H_{IJK}=\frac{2\pi}{L_I L_J L_K}\,.
}
The formula (\ref{skok}) can be also written as
\eqn{skokk}{
\frac 16  \oint H_{IJK}dV^{IJK}\in 2\pi\cdot \mathbbm{Z}\,,
}   
or (using antiselfduality) as
\eqn{skokl}{
 \Delta \int d^5\sigma H^{0IJ}= 2\pi L^I L^J\,,
}
in accord with the interpretation of $H$ as the current of dissolved
membranes (the integral in (\ref{skokl}) is the total membrane charge).

\subsection{Normalization of the current}

We can also work out the value of $c_1$ in (\ref{eqh}).
Let us write this equation for $\alpha\beta\gamma\delta=1234$.
\eqn{dhnum}{
  \partial_{[1}H_{234]}
  =\frac 14(\partial_1 H_{234}
  -\partial_2 H_{341}+\partial_3 H_{412}-\partial_4 H_{123})
  =\frac 14\partial_\alpha H^{05\alpha}
  =J_{1234}\,.}
We see from (\ref{symid}) that
  \eqn{jone}{J_{1234}=\frac 1{4g\sqrt{L_5/(2\pi)}}
\sum_{i=1}^4 \partial_i E_i.}
The integral of $\partial_i E_i$ should be an integer
multiple of $g^2$ (in these conventions) and
because of (\ref{jone}), the integral of $J_{1234}$
should be an integer multiple of
$\pi/2$ which was the way we determined the coefficient
in (\ref{eqh}).

\section{Appendix C: Identities}

\subsection{Identities for gamma matrices}
\begin{eqnarray}
\Ga^\a\Ga^\b &=& \Ga^{\a\b}+\eta^{\a\b}\\
\Ga^{\a'}\Ga^{\b'\g'}
&=&\Ga^{\a'\b'\g'} + \Ga^{\g'}\eta^{\b'\a'}
   - \Ga^{\b'}\eta^{\g'\a'}\,,
\\
\Ga^{\b'\g'}\Ga^{\a'} 
&=&\Ga^{\a'\b'\g'} - \Ga^{\g'}\eta^{\b'\a'}
   + \Ga^{\b'}\eta^{\g'\a'}\,,
\end{eqnarray}

\be
\Ga^{\a'}\Ga^{\b'\g'}
+\Ga^{\b'\g'}\Ga^{\a'} = 2\Ga^{\a'\b'\g'}
\ee
\be
\Ga^{\a'}\Ga_{\a\b} -\Ga_{\a\b}\Ga^{\a'}
    = 4\del_{\lbr\a}^{\a'}\Ga_{\b\rbr}
\ee
\be
\Ga^\d\Ga^{\a\b\g} = \Ga^{\d\a\b\g} + 3\eta^{\d[\a}\Ga^{\b\g]}
\ee
\be
\Ga^\d\Ga^{\a\b\g} + \Ga^{\a\b\g}\Ga^\d
   = 6 \eta^{\d[\a}\Ga^{\b\g]}
\ee
\be
\Ga^{\a'\b'\g'} \Ga_{\a\b} -
\Ga_{\a\b} \Ga^{\a'\b'\g'}
=
  12 \Ga^{\d'\lbr\b'\g'}
    \del_{\lbr\a}^{\a'\rbr}\eta_{\b\rbr\d'}
\ee
\be
\Ga^{\a'\b'\g'} \Ga_{\a\b} +
\Ga_{\a\b} \Ga^{\a'\b'\g'}
=
  -12 \del_{\lbr\a}^{\lbr\a'}\del_{\b\rbr}^{\b'}
\Ga^{\g'\rbr}+2\Ga^{\a'\b'\g'}{}_{\a\b}
\ee
\be
\GPR\Ga_{\u_1\u_2\cdots\u_k}
=(-1)^{k(k+1)/2}
{1\over {(6-k)!}} \ept{\u_1\u_2\cdots\u_k}{\nu_1\nu_2\cdots\nu_{6-k}}
  \Ga_{\nu_1\nu_2\cdots\nu_{6-k}}
\ee
\be
\ept{\u_1\u_2\cdots\u_k}{\nu_1\nu_2\cdots\nu_{6-k}}
= (-1)^{k}
  {\epsilon^{\nu_1\nu_2\cdots\nu_{6-k}}}_{\u_1\u_2\cdots\u_k}
\ee
\be
\ept{\u_1\cdots\u_{6-k}}{\nu_1\cdots\nu_{k}}
         \Ga^{\u_1\cdots\u_{6-k}}
= (-1)^{k(k-1)/2}(6-k)!\cdot\GPR\Ga^{\nu_1\cdots\nu_k}
\ee
\be
\GPR^2 = +1
\ee
\begin{eqnarray}
\Ga^{\a\b}\Ga_\b &=& 5\Ga^\a\\
\Ga^{\a}\Ga_{\a} &=& 6\Id,\\
\Ga^{\a}\Ga_{\u}\Ga_{\a} &=& -4\Ga_{\u},\\
\Ga^{\a}\Ga_{\u_1\u_2\cdots\u_k}\Ga_{\a} &=&
       (-1)^k (6-2k)\Ga_{\u_1\u_2\cdots\u_k},\\
\Ga^{\a_1\cdots\a_l}\Ga_{\a_1\cdots\a_l} &=&
       {{6!}\over {(6-l)!}} (-1)^{l(l-1)/2}\Id,\\
\Ga^{\a _1\cdots\a_l}\Ga_{\u}\Ga_{\a_1\cdots\a_l} &=&
(-1)^{l(l+1)/2}{{(6-2l)5!}\over {(6-l)!}}\Ga_\u\,.
\end{eqnarray}
\begin{equation}
\begin{array}{rclcrcl}
\Ga^{\a\b}\Ga_{\a\b} &=& -30\Id,
& &\Ga^{\a\b\g}\Ga_{\a\b\g} &=& -120\Id,\\
\Ga^{\a\b}\Ga_\u\Ga_{\a\b} &=& -10\Ga_\u,
& &\Ga^{\a\b\g}\Ga_{\u}\Ga_{\a\b\g} &=& 0,\\
\Ga^{\a\b}\Ga_{\u\nu}\Ga_{\a\b} &=& 2\Ga_{\u\nu}\,,
& &\Ga^{\a\b\g}\Ga_{\u\nu}\Ga_{\a\b\g} &=& 24\Ga_{\mu\nu}\,,\\
\Ga^{\a\b}\Ga_{\u\nu\s}\Ga_{\a\b} &=& 6\Ga_{\u\nu\s}\,,
& &\Ga^{\a\b\g}\Ga_{\u\nu\s}\Ga_{\a\b\g} &=& 0.
\end{array}
\end{equation}
Derivation for last equations:
\begin{equation}\begin{array}{rcl}
\Ga^{\a\b\g}\Ga_{\u\nu(\s)}\Ga_{\a\b\g} &=&
(\Ga^{\b\g}\Ga^{\a}
+ \Ga^{\g}\eta^{\b\a} - \Ga^{\b}\eta^{\g\a})
\Ga_{\u\nu(\s)}\Ga_{\a\b\g} =
\Ga^{\b\g}\Ga^{\a}\Ga_{\u\nu(\s)}\Ga_{\a\b\g}
\\&=&
\Ga^{\b\g}\Ga^{\a}\Ga_{\u\nu(\s)}
(\Ga_{\a}\Ga_{\b\g}+\Ga_\b\eta_{\a\g}-\Ga_\g\eta_{\a\b})
\\ 
&=& \Ga^{\b\g}\Ga^{\a}\Ga_{\u\nu(\s)}\Ga_{\a}\Ga_{\b\g} +
2 \Ga^{\g\b}\Ga_\b\Ga_{\u\nu(\s)}\Ga_\g.
\end{array}\end{equation}
   
\begin{equation}\begin{array}{rcl}
\Ga^A\Ga_A &=& 5\Id\,,\\
\Ga^A\Ga_B\Ga_A &=& -3\Ga_B\,,\\
\Ga^A\Ga_{BC}\Ga_A &=& \Ga_{BC}\,,\\
\Ga^A\Ga_{B_1 B_2 \cdots B_k}\Ga_A &=&
(-1)^k (5-2k)\Ga_{B_1 B_2 \cdots B_k}\,.
\end{array}
\end{equation}

\be
\tr{\Ga^{\u_1\u_2\cdots\u_k}\Ga_{\nu_1\nu_2\cdots\nu_k}}
 = 32 k! (-1)^{{{k(k-1)}\over {2}}}
\del_{\lbr\nu_1}^{\lbr\u_1}\del_{\nu_2}^{\u_2} \cdots
\del_{\nu_k\rbr}^{\u_k\rbr}.
\ee
\begin{eqnarray}
\lefteqn{\tr{\Ga^{\u_1\cdots\u_k}\Ga^{A_1 \cdots A_l}
    \Ga_{\nu_1\cdots\nu_k}\Ga_{B_1\cdots B_l}}}\qquad\qquad\nn\\
 &=& 32 k! l! (-1)^{{{(k+l)(k+l-1)}\over {2}}} 
\del_{\lbr\nu_1}^{\lbr\u_1}\del_{\nu_2}^{\u_2} \cdots
\del_{\nu_k\rbr}^{\u_k\rbr}
\del_{\lbr B_1}^{\lbr A_1}\del_{B_2}^{A_2} \cdots
\del_{B_k\rbr}^{A_k\rbr}\,.
\end{eqnarray}

\subsection{Fierz rearrangments}
We need an identity of the form
\be
M_{mn} \equiv (\e_1)_m (\eb_2)_n
 = (\sum_{k=0}^6 \sum_{l=0}^2 C_{\u_1\cdots\u_k A_1\cdots A_l}
         \Ga^{\u_1\cdots\u_k}\Ga^{A_1\cdots A_l})_{mn}\,.
\ee
($l\le 2$ because $\Ga^{01\dots 10} = 1$.)
We then get:
\be
C_{\u_1\cdots\u_k A_1\cdots A_l}
 = {{(-1)^{{{(k+l)(k+l-1)}\over{2}}}}\over {32 k! l!}}
   \tr{M \Ga_{\u_1\cdots \u_k}\Ga_{A_1\cdots A_l}}\,.
\ee
Now we take
\be
\e_2 = -\GPR\e_2\,,\qquad
\e_1 = -\GPR\e_1\,.
\ee
and rearrange $M = \e_1\eb_2$.
Now $\GPR M = -M = -M \GPR$ and we see that only terms with
odd $k$ survive.
\begin{eqnarray}
M &\equiv& \e_1\eb_2
\nn\\
 &=& \left(
-\frac{(\eb_2\Ga_\u\e_1)}{32}\Ga^\u
      +\frac{(\eb_2\Ga_\u\Ga_{A}\e_1)}{32}\Ga^\u\Ga^{A}
      +\frac{(\eb_2\Ga_\u\Ga_{AB}\e_1)}{64}\Ga^\u\Ga^{AB}\right)
     (1+\GPR)
\nn\\&&
      +{{1}\over {192}}(\eb_2\Ga_{\u\nu\s}\e_1)\Ga^{\u\nu\s}
      -{{1}\over
{192}}(\eb_2\Ga_{\u\nu\s}\Ga_{A}\e_1)\Ga^{\u\nu\s}\Ga^A\nn\\
      &&-{{1}\over
{384}}(\eb_2\Ga_{\u\nu\s}\Ga_{AB}\e_1)\Ga^{\u\nu\s}\Ga^{AB}\,.
\\ 
N &\equiv& \e_1\eb_2 - \e_2\eb_1
\nn\\
 &=& \left(-{{1}\over {16}}(\eb_2\Ga_\u\e_1)\Ga^\u
      +{{1}\over {16}}(\eb_2\Ga_\u\Ga_{A}\e_1)\Ga^\u\Ga^{A}\right)
     (1+\GPR)
\nn\\&&
-{{1}\over{192}}(\eb_2\Ga_{\u\nu\s}\Ga_{AB}\e_1)\Ga^{\u\nu\s}\Ga^{AB}\,.
\\
L &\equiv& \e_1\eb_2 + \e_2\eb_1
\nn\\
 &=&
     {{1}\over {32}}(\eb_2\Ga_\u\Ga_{AB}\e_1)\Ga^\u\Ga^{AB}
     (1+\GPR)
\nn\\&&
      +{{1}\over {96}}(\eb_2\Ga_{\u\nu\s}\e_1)\Ga^{\u\nu\s}
      -{{1}\over {96}}(\eb_2\Ga_{\u\nu\s}\Ga_{A}\e_1)\Ga^{\u\nu\s}\Ga^A\,.
\end{eqnarray}
where we have used, e.g.
\be
\eb_2\Ga_{\u\nu\s}\e_1 = \eb_1\Ga_{\u\nu\s}\e_2\,.
\ee
For opposite chirality spinors we have to replace $\GPR$ by $-\GPR$.

For, perhaps, future use, we will also calculate this for
$M = \psi_1\eb_2$ with
\be
\e_2 = -\GPR\e_2,\qquad
\psi_1 = \GPR\psi_1\,.
\ee
\begin{eqnarray}   
M &\equiv& \psi_1\eb_2 = \left(-{{1}\over {32}}(\eb_2\psi_1)\Id
      -{{1}\over {32}}(\eb_2\Ga_{A}\psi_1)\Ga^{A}
      +{{1}\over {64}}(\eb_2\Ga_{AB}\psi_1)\Ga^{AB}\right. 
\nn\\&&\left.   
      +{{1}\over {64}}(\eb_2\Ga_{\u\nu}\psi_1)\Ga^{\u\nu}
      +{{1}\over {64}}(\eb_2\Ga_{\u\nu}\Ga_{A}\psi_1)\Ga^{\u\nu}\Ga^A
     \right.\nn\\ &&\left. -{{1}\over
{128}}(\eb_2\Ga_{\u\nu}\Ga_{AB}\psi_1)\Ga^{\u\nu}\Ga^{AB}
\right)
      (1+\GPR)\,.
\end{eqnarray}
we also need
\begin{eqnarray}
N &\equiv& \Ga_{\a\b}\psi_1\eb_2\Ga^{\a\b}
= \left({{15}\over {16}}(\eb_2\psi_1)\Id
      +{{15}\over {16}}(\eb_2\Ga_{A}\psi_1)\Ga^{A}
      -{{15}\over {32}}(\eb_2\Ga_{AB}\psi_1)\Ga^{AB}\right. 
\nn\\&&\left.
      +{{1}\over {32}}(\eb_2\Ga_{\u\nu}\psi_1)\Ga^{\u\nu}
      +{{1}\over {32}}(\eb_2\Ga_{\u\nu}\Ga_{A}\psi_1)\Ga^{\u\nu}\Ga^A
      \right. \nn\\&&\left.
-{{1}\over{64}}
(\eb_2\Ga_{\u\nu}\Ga_{AB}\psi_1)\Ga^{\u\nu}\Ga^{AB}\right)
      (1+\GPR).
\end{eqnarray}
and
\begin{eqnarray}
K &\equiv& \Ga_A\psi_1\eb_2\Ga^A = \left(-{{5}\over {32}}(\eb_2\psi_1)\Id
      +{{3}\over {32}}(\eb_2\Ga_{A}\psi_1)\Ga^{A}
      +{{1}\over {64}}(\eb_2\Ga_{AB}\psi_1)\Ga^{AB}  \right.
\nn\\&& \left.
      +{{5}\over {64}}(\eb_2\Ga_{\u\nu}\psi_1)\Ga^{\u\nu}
      -{{3}\over {64}}(\eb_2\Ga_{\u\nu}\Ga_{A}\psi_1)\Ga^{\u\nu}\Ga^A
      \right. \nn\\&&\left. -{{1}\over
{128}}(\eb_2\Ga_{\u\nu}\Ga_{AB}\psi_1)\Ga^{\u\nu}\Ga^{AB}
\right)
      (1+\GPR)\,.
\end{eqnarray}
     
\subsection{A few notes about $\mfrak{spin}(5,1)$}

We use the eleven-dimensional language for the spinors. But nevertheless
one could be confused by some elementary facts concerning
the reality condition
for the spinor $(4,4)$ of $\mfrak{spin}(5,1)\times \mfrak{spin}(5)$.
The spinor
representation
$4$ of $\mfrak{spin}(5)$ is quaternionic (pseudoreal). Therefore
$(4,4)$ of
$\mfrak{spin}(5)\times \mfrak{spin}(5)$ is a real 16-dimensional
representation.
But one might think that spinor $4$ of $\mfrak{spin}(5,1)$ is {\it
complex}
so that we cannot impose a reality condition for the $(4,4)$   
representation.
      
But of course, this is not the case. The spinor representation $4$ of
$\mfrak{spin}(5,1)$ is {\it quaternionic} as well since the algebra
$\mfrak{spin}(5,1)$
can be
understood also as $\mfrak{sl}(2,\mathbbm H)$ of $2\times 2$
quaternionic matrices with
unit determinant of its $8\times 8$ real form. This has the right
dimension   
\eqn{fifteen}{4\cdot 4 - 1 = 15 = \frac{6\cdot 5}{2\cdot 1}\,.}
In the language of complex matrices, there is a matrix $j_1$ so that
\eqn{struct}{(j_1)^2=-1\,,\qquad  j_1M_1=\bar M_1 j_1}
for all $4\times 4$ complex matrices $M_1$ of $\mfrak{spin}(5,1)$. Of
course,
for the $4\times 4$ matrices $M_2$ in $\mfrak{spin}(5)$ there is also
such
a matrix $j_2$ that
\eqn{structwo}{(j_2)^2=-1\,,\qquad  j_2M_2=\bar M_2 j_2\,.}
An explicit form for the equations (\ref{struct})-(\ref{structwo}) is
built from $2\times 2$ blocks
\eqn{struexam}{j_1=\tb{{rr}\circ&1\\-1&\circ}\,,\qquad
M_1=\tb{{rr}\alpha&\beta\\ -\bar\beta&\bar\alpha}\,.}
In the $(4,4)$ representation of $\mfrak{spin}(5,1)\times
\mfrak{spin}(5)$ the
matrices
are given by $M=M_1\otimes M_2$ and therefore we can define a matrix $j$
that shows that $M$ is equivalent to a real matrix
\eqn{realcond}{j=j_1\otimes j_2\,,\quad
j^2=1\,,\quad
jM=j_1M_1\otimes j_2M_2=\bar M_1j_1\otimes \bar M_2j_2=\bar M j\,.}
The algebra $\mfrak{spin}(5,1)$ is quite exceptional between the other
forms
of $\mfrak{spin}(6)$. The algebra $\mfrak{so}(6)$ is isomorphic to
$\mfrak{su}(4)$,
algebra $\mfrak{so}(4,2)$ to $\mfrak{su}(2,2)$ and algebra
$\mfrak{so}(3,3)$ to $\mfrak{su}(3,1)$.
The other form of $\mfrak{su}(4)$ isomorphic to $\mfrak{so}(5,1)$ is
sometimes
denoted
$\mfrak{su}^*(4)$ but now we can write it as $\mfrak{sl}(2,\mathbbm H)$ as
well
(the generators are  
$2\times 2$ quaternionic matrices with vanishing real part of the trace).
From the notation $\mfrak{sl}(2,\mathbbm H)$ it is also obvious that
$\mfrak u(2,\mathbbm H)=\mfrak{usp}(4)$ forms
a subgroup (which is isomorphic to $\mfrak{so}(5)$).

%% file: chap5-nonsusy.tex
\chapter{Nonsupersymmetric matrix models}

In this chapter I construct the matrix description for a twisted version
of the IIA string theory on $S^1$ with fermions antiperiodic around a
spatial circle.  The result is a $2+1$-dimensional $\U(N) \times \U(N)$
nonsupersymmetric Yang-Mills theory with fermionic matter transforming
in the $({\bf N},{\bf \bar N})$. The the two $\U(N)$'s are exchanged
if one goes around a twisted circle of the worldvolume. Relations with
type~0 theories are explored and we find type~0 matrix string limits
of our gauge theory.  We argue however that most of these results are
falsified by the absence of SUSY nonrenormalization theorems and that
the models do not in fact have a sensible Lorentz invariant space time
interpretation.

\setlength{\unitlength}{1mm}

\section{The conformal field theory description}

Matrix Theory~\cite{bfss} has been used to describe M-theory with 32
supercharges in 8,9,10 and 11 dimensions as well as various
projections of this theory.  In this chapter we would like to study a
nonsupersymmetric Matrix model in order to obtain a better
understanding of SUSY breaking in string theory. The problems of the
model that we study show that SUSY breaking leads to rather disastrous
consequences.  However, we point out in the conclusions that the
restriction to the light cone frame prevents us from abstracting
completely clearcut lessons from this exercise.

Before proceeding, we note that after this chapter was completed (but
before we had become convinced that the results were worth publishing)
another paper on Matrix models of nonsupersymmetric string theories
appeared~\cite{jesus}.  We do not understand the connection between
the model presented there and the one we study.  Another recent paper
on nonsupersymmetric compactifications, with considerations related to
ours is~\cite{lust}.  Our results do not agree with the suggestion of
these authors that nonsupersymmetric compactifications lead to
Poincar\'e invariant physics.

Let us start with the description of the conformal field theory that
describes the model we are interested in. We start with the type~IIA
theory.  Compactification on a circle of radius $R$ can be described
as modding out the original theory by a symmetry isomorphic to $\IZ$,
consisting of the displacements by $2\pi k R$, $k\in\IZ$, in the
chosen direction.  We can write those displacement as $\exp(2\pi i
kR\hat p)$. A ``GSO-like'' projection by this operator now guarantees
that the total momentum of a string is a multiple of $1/R$. We also
have to add ``twisted sectors'' where the trip around the closed
string is physically equivalent to any element of the group that we
divided by. Those sectors are wound strings,
$X(\sigma+2\pi)=X(\sigma)+2\pi Rw$, where $w\in\IZ$ is the winding
number.

Such a compactification preserves all 32 supercharges. We will study a
more complicated model which breaks the supersymmetry completely. The
symmetry isomorphic to $\IZ$ will be generated by (the direction of
the circle is denoted by the index 2) 
\eqn{generator}{G=\exp(2\pi iR\hat p_2)(-1)^F\,,}
\looseness=1 where $(-1)^F$ counts the spacetime statistics (or spin; in the
Green-Schwarz formalism it is also equivalent to the worldsheet
spin). Because of this extra factor of $(-1)^F$ the physics becomes
very different. The fermionic fields of the spacetime effective field
theory become antiperiodic with the period $2\pi R_2$ while bosons are
still periodic. Such a boundary condition of course breaks
supersymmetry completely because it is impossible to define the (sign
of the) supercharge everywhere. Those antiperiodic boundary conditions
for the fermions are the same as those used in finite temperature
calculations, with Euclidean time replaced by a spatial circle.  This
compactification, introduced in~\cite{rohm} is motivated by
Scherk-Schwarz compactifications of supergravity~\cite{sesva}.  The
physical spectrum is obtained by requiring $G\ket\psi=\ket\psi$ and in
the twisted sectors corresponding to $G^w$, $w\in\IZ$, a trip around
the closed string is equivalent the shift by $2\pi w R_2$ \emph{times}
$(-1)^{wF}$. Because $(-1)^{wF}$ for odd $w$ anticommutes with
fermions in the Green-Schwarz formalism, the fermions $\theta$ must be
antiperiodic in the sectors with odd winding number.  Similarly, the
GSO projection $G\ket\psi=\ket\psi$ now does not imply that $p_2$ must
be a multiple of $1/R_2$. Looking at~(\ref{generator}) we see that
there are two possibilities. Either $(-1)^F$ is equal to $+1$ and
$p_2=n/R_2$ \emph{or} $(-1)^F=-1$ and $p_2=(n+1/2)/R_2$. We have
sectors with $p_2R_2$ both integer or half-integer, but for integer
$p_2R_2$ we project out all the fermions and for half-integer $p_2R_2$
we project out all the bosonic states.

Thus we have four kinds of sectors; odd or even $w$ can be combined
with integer or half-integer $p_2R_2$.  For even $w$ the boundary
conditions are as in the untwisted theory but for odd values of $w$ we
must impose \emph{antiperiodic} boundary conditions for the
Green-Schwarz fermions $\theta$.

Since the sectors with even values of $w$ are well-known (we just keep
only bosons or only fermions according to $p_2$), we note only that in
the sectors with odd values of $w$ the ground state has $8\times
(-1/24-1/48)=-1/2$ excitations both in the left-moving and
right-moving sector (the same as the ground state of a NS-NS sector in
the RNS formalism). In other words, the ground state is a
nondegenerate bosonic tachyon. We must also take into the account the
condition ``$L_0=\tilde L_0$'', more precisely ($N_L=N_R=0$ for the
ground state of even $w$ and $N_L=N_R=-1/2$ for odd $w$ and we define
$n=p_2R_2$)
\eqn{levelmat}{N_L=N_R+nw\,.} 

Furthermore for integer $n$ we must project all the fermions out of
the spectrum. For even $w$ (which means also even $nw$) this leaves us
with the bosonic states of the untwisted IIA theory. For odd $w$
(which implies integer $nw$) the fermionic modes are half-integers and
we see that due to~(\ref{levelmat}) the number of left-moving and
right-moving fermionic excitations must be equal ${}\bmod 2$. Therefore the
level matching condition~(\ref{levelmat}) automatically projects out
all the fermionic states.

Similarly for half-integer $n$ we must get rid of all the bosonic states.
If $w$ is even, $nw$ is integer and odd and apart from~(\ref{levelmat}) we
must also independently impose the condition $(-1)^F=-1$ and we get just
the fermionic part of the spectrum of the type~IIA string theory. However
for $w$ odd (which implies that $nw$ is half-integer and the fermions are
antiperiodic), we see a mismatch $1/2$ modulo $1$ between $N_L$ and $N_R$
in~(\ref{levelmat}), so the bosons are projected out automatically as a
result of the level-matching condition~(\ref{levelmat}).

In both cases we saw that the $(-1)^F$ projection was automatic in
sectors with odd values of $w$. It is a general property of orbifolds
that in the twisted sectors where a trip around the closed string is
equivalent to the symmetry $g$, the GSO projection
$g\ket\psi=\ket\psi$ is a direct consequence of the level-matching
condition.

What about the tachyons? For even values of $w$ we have a part of the
spectrum of type~IIA strings, so there is no tachyon. However for odd
values of $w$, we can find a tachyon. Recall that
\eqn{masstach}{m^2=\frac{4}{\alpha'}N_L+
\left(\frac{n}{R_2}-\frac{wR_2}{\alpha'}\right)^2
=\frac{4}{\alpha'}N_R+\left(\frac{n}{R_2}+\frac{wR_2}{\alpha'}\right)^2.}

For $n=0$, $w=\pm 1$, we see that the ground level $N_L=N_R=-1/2$ is
really tachyonic for $R_2^2/ \alpha' <2$. For sufficiently small
radius $R_2$ there is a bosonic tachyon in the spectrum. If $R_2$ is
really small, also states with $\pm w=3,5,\dots$ (but always $n=0$)
may become tachyons.

However for $n=1/2$ (fermionic sector) and $w=1$ we see from
(\ref{levelmat}) that the lowest possible state has $N_R=-1/2$
and $N_L=0$ (one $\theta_{-1/2}$ left-moving excitation) which means that
$m^2$ expressed in~(\ref{masstach}) is never negative. This means that the
tachyons can appear only in the bosonic spectrum (as scalars).

There are many interesting relations of such a
nonsupersymmetric theory with other theories of this kind. For
example, by a Wick rotation we can turn the twisted spatial circle
\pagebreak[3]
into a time circle. The antiperiodic boundary conditions for the
fermionic field then describe a path integral at finite
temperature. The appearance of the tachyon in the spectrum for
$R_2<\sqrt{2\alpha'}$ is related to the Hagedorn phase transition. The
infinite temperature, or zero $R_2$ limit of our model gives the
type~0 theories.

The type~0 theories (-0A and -0B) are modifications of the type~II
theories containing bosonic states only in ``diagonal sectors'' NS-NS
and RR; we have also only one GSO projection counting the number of
left-moving minus right-moving fermionic excitations. The type~II
theories can be obtained as an $\IZ_2$ orbifold (making separate
projections on left-moving fermions) of the type~0 theories in the
R-NS formalism; the difference between type~IIA and type~IIB theories
is a sign of the projection in the RR sector; type~IIA is in a sense
type~IIB with a discrete torsion. Equivalently, we can also obtain
type~0 theories by orbifolding type~II theories but in the
Green-Schwarz formalism: type~0A and 0B theories can be described in
the Green-Schwarz formalism by the same degrees of freedom as the
corresponding type~II theories, but we must include both PP and AA
sector and perform the corresponding (diagonal) GSO-projection.

\section{The matrix model}

In a first naive attempt to construct a model describing the type~0A
theory, we would probably make a local orbifold (orbifolding in Matrix
theory was described in~\cite{lmztwo}) of the original Matrix Theory,
corresponding to the $\IZ_2$ orbifold of the worldsheet theory in the
Green-Schwarz formalism. We would represent the operator $(-1)^F$ by a
gauge transformation e.g.\ $\sigma_3\otimes {\mathbbm 1}$ and the bosonic
matrices would then be restricted to the block diagonal form, reducing the
original group $\U(2N)$ into $\U(N)\times \U(N)$ with fermions in the
off-diagonal blocks i.e.\ transforming as $({\bf N},{\bf \bar N})$. However
the coordinates $X$ of the two blocks would suffer from an instability
forcing the eigenvalues of the two blocks to escape from each other: the
negative ground-state energy of the ``off-diagonal'' fermions is not
cancelled by a contribution of bosons and therefore the energy is
unbounded from below even for finite $N$.

Bergman and Gaberdiel however pointed out~\cite{bg} that it is more
appropriate to think about type~0A string theory as a type~IIA theory
orbifolded by a $\IZ_2$ group generated by the usual $(-1)^F$
\emph{times} the displacement by half of the circumference of the
corresponding M-theoretical circle. Of course, this displacement could
not be seen perturbatively. We are clearly led to the Scherk-Schwarz
compactification of M-theory. We will thus attempt to construct a more
sophisticated matrix model.  We will find a model which naively
incorporates all of the duality conjectures of Bergman and Gaberdiel
and has type~0A,B and Rohm compactified type~IIA,B matrix string
limits.  In the end, we will find that many of our naive arguments are
false, due to the absence of SUSY nonrenormalization theorems, and
that the model we construct does not have a Lorentz invariant large-$N$
limit.  We argue that this implies that Scherk-Schwarz compactified
\mth\ does not have a Lorentz invariant vacuum.

Let us start with a review of untwisted M-theory compactified on a circle.
The algorithm~\cite{lmztwo} to mod out the BFSS model by a group of
physical symmetries $H$ is to enlarge the gauge group $\U(N)$ and identify
elements of $H$ with some elements $g$ of the gauge group. It means that
the matrices $Y=X,\Pi,\theta$ are constrained to satisfy
\eqn{orbing}{h(Y)=g_hYg_h^{-1}\,,\qquad g_h\in \U(N)\,,\quad h\in H\,.}

We wrote $gYg^{-1}$ because $Y$ transform in the adjoint representation
and the physical action of the symmetries on $Y$ is denoted $h(Y)$. To
obtain the $O(N)$ matrix model describing a single Ho\hacek rava-Witten
domain wall, we can set $g_h={\mathbbm 1}$ and just postulate $Y$ to be
symmetric with respect to the symmetry (consisting of the reflection
of $X^1,\Pi^1$, multiplying spinors $\theta$ by $\gamma_1$ and transposing
all the matrices). Therefore $X^1$ and half $\theta$'s
become antisymmetric hermitean matrices in the adjoint of $O(N)$ while
the other $X$'s and $\theta$'s become symmetric real matrices. The
naive matrix description of the heterotic strings on tori together with
the sectors and GSO-like projections on the heterotic matrix strings
was obtained in~\cite{bamo}.

Compactification of $X^2$ on a circle with radius $R_2$ can be done in
a similar way. We just postulate the set of possible values of the
$\U(N)$ indices to be $\{1,2,\dots, N\}\times (0,2\pi)_{\rm circle}$
and represent the physical symmetry $\exp(2\pi i R_2\hat p)$ by the
gauge transformation ${\mathbbm 1}\otimes\exp(i\sigma_2)$. Note that the
matrices now have two discrete and two continuous
``indices''. Postulating~(\ref{orbing}) tells us that the matrices
must commute with any function of $\sigma$:
\eqn{gaufil}{X^2_{mn}(\sigma_2,\sigma'_2)=X^2_{mn}(\sigma_2)\delta(\sigma_2
-\sigma'_2)-i\delta'(\sigma_2-\sigma'_2)}
and similarly for the other matrices $X^i$ and $\theta$ (without the
$\delta'(\sigma_2-\sigma'_2)$ term). Now if we understand the
summation over the sigma index as integration and ignore the factor
$\delta(0)$ in the trace, the BFSS hamiltonian becomes precisely the
hamiltonian of SYM theory with $\sigma_2$ being an extra coordinate.

``Matrices'' of the form~(\ref{gaufil}) can be also expressed in the
terms of the Fourier modes as done first by Taylor~\cite{taylor}.  The
extra Fourier mode indices replacing $\sigma_2,\sigma'_2$ are denoted
$M,N$ and~(\ref{gaufil}) becomes
\eqn{gaudva}{(X^2_{M+1,N+1})_{mn}=(X^2_{M,N})_{mn}+2\pi R_2
\delta_{M,N}\delta_{mn}}
and similarly for the other matrices without the last term.

We will study a compactification of M-theory on a twisted $T^2$ so we
will use two worldvolume coordinates $\sigma_1,\sigma_2$ to represent
those two circles.  What about the $(-1)^F$ twist which modifies the
compactification of $X^2$? Shifting both ends of the open strings
$M,N\to M+1,N+1$ as in~(\ref{gaudva}) must be accompanied by $(-1)^F$
which commutes with bosons but anticommutes with spacetime
fermions. In the Green-Schwarz formalism it also anticommutes with the
$\theta$'s. So the structure of the bosonic matrices is unchanged and
the condition for $\theta$'s will be twisted:
\eqn{gautheta}{(\theta_{M+1,N+1})_{mn}=-(\theta_{M,N})_{mn}.}
In the continuous basis this is translated to
\eqn{gauc}{\theta_{mn}(\sigma_2,\sigma'_2)=\theta_{mn}(\sigma_2)\delta(\sigma_2
-\sigma'_2+\pi).}
This can be described by saying that $\theta$ has nonzero matrix
elements between opposite points of the $\sigma_2$ circle. We will
often use this ``nonlocal'' interpretation of the resulting theory
even though the theory can be formulated as a conventional
nonsupersymmetric gauge theory with fermionic matter, as we will show
in a moment.

\subsection{The $\U(N)\times \U(N)$ formalism}

In order to get rid of the nonlocality, we must note that if we
identify the opposite points with $\sigma_2$ and $\sigma_2+\pi$, so
that $\sigma_2$ lives on a circle of radius $\pi$, everything becomes
local.  By halving the circle, we double the set of bosonic
fields. The two $\U(N)$ groups at points $\sigma_2$ and $\sigma_2+\pi$
are completely independent, so that the gauge group becomes
$\U(N)\times \U(N)$. We should also note that if we change $\sigma_2$
by $\pi$, the two factors $\U(N)$ exchange; this is an important
boundary condition.

The bosonic fields thus transform in the adjoint representation of
$\U(N)\times \U(N)$. What about the fermions $\theta$? We saw that the
two ``matrix indices'' $\sigma_2$ and $\sigma'_2$ differ by $\pi$. One
of them is thus associated with the gauge group $\U(N)$ at point
$\sigma$, the other with the $\U(N)$ at point $\sigma+\pi$ which is
the other $\U(N)$ factor in the $\U(N)\times \U(N)$ formulation. In
other words, $\theta$'s transform as $({\bf N}, {\bf \bar N})$ under
the $\U(N)\times \U(N)$. This is a complex representation of (complex)
dimension $N^2$ and the complex conjugate $\theta^\dagger$s transform
as $({\bf \bar N},{\bf N})$.  In the old language, $\theta$ and
$\theta^\dagger$ differed by $\pi$ in $\sigma_2$ or in other words,
they corresponded to the opposite orientations of the arrow between
$\sigma_2$ and $\sigma_2+\pi$.  The number of real components is
$2N^2$ (times the dimension of the spinor 16), the same as the
dimension of the adjoint representation. This should not surprise us
since for $R_2\to\infty$ we expect that our nonsupersymmetric model
mimics the physics of the supersymmetric model.

We could also check that the commutation relations derived from the
``nonlocal'' orbifold formulation agree with the canonical commutation
relations of the $\U(N)\times \U(N)$ nonsupersymmetric gauge theory
with the matter in $({\bf N},{\bf \bar N})$. These theories are very
similar to the ``quiver'' theories of Moore and Douglas~\cite{quiver},
but with a peculiar boundary condition that exchanges the two $\U(N)$
groups as we go around the twisted circle.

\subsection{Actions for the local and nonlocal formulations}

We will be considering both descriptions. In one of them, the
Yang-Mills theory has gauge group $\U(N)$ and is defined on a time
coordinate multiplied by a two-torus with circumferences $1/R_1,1/R_2$
(instead of $2\pi$ employed in the previous section) where $R_1,R_2$
are the radii of the spacetime circles in Planck units\footnote{To
simplify the presentation, we choose the convention for the numerical
constants in various dimensionful quantities to agree with these
statements.}  and the fermions are nonlocal degrees of freedom
(arrows) pointing from the point $(\sigma_1,\sigma_2)$ to the point
$(\sigma_1,\sigma_2+1/2R_2)$.  We will call this picture ``nonlocal''.

We will also sometimes use a ``local'' picture where the coordinate
$\sigma_2$ is wrapped twice and its circumference is only $1/2R_2$.
In the local picture, the gauge group is $\U(N)\times \U(N)$ and these two
factors exchange when we go around the $\sigma_2$ circle so that
the ``effective'' period is still equal to $1/R_2$:
\eqn{boundcond}{A^{\alpha}_{ij}\left(\sigma_0,\sigma_1,\sigma_2+\frac 1{2R_2}\right)
=A^{1-\alpha}_{ij}(\sigma_0,\sigma_1,\sigma_2)\,.}
Here $\alpha=0,1$ is an index distinguishing the two factors in
$\U(N)\times \U(N)$. We suppresed the worldvolume vector index
$\mu=0,1,2$.  Indices $i,j$ run from $1$ to $N$; here $i$ spans ${\bf
N}$ and $j$ belongs to ${\bf \bar N}$.  Similar boundary conditions
are imposed on the scalars $X$ which also transform in the adjoint of
$\U(N)\times \U(N)$.  Both satisfy the usual hermiticity conditions.
Fermions $\theta$ (whose spacetime transformation rules are the same
as in the supersymmetric theory) transform in $({\bf N},{\bf \bar
N})$.  Writing them as $\theta_{ij}$, the index $i$ belongs to ${\bf
N}$ of the first $\U(N)$ and the index $j$ belongs to ${\bf \bar N}$
of the second $\U(N)$. In the same way, in
$(\theta^\dagger)_{ij}=(\theta_{ji})^\dagger$ the first index $i$
belongs to ${\bf N}$ of the second $\U(N)$ and the second index $j$
belongs to ${\bf \bar N}$ of the first $\U(N)$ so that $\Tr
\theta^\dagger\theta=\theta^\dagger_{ij}\theta_{ji}$ is invariant.
The boundary condition for $\theta$s reads
\eqn{boundth}{\theta_{ij}\left(\sigma_0,\sigma_1,\sigma_2+\frac 1{2R_2}\right)=
(\theta^\dagger)_{ij}(\sigma_0,\sigma_1,\sigma_2)\,.}
Of course, $\theta$ matrices are complex, they do not obey a hermiticity
condition. The lagrangian is $(i=1,\dots, 7)$
\begin{eqnarray}
{\cal L}&=& \sum_{\alpha=0,1}\Tr\left[-\frac{1}{4}
F_{(\alpha)}^{\mu\nu}F_{(\alpha),\mu\nu}
-\frac{1}{2}D_\mu X_{(\alpha)}^i D^\mu X_{(\alpha)}^i +\frac 14
[X^i_{(\alpha)},X^j_{(\alpha)}]^2 \right]+
\label{lagnas}\\\nonumber&&
+\Tr \left[i\theta^\dagger\gamma^i X^i_{(\alpha=0)} \theta
+i\theta\gamma^i X^i_{(\alpha=1)}\theta^\dagger 
+\theta^\dagger\gamma_\mu \partial^\mu \theta
+i\theta^\dagger\gamma_\mu A^\mu_{(\alpha=0)} \theta
+i\theta\gamma_\mu A^\mu_{(\alpha=1)}\theta^\dagger\right].
\end{eqnarray}
The trace always runs over $N\times N$ matrices.
We have put the dimensionful quantity $g_{YM}$ equal to one. The action is
simply
\eqn{akce}{{\cal A}=\int d\sigma^0\int_{0}^{1/R_1}d\sigma^1
\int_{0}^{1/(2R_2)}d\sigma^2 {\cal L}(\sigma^0,\sigma^1,\sigma^2)\,.}
For the purposes of the calculations of Feynman diagrams it is also useful
to write the action in the nonlocal (nl) formulation of the theory.
In this formulation, the period of $\sigma^2$ is doubled and equal
to $1/R_2$. The fields can be identified as follows (the dependences
on $\sigma^0,\sigma^1$ and indices $\mu,i$ are suppressed):
\begin{eqnarray}
X_{(\alpha)}(\sigma^2)&=&\!
X_{nl}\left(\sigma^2\!+\frac{\alpha}{(2R_2)}\right),\qquad
A_{(\alpha)}(\sigma^2)=
A_{nl}\left(\sigma^2\!+\frac{\alpha}{(2R_2)}\right),\quad \alpha=0,1,
\label{nelok}\\
\theta(\sigma^2)&=&\theta_{nl}(\sigma^2)\,,\qquad
\theta^\dagger(\sigma^2)=\theta^\dagger_{nl}(\sigma^2)=
\theta_{nl}\left(\sigma^2\!+\frac 1{(2R_2)}\right),\quad
0\leq\sigma^2\leq \frac1{2R_2}\,.\quad~~
\nonumber
\end{eqnarray}
All the equalities are $N\times N$ matrix equalities.
In this nonlocal language the action can be written as
\eqn{nonlocakce}{{\cal A}=\int d\sigma^0\int_{0}^{1/R_1}d\sigma^1
\int_{0}^{1/R_2}d\sigma^2 {\cal L}_{nl}(\sigma^0,\sigma^1,\sigma^2)\,,}
where (the subscript ``nl'' of all fields is suppressed)
\begin{eqnarray}
{\cal L}_{nl}&=&
\Tr\left[-\frac{1}{4}F^{\mu\nu}F_{\mu\nu}
-\frac{1}{2}D_\mu X^i D^\mu X^i +\frac 14 [X^i,X^j]^2 \right]+
\nonumber\\&&
+\Tr \left[\theta^\dagger\gamma_\mu \partial^\mu \theta
+i\theta^\dagger(\gamma^i X^i(\sigma^2)+\gamma_\mu A^\mu(\sigma^2))
\theta \right]+
\nonumber\\&&
+\Tr\left[i\theta \left(\gamma^i
X^i(\sigma^2+\frac{1}{2R_2})+\gamma_\mu
A^\mu(\sigma^2+\frac{1}{2R_2})\right) \theta^\dagger\right].
\label{nonloclag}
\end{eqnarray}
We denoted the $\sigma^2$ dependence in which one of the fermionic terms
is nonlocal.

\section{Alternative derivation and connection with type~0 theories}

\looseness=1 It has long been known~\cite{seibwit,dixonhar} that the
type~0$_{A,B}$
string theories in ten dimensions can be viewed as infinite
temperature limits of type~II$_{B,A}$ theories.  Rotating the
Euclidean time to a spacelike direction, this means that the zero
radius limits of Rohm compactifications are the type~0 theories.  It
is less well known (but, we believe, known to many experts) that the
finite radius Rohm compactifications are compactifications of the
type~0 theories on dual circles with a certain orbifold projection.
Indeed, both type~0 theories have two types of Ramond-Ramond fields
which are related by a discrete symmetry. This doubled number is a
consequence of having one GSO-projection only (the diagonal one). More
precisely, the operator \calr$=(-1)^{F_R}$ which counts the
right-moving fermionic excitations has eigenvalues $(+1)$ for half of
the RR-fields and $(-1)$ for the other half. Therefore $(-1)^{F_R}$ is
a generator of a $\IZ_2$ symmetry that exchanges the RR-fields in a
basis rotated by 45 degrees, i.e.\ ${\rm RR}_{+1}+{\rm RR}_{-1}$ with
${\rm RR}_{+1}-{\rm RR}_{-1}$ where ${\rm RR}_{\pm1}$ denotes the
fields with $(-1)^{F_R}=\pm1$.

Twisted compactification of type~0A or -0B with monodromy \calr\ i.e.\ the
orbifold of type~0 string theory on a circle of circumference $2L$ by the
symmetry ${\cal R}\,\exp(iLp)$ gives a string model T-dual to the Rohm
compactified type~IIB or type~IIA string, respectively.  We will refer to
the twisted circle as the Scherk-Schwarz circle when describing it from
the type~II point of view and as the \calr circle from the type~0 point of
view.

This T-duality is not hard to understand at the level of the string
spectrum. Because of the GSO projection, Scherk-Schwarz compactified
type~II theory contains bosonic states of integer momenta and
\pagebreak[3]
fermionic states of half-integer momenta (in appropriate units). The
dual type~0 string theory initially had bosonic excitations only (in
NS-NS and RR sectors).  But because of the extra orbifold by ${\cal
R}\,\exp(iLp)$, we obtain also fermions in NS-R and R-NS twisted
sectors with a half-integer winding. This agrees with the assumption
of T-duality. Apart from T-duality between type~IIA/IIB on a
Scherk-Schwarz circle and type~0B/0A on an \calr\ circle which we just
mentioned, we should be aware of the T-duality between type~0A and
type~0B string theory on a usual circle.

Now consider the DLCQ of \mth\ compactified on a Scherk-Schwarz
circle.  Using the logic of~\cite{whycorrect}, this is a zero coupling
limit of type~IIA string theory compactified on a Scherk-Schwarz
circle of Planck size, in the presence of $N$ D0-branes.  Using the
T-duality adumbrated in the previous paragraphs, this is weakly
coupled type~0B string theory on an \calr\ circle in the presence of
$N$ D-strings of the first kind; since the \calr\ monodromy exchanges
two types of D-strings, there must be an equal number of D-strings of
the other type. In other words, the D-strings are compactified on a
circle dual to the \mth\ Scherk-Schwarz circle, with \calr\ twisted
boundary conditions.

Now we can use the description of D-branes in type~0 theories
discovered by Bergman and Gaberdiel~\cite{bg}. As we have said, there
are two types of D-strings in type~0B theory, each of which has a
bosonic $1+1$ dimensional gauge theory on its world volume: open
strings stretched between two like D-strings contain bosonic states
only. These two types are related (exchanged) by the \calr\
symmetry. In the presence of closely spaced D-strings of both types,
there are additional fermionic degrees of freedom which transform in
the $(N,\bar{M}) [\oplus (\bar{N},M)]$ of the $\U(N) \times \U(M)$
gauge group: open strings stretched between two unlike D-strings
contain fermions only.  These fermions are spacetime spinors. In the
corresponding Seiberg limit, all the closed string states (including
tachyons) are decoupled and in the corresponding DKPS energy scale
only the massless open string states survive.

The result is a $1+1$ dimensional $\U(N) \times \U(N)$ gauge theory
with fermions in the bifundamental and a boundary condition that
exchanges the two $\U(N)$ groups as we go around the circle, a result
of the \calr\ monodromy. This is the same gauge theory we arrived at
by the orbifolding procedure of the previous section.

It is now easy to compactify an extra dimension on an ordinary circle
and obtain the $2+1$ dimensional gauge theory of the previous section
as the matrix description of Rohm compactification.  The double
T-duality in Seiberg's derivation can be done in two possible orders,
giving always the same result. In the next section, we will see that
at least formally we can rederive the various string theories as
matrix string limits of the gauge theory.  In particular, this will
provide a derivation of the Bergman-Gaberdiel duality relation between
Scherk-Schwarz compactification of \mth, and type~0A strings.

\section{The matrix string limits}

\subsection{Rohm compactified type~IIA strings}

In the limit where the spacetime radius $R_2$ goes to infinity, the
radius of the worldvolume torus $1/R_2$ goes to zero so that we also
have $1/R_1\gg 1/R_2$. Therefore the fields become effectively
independent of $\sigma^2$ up to a gauge transformation. Furthermore,
because of the boundary conditions exchanging the two $\U(N)$'s, the
expectation values of scalars in both $\U(N)$'s must be equal to each
other (up to a gauge transformation). This can be also seen in the
nonlocal formulation: in the limit $R_2\to\infty$ the fields must be
constant (up to a gauge transformation) on the long circle of
circumference $1/R_2$.

Thus in this limit we can classify all the fields according to how
they transform under the $\sigma^2$ independent gauge symmetry
$\U(N)$. In the nonlocal language this $\U(N)$ is just a ``global''
(but $\sigma^1$ dependent) symmetry. In the $\U(N)\times \U(N)$
language this is the diagonal symmetry $\U(N)$. In both cases, we find
that not only bosons but also fermions (transforming originally in
$({\bf N},{\bf \bar N})$) transform in the adjoint (the same as ${\bf
N}\otimes {\bf \bar N}$) of this $\U(N)$. There is only one set of
fields: the $\sigma^2$ independence causes the bosons in both
$\U(N)$'s to be equal and the complex matrices $\theta$ to be
hermitean.

In the matrix string limit we expect to get a matrix description of
the Scherk-Schwarz compactification of type~IIA strings on a long
circle. The appearance of the matrix strings (at a naive level) can be
explained as usual: most things work much like in the supersymmetric
matrix string theory~\cite{motlb,bs,dvv}.

In the nonlocal formulation, $\U(N)$ gauge group is broken completely
down to a semidirect product of $\U(1)^N$ and the Weyl group, $S_N$,
of $\U(N)$. Therefore the classical configurations around which we
expand are diagonalizable $N\times N$ matrices where the basis in
which they can be diagonalized can undergo a permutation $p\in S_N$
for $\sigma^1\to\sigma^1+1/R_1$:
\eqn{screwing}{X_i(\sigma^1)=U(\sigma^1)
\diag(x_i^1,x_i^2,\dots x_i^n)U^{-1},\qquad
\U\left(\sigma^1+\frac 1{R_1}\right)=U(\sigma^1)p\,.}

This is the mechanism of matrix
strings~\cite{motlb,bs,dvv}. Every permutation $p$ can be
decomposed into a product of cycles and each cycle of length $k$ then
effectively describe a ``long string'' with the longitudinal momentum
equal to $p^+=k/R^-$. For instance, a single cyclic permutation of $k$
entries (written as a $k\times k$ matrix $p$) describes a single
string:
\eqn{permutace}{p=
\pmatrix{ \circ& 1& \circ&\dots&\,\circ \cr
 \circ& \circ& 1&\dots&\circ\cr
\vdots&\vdots&\vdots&\ddots&1\cr
 1& \circ& \circ&\dots&\circ}.}
The definition~(\ref{screwing}) of $X_i$ creates effectively a string
of length $k$ (relatively to the circumference $1/R_1$). We can write
the eigenvalues as
\eqn{srouby}{x_i^m(\sigma^1)=x^{\rm long}_i
\left(\sigma^1+\frac{(m-1)}{R_1}\right), \qquad m=1,2,\dots, k}
where $x^{long}_i$ has period $k/R_1$. Assuming the $k/R_1$
periodicity of $x^{long}_i$ we can show $1/R_1$ periodicity of the
matrix (\ref{screwing}) with $p$ defined in~(\ref{permutace}).

The matrix origin of the level-matching conditions was first explained
in~\cite{dvv}: the residual symmetry $\IZ_k$ rotating the ``long''
string is a gauge symmetry and because the states must be invariant
under the gauge transformations, we find out that $L_0-\tilde L_0$
must be a multiple of $k$ (the length of the string) because the
generator of $\IZ_k$ can be written as $\exp(i(L_0-\tilde L_0)/k)$.
In the large-$N$ limit such states are very heavy unless $L_0=\tilde
L_0$ and we reproduce the usual level-matching conditions. In this
limit the discrete group $\IZ_k$ approximates the continuous group
quite well.

\subsection{Dependence on the fluxes}

In the conformal field theory of the Rohm compactification, sectors
with odd or even winding numbers should have antiperiodic or periodic
spinors $\theta$, respectively.  We want to find the analog of this
statement in the Matrix formulation.

Let us put $w$ units of the magnetic flux in the nonlocal representation
of our theory. The corresponding potential can be taken to be
\eqn{magflux}{A_{\mu=1}=2\pi R_1R_2\sigma_2\frac wN\,, \qquad
A_{\mu=2}=0}
Recall that the periods of $\sigma_1,\sigma_2$ are $1/R_1,1/R_2$.  In
the local $\U(N)\times \U(N)$ formulation, the fields in the region
$0\leq \sigma_2 \leq 1/2R_2$ from~(\ref{magflux}) define the block of
the first $\U(N)$ and the region $1/2R_2 \leq \sigma_2 \leq 1/R_2$
defines the second $\U(N)$. Note that for $\sigma_2\to\sigma_2+1/R_2$,
$\Tr A_{\mu=1}$ changes by $2\pi R_1 w$ which agrees with the
circumference of~$X^1$.

Now if we substitute the background~(\ref{magflux}) into~(\ref{lagnas})
we see that the contributions of the form $\theta A\theta$ from the last
two terms give us a contribution coming from the 
difference $\sigma_2\to\sigma_2+1/2R_2$ which is equal to
\eqn{magff}{i\Tr[\theta^\dagger \gamma_1\theta]\frac{\pi w R_1}{N}\,.}
Such a term without derivatives would make the dynamics nonstandard.
However it is easy to get rid of it by a simple redefinition (we
suppress $\sigma_0,\sigma_2$ dependence)
\eqn{magredef}{\theta(\sigma_1) \to \theta(\sigma_1) 
\exp\left(\frac{i\sigma_1 R_1\pi w}{N}\right).}
Note that under $\sigma_1\to\sigma_1+N/R_1$ which corresponds to a
loop around a matrix string of length $N$, $\theta$ changes by a
factor of $(-1)^w$. This confirms our expectations: in the sectors
with an odd magnetic flux (=\,winding number) the fermions $\theta$ are
antiperiodic.

We might also wonder about the electric flux (=\,compact momentum $p_2$)
in the direction of $\sigma_2$. As we have explained in the beginning,
this flux should be allowed to take $1/2$ of the original quantum so
that the sectors with half-integer electric flux contain just fermions
and the usual sectors with integer electric flux contain bosons only,
the other being projected out by the GSO conditions in both
cases. This behaviour should be guaranteed ``by definition'': the
operator $\exp(2\pi R_2\hat p_2)(-1)^F$ is identified with a gauge
transformation (namely $\exp(2\pi R_2\sigma_2)\otimes {\mathbbm 1}_{N\times
N}$).

However it might seem a little strange that the together with the
$\theta$ excitations one must also change the electric flux; it might
be useful to see the origin of the sectors of various electric flux
``microscopically''. We propose the following way to think about this
issue. The $\theta$ excitations in a compact space carry charge $\pm$
with respect to groups $\U(1)$ at the opposite points of
$\sigma_2$. The total charge vanishes therefore we do not have an
obstruction to excite $\theta$. However the charge does not vanish
locally, therefore we should accompany the excitation by an electric
flux tube running between $\sigma_2$ and $\sigma_2+1/2R_2$ in a chosen
direction (it is useful to think about it as a ``branch-cut'') and the
total electric flux induced by this excitation equals one half of the
quantum in the supersymmetric (untwisted) theory.

As a consequence of these observations, we see that our model contains
the string field theory Hilbert space of the Rohm compactification in
the large-$N$ limit.  At a very formal level, the dynamics of the
model in an appropriate limit of small radii and large Yang-Mills
coupling, appears to reduce to that of free Rohm strings.  However,
this is not necessarily a correct conclusion.  The analysis of the
moduli space lagrangian is done at the classical level, but the
apparent free string limit corresponds to a strongly coupled YM
theory.  In~\cite{bs} it was emphasized that the derivation of
Matrix string theory depends crucially on the nonrenormalization
theorem for the moduli space lagrangian.  We do not have such a
theorem here and cannot truly derive the free Rohm string theory from
out model.  This is only the first of many difficulties.

A further important point is that the $\U(1)$ gauge theory (or
$\U(1)\times \U(1)$ in the local formalism) leads to a free theory
which is identical to the conformal field theory in the matrix string
limit $1/R_1 \gg 1/R_2$. In particular we can see that the ground
state in the sectors with an odd magnetic flux has negative light cone
energy, and would have to be interpreted as a tachyon in a
relativistic theory.  Even if we assumed the clustering property to be
correct (in the next section we show that this property is likely to
be broken at the two-loop level), this tachyon would lead to
inconsistency in the large-$N$ limit: it would be energetically
favoured for a configuration in the $\U(N)$ theory to emit the $N=1$
tachyonic string --- and compensate the magnetic flux by the opposite
value of the flux in the remaining $\U(N-1)$ theory. The energy of
tachyon is of order $-N^0$ which is negative and $N$ times bigger than
the scale of energies we would hope to study in the large-$N$ limit
(only states with energies of order $1/N$ admit a relativistic
interpretation in the large-$N$ limit).

To make this more clear: in order to establish the existence of a
relativistic large-$N$ limit we would have to find states with dispersion
relation ${{\bf p}^2 + m^2\over N}$ in the model, as well as
multiparticle states corresponding to separated particles which scatter
in a manner consistent with relativity.  The observation of the previous 
paragraph shows that such states would generally be unstable to emission
of tachyons carrying the smallest unit of longitudinal
momentum.\footnote{Note that in SUSY Matrix Theory the excitations along
directions where the gauge group is completely broken down to $\U(1)$
factors have \emph{higher} energy than the states with large longitudinal
momentum.}  The only way to prevent this disaster is to lift the
moduli space.  However, once we imagine that the moduli space is lifted
it is unlikely that multiparticle states of any kind exist and the
model loses all possible spacetime interpretation.  We will investigate
the cluster property of our model below.  However, we first want to
investigate the type~0 string limits of our model.  As above, we will
work in a purely classical manner and ignore the fact that the moduli
space is lifted by quantum corrections.

\subsection{Type~0 matrix strings}

The Rohm compactified IIA string is the formal limit of our $2+1$
dimensional gauge theory when the untwisted circle of the Yang-Mills
torus is much
larger than the scale defined by the gauge coupling, while the twisted
circle is of order this scale or smaller.  We will now consider three
other limits.  The relation between the Yang-Mills parameters and 
the \mth\ parameters is
\begin{equation}
g_{YM}^2 = \frac{R}{L_1 \, L_2}
\qquad \Sigma_i = \frac{l_{pl}^3}{R\, L_i} \,,
\end{equation}
where $\Sigma_{1,2}$ is the untwisted (twisted) YM radius, $L_{1,2}$
are the corresponding \mth\ radii, and $R$ is the lightlike
compactification length.  In the type~0 string limit, we want to take
$L_2 \rightarrow 0$, with $\Sigma_2$ fixed (it is the string length
squared divided by $R$) and $L_1$ of order the string length).  The
latter restriction means that $g_{YM}^2\Sigma_1 $ is fixed.  The limit
is thus a $1+1$ dimensional gauge theory on a fixed length twisted
circle, with gauge coupling going to infinity.
\pagebreak[3]

\looseness=1 Restricting ourselves to classical considerations, we are led to the
classical moduli space of this gauge theory.  The bosonic sector of
the moduli space consists of two sets of independent $N \times N$
diagonalizable matrices.  However, in order to obtain configurations
which obey the twisted boundary conditions and have energy of order
$1/N$, one must consider only topological sectors in which the
matrices in the two gauge groups are identical.  Note however, that
since the bosonic variables are in the adjoint representation, they
are not affected by gauge transformations which are in the $\U(1)$
subgroup.  This additional freedom becomes important when we consider
the fermionic variables.  The boundary conditions on these allow one
other kind of configuration with energy of order $1/N$: considering
the fermions as $N \times N$ matrices, we can allow configurations in
which the diagonal matrix elements come back to minus themselves
(corresponding to the gauge transformation $\pm (1, - 1)$ in
$\U(N)\times \U(N)$) after a cycle with length of order $N$.  The
resulting low energy degrees of freedom are fermion fields on the
``long string'' with either periodic or antiperiodic boundary
conditions.  The gauge fields are vector like so in terms of left and
right moving fields we get only the PP and AA combinations of boundary
conditions.  The $O(8)$ chirality of the fermions is correlated to the
world sheet chirality as in IIA matrix string theory.  One also
obtains a GSO projection on these fermionic degrees of freedom by
imposing the gauge projection corresponding to the $(1, -1)$
transformation.  The resulting model is thus seen to be the type~0A
string theory, written in light cone Green-Schwarz variables.
Remembering that the $1+1$ twisted gauge theory was the matrix
description of Scherk-Schwarz compactification of M-theory, we
recognize that we have derived the conjecture of Bergmann and
Gaberdiel.

To obtain the 0B matrix string limit and the T-duality (on an
untwisted circle) between the two type~0 theories, we simply follow
the results of one of the present authors and Seiberg~\cite{bs} and
first take the strongly coupled Yang-Mills limit by going to the
classical moduli space and performing a $2+1$ dimensional duality
transformation.  This corresponds to both directions of the Yang-Mills
torus being much larger than the Yang-Mills scale.  We then do a
dimensional reduction to a $1+1$ dimensional theory to describe the 0B
and IIB string limits.  In the former, the twisted circle is taken
much larger than the untwisted one, while their relative sizes are
reversed in the latter limit.  After the duality transformation and
dimensional reduction the manipulations are identical to those
reported above.

A serious gap in the argument is the absence of $2+1$ superconformal
invariance.  In~\cite{bs} this was the crucial fact that enabled
one to show that the interacting type~IIB theory was Lorentz
invariant.  Here that argument fails.  We view this as an indication
that the spacetime picture derived from free type~0 string theory is
misleading.  We will discuss this further below.  Indeed, in the next
subsection we show that the cluster property which is at the heart of
the derivation of spacetime from Matrix Theory fails to hold in our
model.

\subsection{Breakdown of the cluster property}

The easiest way to derive the Feynman rules is to use the nonlocal
formulation~(\ref{nonloclag}). It looks similar to a local lagrangian
except that the gauge field in the last term (i.e.\ the whole third
line) is taken from $\sigma^2+1/2R^2$. In the Feynman diagrams the
propagators have (worldvolume) momenta in the lattice corresponding to
the compactification, i.e.\ $P_1,P_2$ are multiples of $2\pi R_1$ or
$2\pi R_2$ respectively. The last term in~(\ref{nonloclag}) gives us a
vertex with two fermions and one gauge boson and the corresponding
Feynman vertex contains a factor $(-1)^{P^2/2\pi R_2}$.

In order to determine the cluster properties of our theory, we must
calculate the effective action along the flat directions in the
classical moduli space of the gauge theory.  We will concentrate on a
single direction in which (in the nonlocal formulation) the gauge
group is broken to $\U(N_1) \times \U(N_2)$.  That is, we calculate
two body forces, rather than general $k$ body interactions.  There is
a subtlety in this calculation which has to do with our lack of
knowledge of the spectrum of this nonsupersymmetric theory.

\looseness=1 In general, one may question the validity of the Born-Oppenheimer
approximation for the flat directions because the individual
nonabelian gauge groups appear to give rise to infrared divergences in
perturbation theory.  In the SUSY version of Matrix Theory this
problem is resolved by the (folk) theorem that the general $\U(N)$
theory (compactified on a torus) has threshold bound states.  These
correspond to wave functions normalizable along the flat directions
and should cut off the infrared divergences.  In our SUSY violating
model, we do not know the relevant theorems.

The most conservative way to interpret our calculation is to take $N_1
= N_2 = 1$ in the $\U(2)$ version of the model.  If one finds an
attractive two body force then it is reasonable to imagine that in
fact the general $\U(N)$ theory has a normalizable ground state, thus
justifying the Born-Oppenheimer approximation in the general case.

So let us proceed to calculate the potential in the $\U(2)$ case, and
let $R$ be the field which represents the separation between two
excitations of the $\U(1)$ model.  From the point of view of $2+1$
dimensional field theory, $R$ is a scalar field, with mass dimension
$1/2$.  It is related to the distance measured in M-theory by powers
of the eleven dimensional Planck scale.  At large $R$, the charged
fields of the $\U(2)$ model are very heavy.  To integrate them out we
must understand the UV physics of the model.  The formulation in terms
of a $\U(2)\times \U(2)$ theory with peculiar boundary conditions
shows us that the UV divergences are of the same degree as those of
the SUSY model, though some of the SUSY cancellations do not occur, as
we will see below.  Ultraviolet physics is thus dominated by the fixed
point at vanishing Yang-Mills coupling and we can compute the large
$R$ expansion of the effective action by perturbation theory.


\qquad\begin{figure}[t]
\begin{minipage}[l]{6.5cm}
\quad\epsfig{file=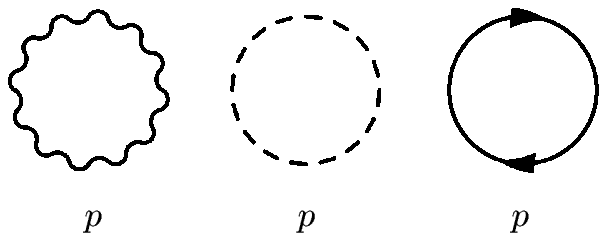}
\caption{One-loop diagrams.}\label{fignonsusy1}
\end{minipage}
\begin{minipage}[l]{7.8cm}\vspace{2mm}
\quad\epsfig{file=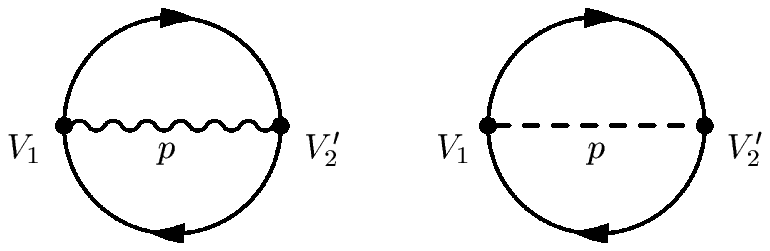}
\caption{Non-vanishing two-loop contributions to the effective
potential. Here $V'_2\equiv (\epsilon(p)-1)V_2$.}
\label{fignonsusy2}\end{minipage}\end{figure}

The one loop contribution, figure~\ref{fignonsusy1}, to the effective
potential vanishes because it is identical to that in the SUSY model.
The only difference between the models in the nonlocal formulation is
the peculiar vertex described above.  The leading contribution comes
from two loops and is of order $g_{YM}^2$ (the squared coupling has
dimensions of mass).  It comes only from the diagrams containing
fermion lines shown in figure~\ref{fignonsusy2}.  These diagrams should be
evaluated in the nonlocal model and then their value in the SUSY model
should be subtracted.  The rest of the two loop diagrams in the model
are the same as the SUSY case and they cancel (for time independent
$R$) against the SUSY values of the diagrams shown.  Taking $R$ very
large in the diagrams is, by dimensional analysis, equivalent to
taking the volume large, and the potential is extensive in the volume
in the large volume limit.  The massive particles in the loops have
masses of order $g_{YM} R$ and this quantity is kept fixed in the loop
expansion.

Our lagrangian has two gauge boson fermion vertices $V_1 + \epsilon
(p) V_2$.  $p$ is the momentum.  In the SUSY theory, these are the two
terms in the commutator. In our lagrangian, the first is identical to
that in the SUSY theory while the second differs from it by the sign
$\epsilon (p)$, which is negative for odd values of the loop momentum
around the twisted circle.  Schematically then, the nonvanishing two
loop contribution has the form
\eqn{twoloop}{\langle(V_1 + \epsilon (p) V_2)^2\rangle - \langle (V_1 +
V_2)^2 \rangle}
This can be rewritten as
\eqn{twolooptwo}{2 \langle (\epsilon (p) - 1) V_1 V_2 \rangle\,.}

\looseness=-1 The resulting loop integral is quadratically ultraviolet divergent.
The leading divergence is independent of $R$, but there are subleading
terms of order $g_{YM}^2 \Lambda \left\vert g_{YM} R\right\vert$ and
$g_{YM}^2 {\ln}(\Lambda) (g_YM R)^2$.  Corrections higher order in the
Yang-Mills coupling, as well as those coming from finite volume of the
Yang-Mills torus, are subleading both in $R$ and $\Lambda$. Thus, the
leading order contribution to the potential is either confining, or
gives a disastrous runaway to large $R$. Which of these is the correct
behavior is determined by our choice of subtractions.  It would seem
absurd to choose the renormalized coefficient of $R^2$ to be negative,
and obtain a hamiltonian unbounded from below.  If that is the case,
then a confining potential prevents excitations from separating from
each other in the would-be transverse spacetime.  In other words, the
theory does not have a spacetime interpretation at all, let alone a
relativistically invariant one.
\pagebreak[3]

We also see that the fear expressed in the previous chapter that all
excitations will decay into tachyons of minimal longitudinal momentum
was ill founded.  Instead it would appear that the entire system will
form a single clump in transverse space.  The $\U(1)$ part of the
theory decouples, so we can give this clump transverse momentum and
obtain an energy spectrum
\eqn{enspec}{P^- \sim {R {\bf P}^2 \over N} + \Delta\,.}
where $\Delta$ is the ground state energy of our nonlocal $\SU(N)$
Yang-Mills theory.

If $\Delta$ were to turn out positive and of order $1/N$, this
dispersion relation would look like that of a massive relativistic
particle.  We might be tempted to say that the system looked like a
single black hole propagating in an asymptotically flat spacetime.
The stability of the black hole would be explained if its mass were
within the Planck regime.  This interpretation does not appear to be
consistent, for semiclassical analysis of such a system indicates that
it has excitations corresponding to asymptotic gravitons propagating
in the black hole background.  Our result about the lifting of the
moduli space precludes the existence of such excitations.

Since the vacuum energy is divergent, the positivity of $\Delta$ is a
matter of choice.  However, large-$N$ analysis suggests that it scales
like a positive power of $N$, so we have another reason that the black
hole interpretation does not seem viable.

\section{Conclusions}

What are we to make of all these disasters?  We believe that our work
is solid evidence for the absence of a Lorentz invariant vacuum of
M-theory based on the Rohm compactification.  The Rohm strings are
certainly degrees of freedom of our matrix model, even if we cannot
derive the (apparently meaningless because of the tachyon and
unbounded effective potential) string perturbation expansion from it
(as a consequence of the absence of a nonrenormalization theorem).

\looseness=1 However, naive physical intuition based on the string perturbation
series, suggest that if there is a stable solution corresponding to
the Rohm model, it is not Lorentz invariant.  While we cannot trust
the perturbative calculations in detail, they at least imply that the
vacuum energy of the system is negative at its (hypothetical) stable
minimum.  (We remind the reader that even at large radius, before the
tachyon appears, the potential calculated by Rohm is negative and the
system wants to flow to smaller radius).  Perhaps there is a
nonsupersymmetric anti-de~Sitter solution of M-theory to which the
Rohm model ``flows''.  There are many problems with such an
interpretation, since it involves changing asymptotic boundary
conditions in a generally covariant theory.  Normally one would
imagine that M-theory with two different sets of asymptotic boundary
conditions breaks up into two different quantum mechanical systems
which simply do not talk to each other.  The finite energy states with
one set of boundary conditions simply have no overlap with the finite
energy states of another (the definition of energy is completely\linebreak
different).

\looseness=1 As an aside we note that an extremely interesting question arises for
systems (unlike the Rohm compactification) which have a
\emph{metastable} Minkowski vacuum.  In the semiclassical
approximation~\cite{cdl} one can sometimes find instantons which
represent tunneling of a Minkowski vacuum into a ``bubble of
anti-de~Sitter space''.  Does this idea make any sense in a fully
quantum mechanical theory, particularly if one believes in the
holographic principle?  Coleman and De~Luccia argue that the system
inside the AdS bubble is unstable to recollapse and interpret this as
a disaster of cosmic proportions: Minkowski space fills up with
bubbles, expanding at the speed of light, the interior of each of
which becomes singular in finite proper time.  It is hard to imagine
how such a scenario could be described in a holographic
framework.\footnote{This should not be taken simply as an indication
that a holographic description of asymptotically flat spacetimes is
somehow sick.  The Coleman-De~Luccia instanton also exists in an
asymptotically AdS framework, with two negative energy vacua.  If the
higher energy state has very small vacuum energy the semiclassical
analysis is practically unchanged.  So, if the Coleman-De~Luccia
phenomenon really exists in M-theory we should be able to find a
framework for studying it within AdS/CFT.}

\looseness=1 At any rate, it is clear that the fate of the Rohm compactification
depends crucially on a change in vacuum expectation values.  In this
sense one might argue that our attempt to study it in light cone frame
was ``doomed from the start''.  It is a notorious defect of the light
cone approach that finding the correct vacuum is extremely difficult.
It involves understanding and cancelling the large-$N$ divergences of
the limiting DLCQ, by changing parameters in the light cone
hamiltonian.  If the correct vacuum is a finite distance away in field
space from the naive vacuum from which one constructs the original
DLCQ hamiltonian, this may simply mean that the true hamiltonian has
little resemblance to the one from which one starts.  If our physical
arguments above are a good guide, the problem may be even more severe.
The correct vacuum may not even have a light cone frame hamiltonian
formulation.

We confess to having jumped in to the technical details of our
construction before thinking through the physical arguments above.
Nonetheless, we feel that our failure is a useful reminder that
gravitational physics is very different from quantum field theory, and
an indication of the extreme delicacy of SUSY breaking in M-theory.

%% file: chap6-bfm.tex
\chapter{Asymptotic limits in the moduli space of M-theory}

In this chapter which results from a work done together with Prof. 
Tom Banks and Prof. Willy Fischler from Texas \cite{BFM},
we show that a subgroup of the modular group of M-theory compactified on a
ten torus, implies the Lorentzian structure of the moduli space, that is
usually associated with naive discussions of quantum cosmology based on
the low energy Einstein action. This structure implies a natural division
of the asymptotic domains of the moduli space into regions which
can/cannot be mapped to Type II string theory or 11D Supergravity (SUGRA)
with large radii.  We call these the safe and unsafe domains. The safe
domain is the interior of the future light cone in the moduli space while
the unsafe domain contains the spacelike region and the past light cone.
Within the safe domain, apparent cosmological singularities can be
resolved by duality transformations and we briefly provide a physical
picture of how this occurs.  The unsafe domains represent true
singularities where all field theoretic description of the physics breaks
down. They violate the holographic principle. We argue that this structure
provides a natural arrow of time for cosmology.  All of the Kasner
solutions, of the compactified SUGRA theory interpolate between the past
and future light cones of the moduli space.  We describe tentative
generalizations of this analysis to moduli spaces with less SUSY.

\section{Introduction}
There have been a large number of papers on the application of string
theory and M-theory to cosmology \cite{cosmoa}-\cite{cosmog}.  In the
present chapter 
we will study the cosmology of
toroidally compactified M-theory, and argue that 
some of the singularities encountered in the low energy field theory
approximation can be resolved by U-duality.  We argue that near any of
the singularities we study, new light states appear, which cannot be
described by the low energy field theory.  The matter present
in the original universe decays into these states and the universe
proceeds to expand into a new large geometry which is described by a
different effective field theory.  

In the course of our presentation we will have occasion to investigate the
moduli space of M-theory with up to ten compactified rectilinear toroidal
dimensions and vanishing three form potential.  We believe that the results
of this investigation are extremely interesting.  For tori of dimension, $d
\leq 8$, we find that all noncompact regions of the moduli space can be
mapped into weakly coupled Type II string theory, or 11D SUGRA, at large
volume.  For $d \leq 7$ this result is a subset of the results of Witten
\cite{witten} on toroidal compactification of string theory.  The result
for $ d= 9$ is more interesting.  There we find a region of moduli space
which cannot be mapped into well understood regions.  We argue that a
spacetime cannot be in this regime if it satisfies the Bekenstein bound.

For the ten torus the moduli space can be viewed as a $9+1$ dimensional
Minkowski space.  The interior of the future light cone is the region that
can be mapped into Type II or 11D (we call the region which can be so
mapped the safe domain of the moduli space).  We again argue that the
other regions violate the Bekenstein bound, in the form of the
cosmological holographic principle of \cite{lenwilly}. Interestingly, the
pure gravity Kasner solutions lie precisely on the light cone in moduli
space.  The condition that homogeneous perturbations of the Kasner
solutions by matter lie inside the light cone of moduli space is precisely
that the energy density be positive.

However, every Kasner solution interpolates between the past and future 
light cones.   Thus, M-theory appears to define a natural {\it arrow of 
time} for cosmological solutions in the sense that it is reasonable to 
define the future as that direction in which the universe approaches the
safe domain.  Cosmological solutions appear to  interpolate between a
past where the holographic principle cannot be satisfied and a future
where it can.  

We argue that the $9+1$ dimensional structure, which we derive purely
group theoretically, \footnote{and thus presumably the structure of the
(in)famous hyperbolic algebra $E_{10}$, about which we shall have nothing
to say in this chapter, besides what is implicit in our use of its Weyl
group.} is intimately connected to the De~Witt metric on the moduli space.  
In particular, in the low energy interpretation, the signature of the
space is a consequence of the familiar fact that the conformal factor has
a negative kinetic energy in the Einstein action.  Thus, the fact that the
duality group and its moduli space are exact properties of M-theory tells
us that this structure of the low energy effective action has a more
fundamental significance than one might have imagined.  The results of
this section are the core of the chapter and the reader of limited
patience
should concentrate his attention on them.

In the Section 4 of this chapter we speculatively generalize our arguments
to
moduli spaces with less SUSY.  We argue that the proper arena for the
study of M-theoretic cosmology is to be found in moduli spaces of compact
ten manifolds with three form potential, preserving some SUSY.  The
duality group is of course the group of discrete isometries of the metric
on moduli space.  We argue that this is always a $p+1$ dimensional Lorentz
manifold, where $p$ is the dimension of the moduli space of appropriate,
static, SUSY preserving solutions of 11D SUGRA in 10 compact dimensions,
restricted to manifolds of unit volume.  We discuss moduli spaces with
varying amounts of SUSY, raise questions about the adequacy of the 11D
SUGRA picture in less supersymmetric cases, and touch on the vexing puzzle
of what it means to speak about a potential on the moduli space in those
cases where SUSY allows one. In the Appendix we discuss how the even
self-dual lattices $\Gamma_8$ and $\Gamma_{9,1}$ appear in our
framework.

\section{Moduli, vacua, quantum cosmology and singularities}

\subsection{Some idiosyncratic views on general wisdom}

M-theorists have traditionally occupied themselves with moduli spaces of
Poincar\'e invariant SUSY vacua. It was hoped that the traditional field
theoretic mechanisms for finding stable vacuum states would uniquely pick
out a state which resembled the world we observe.

This point of view is very hard to maintain after the String Duality
Revolution.  It is clear that M-theory has multiparameter families of
exact quantum mechanical SUSY ground states, and that the first
phenomenological question to be answered by M-theory is why we do not live
in one of these SUSY states. It is logical to suppose that the answer to
this question lies in cosmology. That is, the universe is as it is not
because this is the only stable endpoint to evolution conceivable in
M-theory but also because of the details of its early evolution.  To
motivate this proposal, recall that in infinite Poincar\'e invariant space
time of three or more dimensions, moduli define superselection sectors.  
Their dynamics is frozen and solving it consists of minimizing some
effective potential once and for all, or just setting them at some
arbitrary value if there is no potential.  Only in cosmology do the moduli
become real dynamical variables.  Since we are now convinced that
Poincar\'e invariant physics does not destabilize these states, we must
turn to cosmology for an explanation of their absence in the world of
phenomena.

The focus of our cosmological investigations will be the moduli which are
used to parametrize SUSY compactifications of M-theory. We will argue that
there is a natural Born-Oppenheimer approximation to the theory in which
these moduli are the slow variables.  In various semiclassical
approximations the moduli arise as zero modes of fields in compactified
geometries.  One of the central results of String Duality was the
realization that various aspects of the space of moduli could be discussed
(which aspects depend on precisely how much SUSY there is) even in regions
where the notions of geometry, field theory and even weakly coupled string
theory, were invalid.  {\it The notion of the moduli space is more robust
than its origin in the zero modes of fields would lead us to believe.}

The moduli spaces of solutions of the SUGRA equations of motion that
preserve eight or more SUSYs, parametrize exact flat directions of the
effective action of M-theory. Thus they can perform arbitrarily slow
motions. Furthermore, their action is proportional to the volume of the
universe in fundamental units.  Thus, once the universe is even an order
of magnitude larger than the fundamental scale we should be able to treat
the moduli as classical variables.  They provide the natural definition of
a semiclassical time variable which is necessary to the physical
interpretation of a generally covariant theory. In this and the next
section we will concentrate on the case with maximal SUSY.  We relax this
restriction in section 4.  There we also discuss briefly the confusing
situation of four or fewer SUSYs where there can be a potential on the
moduli space.

In this chapter we will always use the term moduli in the sense outlined
above. They are the slowest modes in the Born-Oppenheimer approximation to
M-theory which becomes valid in the regime we conventionally describe by
quantum field theory. In this regime they can be thought of as special
modes of fields on a large smooth manifold.  However, we believe that
string duality has provided evidence that the moduli of supersymmetric
compactifications are exact concepts in M-theory, while the field
theoretic (or perturbative string theoretic)  structures from which they
were derived are only approximations.  The Born-Oppenheimer
approximation for the moduli is likely to be valid even in regimes
where field theory breaks down.
The first task in understanding an M-theoretic cosmology is to discuss 
the dynamics of the moduli.  After
that we can begin to ask when and how the conventional picture of quantum
field theory in a classical curved spacetime becomes a good approximation.

\subsection{Quantum cosmology}

The subject of Quantum Cosmology is quite confusing.  We will try to be
brief in explaining our view of this arcane subject.  There are two issues
involved in quantizing a theory with general covariance or even just time
reparametrization invariance.  The first is the construction of a Hilbert
space of gauge invariant physical states.  The second is describing the
physical interpretation of the states and in particular, the notion of
time evolution in a system whose canonical Hamiltonian has been set equal
to zero as part of the constraints of gauge invariance. The first of these
problems has been solved only in simple systems like first quantized
string theory or Chern-Simons theory, including pure $2+1$ dimensional
Einstein gravity.  However, it is a purely mathematical problem, involving
no interpretational challenges.  We are hopeful that it will be solved in
M-theory, at least in principle, but such a resolution clearly must await
a complete mathematical formulation of the theory.

The answer to the second problem, depends on a semiclassical
approximation.  The principle of time reparametrization invariance forces
us to base our measurements of time on a physical variable.  If all
physical variables are quantum mechanical, one cannot expect the notion of
time to resemble the one we are used to from quantum field theory.  It is
well understood \cite{bfstbruba}-\cite{bfstbrubc} how to derive a
conventional time dependent Schr\"odinger equation for the quantum
variables from the semiclassical approximation to the constraint equations
for a time reparametrization invariant system.  We will review this for
the particular system of maximally SUSY moduli.

In fact, we will restrict our attention to the subspace of moduli space
described by rectilinear tori with vanishing three form.
In the language of 11D SUGRA, we are
discussing metrics of the Kasner form
\eqn{metric}{ds^2 = - dt^2 + L_i^2 (t) (dx^i)^2 }
where the $x^i$ are ten periodic coordinates with period $1$.
When restricted to this class of metrics, the Einstein Lagrangian has
the form
\eqn{lag}{\Lag = V\left[ 
\sum_i {\dot{L}_i^2 \over L_i^2} -
\left(\sum_i {\dot{L}_i \over L_i}\right)^2 
\right],}
where $V$, the volume, is the product of the $L_i$.
In choosing to write the metric in these coordinates, we have lost the
equation of motion obtained by varying the variable $g_{00}$.  This is
easily restored by imposing the constraint of time reparametrization
invariance. The Hamiltonian $E_{00}$ derived from (\ref{lag}) should
vanish on
physical states.  This gives rise to the classical Wheeler-De~Witt equation
\eqn{wd}{
2E_{00} = \left(\sum_i \frac{\dot{L}_i}{L_i}\right)^2 
- \sum_i \left(\frac{\dot{L}_i}{L_i}\right)^2  = 0 ,}
which in turn leads to a naive quantum Wheeler-De~Witt equation:
\eqn{qwd}{{1\over 4V}\left(\sum_i \Pi_i^2 - {2\over 9}(\sum_i \Pi_i)^2 
\right)\Psi = 0.}
That is, we quantize the system by converting the unconstrained phase
space variables (we choose the logarithms of the $L_i$ as canonical
coordinates) 
to operators in a function space.  Then physical states
are functions satisfying the partial differential equation (\ref{qwd}).
There are complicated mathematical questions involved in constructing an
appropriate inner product on the space of solutions, and related
problems of operator ordering.  In more complex systems it is
essential to use the BRST formalism to solve these problems.  
We are unlikely to be able to resolve these questions before discovering
the full nonperturbative formulation of M-theory.   However, for our
present semiclassical considerations these mathematical details are not
crucial. 

We have already emphasized that when the volume of the system is large
compared to the Planck scale, 
the moduli behave classically.  It is then possible to use the time
defined by a particular classical solution (in a particular coordinate
system in which the solution is nonsingular for almost all time).
Mathematically what this means is
that in the large volume limit, the solution to the Wheeler De~Witt
equation takes the form
\eqn{semicsoln}{ \psi_{W\!K\!B} (c) \Psi (q , t[c_0]) }
Here $c$ is shorthand for the variables which are treated by classical
mechanics, $q$ denotes the rest of the variables and $c_0$ is some
function of the classical variables which is a monotonic function of
time.  The wave function $\Psi$ satisfies a time dependent Schr\"odinger 
equation
\eqn{schrod} {i \partial_t \Psi = H(t) \Psi}
and it is easy to find an inner product which makes the space of its
solutions into a Hilbert space and the operators $H(t)$ Hermitian.
In the case where the quantum variables $q$ are quantum fields on the
geometry defined by the classical solution, this approximation is
generally called Quantum Field Theory in Curved Spacetime.  We emphasize
however that the procedure is very general and depends only on the
validity of the WKB approximation for at least one classical variable,
and the fact that the Wheeler De~Witt equation is a second order
hyperbolic PDE, with one timelike coordinate.  These facts are derived
in the low energy approximation of M-theory by SUGRA.  However, we will
present evidence in the next section that they are consequences of the 
U-duality symmetry of the theory and therefore possess a validity beyond
that of the SUGRA approximation.

From the low energy point of view, the hyperbolic nature of the equation
is a consequence of the famous negative sign for the kinetic energy of the
conformal factor in Einstein gravity, and the fact that the kinetic
energies of all the other variables are positive.  It means that the
moduli space has a Lorentzian metric.
 
\subsection{Kasner singularities and U-duality}

The classical Wheeler-De~Witt-Einstein equation for Kasner metrics takes
the form:
\eqn{cwde}{({\dot{V} \over V})^2 - \sum_{i=1}^{10}  ({\dot{L_i} \over
L_i})^2 = 0}
This should be supplemented by equations for the ratios of individual
radii $R_i; \prod_{i=1}^{10} R_i =1 $.  The latter take the form of
geodesic motion with friction 
on the manifold of $R_i$ (which we parametrize {\it
e.g.} by the first nine ratios) 

\eqn{nlsigma}{ \partial_t^2 R_i + 
\Gamma^i_{jk} \partial_t R_j  \partial_t R_k + \partial_t ({\rm 
ln} V) \partial_t R_i}
$\Gamma$ is the Christoffel symbol of the metric $G_{ij}$ on the
unimodular moduli space.  We write the equation in this general form
because many of our results remain valid when the rest of the variables 
are restored to the moduli space, and even
generalize to the less supersymmetric moduli
spaces discussed in section 4.  By introducing a new time variable
through
$V(t) \partial_t = - \partial_s$ we convert this equation into
nondissipative geodesic motion on moduli space.  Since the 
``energy'' conjugate to the variable $s$ is conserved, the energy
of
the nonlinear model in cosmic time (the negative term in the Wheeler
De~Witt equation) satisfies
\eqn{en} {\partial_t E = - 2 \partial_t (\mbox{ln\,} V) E}
whose solution is
\eqn{ensol}{ E = {E_0 \over V^2}}
Plugging this into the Wheeler De~Witt equation we find that $V \sim t$ 
(for solutions which
expand as $t \rightarrow \infty$).  Thus, for this class of solutions we
can choose
the volume as the monotonic variable $c_0$ which defines the time in the
quantum theory.

For the Kasner moduli space, we find that the solution of the equations
for 
individual radii
are
\eqn{kassoln}{R_i (t) = \lpl (t/t_0 )^{p_i}}
where
\eqn{kascond}{\sum p_i^2 = \sum p_i = 1}
Note that the equation (\ref{kascond}) implies that at least one of the
$p_i$ is 
negative 
(we have again restricted attention to the case where the volume expands
as 
time goes to 
infinity).    

It is well known that all of these solutions are singular 
at both infinite and zero time.  
Note that if we add a matter or radiation energy density to
the system
then it dominates the system in the infinite volume limit and changes
the 
solutions for the
geometry there.  However, near the singularity at vanishing volume both 
matter and radiation 
become negligible (despite the fact that their densities are 
becoming infinite) 
and the solutions retain their Kasner form.

All of this is true in 11D SUGRA.  In M-theory we know that many regions
of moduli space which are apparently singular in 11D SUGRA can be
reinterpreted as living in large spaces described by weakly coupled Type
II string theory or a dual version of 11D SUGRA. The vacuum Einstein
equations are of course invariant under these U-duality transformations.
So one is lead to believe that many apparent singularities of the Kasner
universes are perfectly innocuous.

Note however that phenomenological matter and radiation densities which
one might add to the equations are not invariant under duality.  The
energy density truly becomes singular as the volume goes to zero.  How
then are we to understand the meaning of the duality symmetry?  The
resolution is as follows.  We know that when radii go to zero, the
effective field theory description of the universe in 11D SUGRA becomes
singular due to the appearance of new low frequency states.  We also know
that the singularity in the energy densities of matter and radiation
implies that scattering cross sections are becoming large.  Thus, it seems
inevitable that phase space considerations will favor the rapid
annihilation of the existing energy densities into the new light degrees
of freedom.  This would be enhanced for Kaluza-Klein like modes, whose
individual energies are becoming large near the singularity.

Thus, near a singularity with a dual interpretation, the contents of the
universe will be rapidly converted into new light modes, which have a
completely different view of what the geometry of space is\footnote{After
this work was substantially complete, we received a paper, \cite{riotto},
which proposes a similar view of certain singularities. See also
\cite{rama}.}. The
most
effective description of the new situation is in terms of the transformed
moduli and the new light degrees of freedom.  The latter can be described
in terms of fields in the reinterpreted geometry.  We want to emphasize
strongly the fact that the moduli do not change in this transformation,
but are merely reinterpreted.  This squares with our notion that they are
exact concepts in M-theory.  By contrast, the fields whose zero modes they
appear to be in a particular semiclassical regime, do not always make
sense.  The momentum modes of one interpretation are brane winding modes
in another and there is no approximate way in which we can consider both
sets of local fields at the same time.  Fortunately, there is also no
regime in which both kinds of modes are at low energy simultaneously, so
in every regime where the time dependence is slow enough to make a low
energy approximation, we can use local field theory.

This mechanism for resolving cosmological singularities leads naturally to
the question of precisely which noncompact regions of moduli space can be
mapped into what we will call the {\it safe domain} in which the theory
can be interpreted as either 11D SUGRA or Type II string theory with radii
large in the appropriate units.  The answer to this question is, we
believe, more interesting than the idea which motivated it.  We now turn
to the study of the moduli space.

\section{The moduli space of  M-theory on rectangular tori}

In this section, we will study the structure of the moduli space
of M-theory compactified on various tori $T^k$ with $k\leq 10$.  We
are  especially interested in noncompact regions of this space which
might represent either singularities or large universes. As above, 
the three-form potential $A_{MNP}$ will be
set to zero and the circumferences of the cycles of the torus
will be expressed as the exponentials
\eqn{radiiexp}{ {L_i \over \lpl} = t^{p_i} ,\qquad
i=1,2, \dots, k.}

The remaining coordinates $x^0$ (time) and $x^{k+1}\dots x^{10}$ are
considered to be infinite and we never dualize them.

So the radii are encoded in the logarithms $p_i$. We will study limits of
the moduli space in various directions which correspond to keeping
$p_i$ fixed and sending $t\to\infty$ (the change to $t\to 0$
is equivalent to $p_i\to -p_i$ so we do not need to study
it separately).

We want to emphasize that our discussion of asymptotic domains of moduli
space is complete, even though we restrict ourselves to rectilinear tori
with vanishing three form.  Intuitively this is because the moduli we
leave out are angle variables.  More formally, the full moduli space is
a homogeneous space.  Asymptotic domains of the space correspond to
asymptotic group actions, and these can always be chosen in the Cartan
subalgebra.  The $p_i$ above can be thought of as parametrizing a
particular Cartan direction in $E_{10}$.\footnote{We thank E.Witten for a
discussion of this point.}

\subsection{The \rut}

M-theory has dualities which allows us to identify the vacua with
different $p_i$'s.  A subgroup of this duality group is the $S_k$ which
permutes the $p_i$'s.
Without  loss of generality, we can assume that
$p_1\leq p_2\leq \dots \leq p_9$. We will assume this in most of
the text.
The full group that leaves invariant rectilinear tori with
vanishing three form is the Weyl group of the noncompact $E_k$ group
of SUGRA. We will denote it by $\Rut_k$.  We will give an elementary
derivation of the properties of this group for the convenience of
the reader.  Much of this is review, but our results about the
boundaries of the fundamental domain of the action of $\Rut_k$ with $k =
9,10$ on the moduli space, are new.
$\Rut_k$ is generated
by the permutations, and one other transformation which acts as follows:
\eqn{rutdef}{(p_1,p_2,\dots, p_k)
\mapsto
(p_1-{2s\over 3},
p_2-{2s\over 3},
p_3-{2s\over 3},
p_4+{s\over 3},
\dots,
p_k+{s\over 3}).}
where $s=(p_1+p_2+p_3)$.  
Before explaining why this transformation is a
symmetry of M-theory, let us point out several of its properties
(\ref{rutdef}).

\begin{itemize}
 \item The total sum $S=\sum_{i=1}^k p_i$ changes to $S\mapsto
S+(k-9)s/3$. 
So if $s<0$, the sum increases for $k<9$, decreases for $k>9$
and is left invariant for $k=9$.

 \item If we consider all $p_i$'s to be integers which are
equal modulo 3, this property will hold also after
the \rut. The reason is that, due to the assumptions, $s$ is a multiple
of three and the coefficients $-2/3$ and $+1/3$ differ by an integer.
%
As a result, from any initial integer $p_i$'s we get $p_i$'s
which are multiples of $1/3$ which means that all the matrix elements
of matrices in $\Rut_{k}$ are integer multiples of $1/3$. 

 \item The order of $p_1,p_2,p_3$ is not changed (the difference
$p_1-p_2$ remains constant, for instance). Similarly,
the order of $p_4,p_5,\dots, p_k$ is unchanged. However the
ordering between $p_{1...3}$ and $p_{4...k}$ can change in general.
By convention, we will follow each \rut{} by a permutation which places
the $p_i$'s in ascending order.

\item The bilinear quantity $I= (9 - k) \sum (p_i^2) + (\sum p_i )^2 = (10
- k) \sum(p_i^2) + 
 2 \sum_{i < j} p_i p_j$ is left invariant by $\Rut_k$.
\end{itemize}

The fact that \rut{} is a symmetry of M-theory can be proved as follows.
Let us interpret $L_1$ as the M-theoretical circle of a type IIA string
theory. Then the simplest duality which gives us a theory of the same kind
(IIA) is the double T-duality. Let us perform it on the circles $L_2$
and $L_3$. The claim is that if we combine this double T-duality
with a permutation of $L_2$ and $L_3$ and interpret the new $L_1$ as the
M-theoretical circle again, we get precisely (\ref{rutdef}).

Another illuminating way to view the transformation \rut{} is to
compactify M-theory on a three torus.  The original M2-brane and the
M5-brane wrapped on the three torus are both BPS membranes in eight
dimensions.  One can argue that there is a duality transformation
exchanging them \cite{ofer}. In the limit in which one of the cycles of
the $T^3$ is small, so that a type II string description becomes
appropriate, it is just the double T-duality of the previous paragraph.  
The fact that this transformation plus permutations generates $\Rut_k$ was
proven by the authors of \cite{elitzur} for $k \leq 9$, see also
\cite{pioline}.

\subsection{Extreme moduli}

There are three types of boundaries of the toroidal moduli space which
are amenable to detailed analysis.  The first is the limit in which
eleven-dimensional supergravity becomes valid. We will
denote this limit as 11D. The other two limits are weakly coupled
type IIA and type IIB theories in 10 dimensions. We will call the domain
of asymptotic moduli space which can be mapped into one of these limits,
the safe domain.

\begin{itemize}

 \item For the limit 11D, all the radii must be greater than $\lpl$.
Note that for $t\to\infty$ it means that all the radii are much greater
than $\lpl$. In terms of the $p_i$'s , this is the inequality $p_i>0$.

 \item For type IIA, the dimensionless coupling constant
$g_s^{IIA}$ must be smaller than 1 (much smaller for $t\to\infty$)
and all the remaining radii must be greater than $\lst$ (much
greater for $t\to\infty$).

 \item For type IIB, the dimensionless coupling constant
$g_s^{IIB}$ must be smaller than 1 (much smaller for $t\to\infty$)
and all the remaining radii must be greater than $\lst$ (much
greater for $t\to\infty$), including the extra radius whose momentum
arises as the number of wrapped M2-branes on the small $T^2$ in the
dual 11D SUGRA picture.

\end{itemize}

If we assume the canonical ordering of the radii, i.e. $p_1\leq p_2\leq
p_3\leq \dots \leq p_k$, we can simplify these requirements as follows:

\begin{itemize}
 \item 11D:    $0<p_1$
 \item IIA:    $p_1<0<p_1+2p_2$
 \item IIB:    $p_1+2p_2<0<p_1+2p_3$
\end{itemize}
To derive this, we have used the familiar relations:
\eqn{fama}{  {L_1\over \lpl}=(g_s^{IIA})^{2/3}=
\left({\lpl\over\lst}\right)^2=
\left({L_1\over \lst}\right)^{2/3}}
for the 11D/IIA duality ($L_1$ is the M-theoretical circle) and similar
relations for the 11D/IIB case ($L_1<L_2$ are the parameters of the
$T^2$ and $L_{IIB}$ is the circumference of the extra circle):
\begin{eqnarray}
{L_1\over L_2}=g_s^{IIB},\quad
1={L_1\lst^2\over\lpl^3}={g_s^{IIB}L_2\lst^2\over\lpl^3}=
{L_{IIB}L_1L_2\over\lpl^3},\\
\frac{1}{g_s^{IIB}}\left(\frac{\lpl}{\lst}
\right)^4=\frac{L_1L_2}{\lpl^2}=\frac{\lpl}{L_{IIB}}=
(g_s^{IIB})^{1/3}\left(\lst\over L_{IIB}\right)^{4/3}\label{famb}
\end{eqnarray}

Note that the regions defined by the inequalities above cannot overlap,
since the regions are defined by $M,M^c\cap A,A^c\cap B$ where
$A^c$ means the complement of a set.
Furthermore, assuming $p_i<p_{i+1}$ it is easy to show that 
$p_1+2p_3<0$ implies $p_1+2p_2<0$ and $p_1+2p_2<0$ implies
$3p_1<0$ or $p_1<0$. 

This means that (neglecting the boundaries where
the inequalities are saturated) the region outside 
$\mbox{11D}\cup\mbox{IIA}\cup\mbox{IIB}$ is defined simply by
$p_1+2p_3<0$.  The latter characterization of the safe domain 
of moduli space will simplify our discussion considerably.

\vspace{3mm}

The invariance of the bilinear form defined above gives an important
constraint on the action of $\Rut_k$ on the moduli space.  For $k=10$ it
is easy to see that, considering the $p_i$ to be the coordinates of a ten
vector, it defines a Lorentzian metric on this ten dimensional space.  
Thus the group $\Rut_{10}$ is a discrete subgroup of $O(1,9)$.  The
direction in this space corresponding to the sum of the $p_i$ is timelike,
while the hyperplane on which this sum vanishes is spacelike. We can
obtain the group $\Rut_9$ from the group $\Rut_{10}$ by taking $p_{10}$ to
infinity and considering only transformations which leave it invariant.  
Obviously then, $\Rut_9$ is a discrete subgroup of the transverse Galilean
group of the infinite momentum frame. For $k \leq 8$ on the other hand,
the bilinear form is positive definite and $\Rut_k$ is contained in
$O(k)$.  Since the latter group is compact, and there is a basis in which
the $\Rut_k$ matrices are all integers divided by $3$, we conclude that in
these cases $\Rut_k$ is a finite group. In a moment we will show that
$\Rut_9$ and {\it a fortiori} $\Rut_{10} $ are infinite. Finally we note
that the \rut{} is a spatial reflection in $O(1,9)$.  Indeed it squares to
$1$ so its determinant is $\pm 1$.  On the other hand, if we take all but
three coordinates very large, then the \rut{} of those coordinates is very
close to the spatial reflection through the plane $p_1 + p_2 + p_3 = 0$,
so it is a reflection of a single spatial coordinate.

\begin{figure} [t]
\qquad\qquad\qquad\epsfig{file=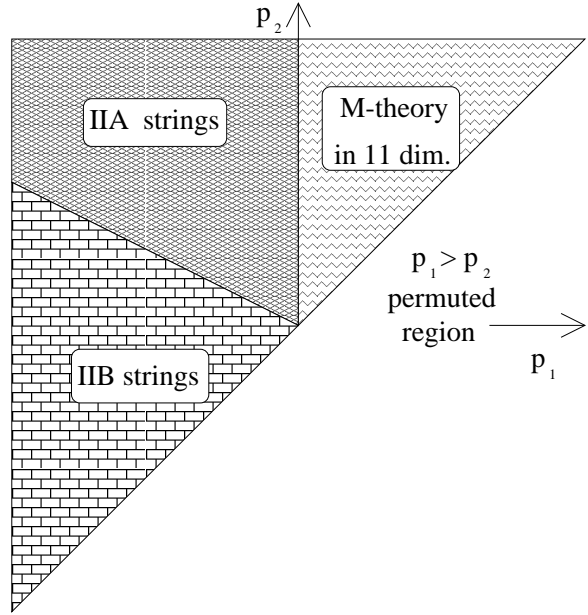}
\caption{The structure of the moduli space for $T^2$.}
\label{myfigure}
\end{figure}


\vspace{3mm}

We now prove that $\Rut_9$ is infinite.
Start with the first vector of $p_i$'s given below and iterate
(\ref{rutdef}) on the three smallest radii (a strategy which we will use
all the time) -- and sort $p_i$'s after each step, so that their index
reflects their order on the real line. We get
\eqn{ninefinite}{
\begin{array}{lcr}
(-1,-1,-1,&-1,-1,-1,&-1,-1,-1)\\
(-2,-2,-2,&-2,-2,-2,&+1,+1,+1)\\
(-4,-4,-4,&-1,-1,-1,&+2,+2,+2)\\
(-5,-5,-5,&-2,-2,-2,&+4,+4,+4)\\
{ }&\vdots&{ }\\
(3\times (2-3n),&3\times (-1),&3\times (3n-4))\\
(3\times (1-3n),&3\times (-2),&3\times (3n-2))
\end{array}
}
so the entries grow (linearly) to infinity.

\subsection{Covering the moduli space}

We will show that there is a useful strategy which can be used to
transform any point $\{p_i\}$ into the safe domain in the case 
of $T^k$, $k<9$. The strategy
is to perform iteratively \rut{s} on the three smallest radii.

Assuming that $\{p_i\}$ is outside the safe domain, i.e.
$p_1+2p_3<0$ ($p_i$'s are sorted so that $p_i\leq p_{i+1}$),
it is easy to see that $p_1+p_2+p_3<0$ (because $p_2\leq p_3$).
As we said below the equation (\ref{rutdef}), 
the \rut{} on $p_1,p_2,p_3$ always increases the total
sum $\sum p_i$ for $p_1+p_2+p_3<0$. But this sum cannot increase
indefinitely because the group $\Rut_k$ is finite for
$k<9$. Therefore the iteration proccess must terminate at some
point. The only way this can happen is that
the assumption $p_1+2p_3<0$ no longer holds, which means that
we are in the safe domain. This completes the proof for $k<9$.

\vspace{3mm}

For $k=9$ the proof is more difficult. The group $\Rut_9$ is infinite
and furthermore, the sum of all $p_i$'s does not change. In fact 
the conservation of $\sum p_i$ is the reason that only points with
$\sum p_i>0$ can be dualized to the safe domain.
The reason is that if $p_1+2p_3\geq 0$, also $3p_1+6p_3\geq 0$
and consequently
\eqn{ninep}{p_1+p_2+p_3+p_4+p_5+p_6+p_7+p_8+p_9 \geq
p_1 +p_1+p_1 + p_3+p_3+p_3+p_3+p_3+p_3\geq 0.}
This inequality is saturated only if all $p_i$'s are equal
to each other. If their sum vanishes, each $p_i$ must then vanish.
But we cannot obtain a zero vector from a nonzero vector
by \rut{s} because they are nonsingular. If the sum $\sum p_i$ is
negative, it is also clear that we cannot reach the safe domain.

However, if $\sum_{i=1}^9 p_i>0$, then we can map the region of moduli space
with $t \rightarrow \infty $ to the safe domain. 
We will prove it for rational $p_i$'s only. This assumption compensates
for the fact that the order of
$\Rut_9$ is infinite.
Assuming $p_i$'s rational is however sufficient
because we will see that a finite product of \rut{s} brings us
to the safe domain. But a composition of a finite number
of \rut{s} is a continuous map from $\IR^9$ to $\IR^9$ so there must be 
at least a
``ray'' part of a neighborhood which can be also dualized to the
safe domain. Because $\IQ^9$ is dense in $\IR^9$, our argument proves the 
result
for general values of $p_i$.

From now on we assume that the 
$p_i$'s are rational numbers. Everything is scale invariant so we
may multiply them by a common denominator to make integers. In fact, we choose
them to be integer multiples of three since in that case we will have integer
$p_i$'s even after \rut{s}. The numbers $p_i$ are now integers equal
modulo 3 and their sum is positive. We will define a critical quantity
\eqn{cq}{C=\sum_{i<j}^{1...9} (p_i-p_j)^2.}
This is {\it a priori} an integer greater than or equal to zero 
which is invariant
under permutations. What happens to $C$ if we make
a \rut{} on the radii $p_1,p_2,p_3$? The differences
$p_1-p_2$, $p_1-p_3$, $p_2-p_3$ do not change and this holds for
$p_4-p_5$, \dots $p_8-p_9$, too. The only contributions to
(\ref{cq}) which are changed are from $3\cdot 6=18$ ``mixed'' terms like
$(p_1-p_4)^2$. Using (\ref{rutdef}),
\eqn{rutcq}{(p_1-p_4) \mapsto (p_1-\frac{2s}3) -(p_4+\frac{s}3)=
(p_1-p_4)-s}
so its square 
\eqn{rs}{(p_1-p_4)^2\mapsto [(p_1-p_4)-s]^2=(p_1-p_4)^2 - 2s(p_1-p_4)
+s^2}
changes by $- 2s(p_1-p_4) +s^2$. Summing over all 18 terms we get
($s=p_1+p_2+p_3$)
\eqn{delta}{\Delta C= -2s[6(p_1+p_2+p_3)-3(p_4+\dots+p_9)]+18s^2
=6s^2 + 6\left((\sum_{i=1}^9 p_i)-s\right)=6s\sum_{i=1}^9 p_i.}
But this quantity is strictly negative because $\sum p_i$ is positive
and $s<0$ (we define the safe domain with boundaries, $p_1+2p_3\geq 0$).

This means that $C$ defined in (\ref{cq}) decreases after each
\rut{} on the three smallest radii. Since it is a non-negative integer,
it cannot decrease indefinitely. 
Thus the assumption $p_1+2p_3<0$ becomes invalid after a finite number
of steps and we reach the safe domain. 

The mathematical distinction between the two regions of the moduli space
according to the sign of the sum of nine $p_i$'s, has a satisfying
interpretation in terms of the holographic principle.  In the safe
domain, the volume of space grows in the appropriate 
Planck units, while in the region with
negative sum it shrinks to zero.  The holographic principle tells us
that in the former region we are allowed to describe many of the states
of M-theory in terms of effective field theory while in the latter
region we are not.  The two can therefore not be dual to each other.

Now let us turn to the fully compactified case.  As we pointed out, the
bilinear form $I \equiv 2\sum_{i < j} p_i p_j$ defines a Lorentzian
signature metric on the vector space whose components are the $p_i$.  The
\rut{} is a spatial reflection and therefore the group
$\Rut_{10}$ consists of orthochronous Lorentz transformations.  
Now consider a vector in the safe domain.  We can write it as
\eqn{safevec}{(-2, -2 + a_1, 1 + a_2, \ldots,  1+a_9 )S,
\qquad S\in\IR^+}
where the $a_i$ are positive.  It is easy to see that $I$ is positive
on this configuration.  This means that only the inside of the light
cone can be mapped into the safe domain.  Furthermore, since $\sum p_i$
is positive in the safe domain and the transformations are
orthochronous, only the interior of the 
future light cone in moduli space can be mapped
into the safe domain.  

We would now like to show that the entire interior of the forward light
cone can be so mapped.  We use the same strategy of rational coordinates
dense in $\IR^{10}$.  If we start outside the safe domain, the sum of the
first three $p_i$ is negative.  We again pursue the strategy of doing a
\rut{} on the first three coordinates and then reordering
and iterating.  For the case of $\Rut_9$ the sum of the coordinates was
an invariant, but here it decreases under the \rut{} of the three
smallest coordinates, if their sum is negative.  
But $\sum p_i$ is (starting from rational values
and rescaling to get integers congruent modulo three as before) a
positive integer and must remain so after $\Rut_{10}$ operations.  
Thus, after a finite number of iterations, the
assumption that the sum of the three smallest coordinates is negative
must fail, and we are in the safe domain.  In fact, we generically enter
the safe domain before this point.  The complement of the safe domain
always has negative sum of the first three coordinates, but there are
elements in the safe domain where this sum is negative.

It is quite remarkable that the bilinear form $I$ is proportional to 
the Wheeler-De~Witt Hamiltonian for the Kasner solutions:
\eqn{wdI}{\frac{I}{t^2}=\left(\sum_i \frac{dL_i/dt}{L_i}\right)^2
- \sum_i\left(\frac{dL_i/dt}{L_i}\right)^2=\frac{2}{t^2}\sum_{i<j}p_ip_j.}
The solutions themselves thus lie precisely on the future light cone in
moduli space. Each solution has two asymptotic regions ($t \rightarrow
0,\infty$ in (\ref{metric})), one of which is in the past light cone and
the other in the future light cone of moduli space.  The structure of the
modular group thus suggests a natural arrow of time for cosmological
evolution.  The future may be defined as the direction in which the
solution approaches the safe domain of moduli space.  All of the Kasner
solutions then, have a true singularity in their past, which cannot be
removed by duality transformations.

Actually, since the Kasner solutions are on the light cone, which is
the boundary of the safe domain,  we must add
a small homogeneous energy density to the system in order to make this
statement correct.  The condition that we can map into the safe domain
is then the statement that this additional energy density is positive.
Note that in the safe domain, and if the equation of state of this matter
satisfies (but does not saturate) 
the holographic bound of \cite{lenwilly}, this energy density
dominates the evolution of the universe, while near the singularity, it
becomes negligible compared to the Kasner degrees of freedom.  The assumption
of a homogeneous negative energy density is manifestly incompatible with
Einstein's equations in a compact flat universe so we see that the
spacelike domain of moduli space corresponds to a physical situation
which cannot occur in the safe domain.

The backward lightcone of the asymptotic moduli space is, as we have
said, visited by all of the classical solutions of the theory.  
However, it violates the holographic principle of \cite{lenwilly} if we
imagine that the universe has a constant entropy density per comoving
volume.  We emphasize that in this context, entropy means the logarithm
of the dimension of the Hilbert space of those states which can be given
a field theoretic interpretation and thus localized inside the volume.

Thus, there is again a clear physical reason why the unsafe domain of
moduli space  cannot be
mapped into the safe domain.  
Note again that matter obeying the holographic bound of
\cite{lenwilly} in the future, cannot alter the nature of the solutions 
near the true singularities.

\vspace{3mm}

To summarize: the U-duality group $\Rut_{10}$ divides the asymptotic
domains of moduli space into three regions, corresponding to the
spacelike and future and past timelike regimes of a Lorentzian manifold.
Only the future lightcone can be understood in terms of weakly coupled SUGRA or
string theory.  The group theory provides an exact M-theoretic meaning for the
Wheeler-De~Witt Hamiltonian for moduli.  Classical solutions of the low
energy effective equations of motion with positive energy density for
matter distributions lie in the timelike region of moduli space and 
interpolate between the past and future light cones.
We find it remarkable that the purely group theoretical considerations
of this section seem to capture so much of the physics of toroidal
cosmologies.

\section{Moduli spaces with less SUSY}

We would like to generalize the above considerations to situations 
which preserve less 
SUSY.  This enterprise immediately raises some questions, the first of which is
what we mean by SUSY.  Cosmologies with compact spatial sections have no
global symmetries in the standard sense
since there is no asymptotic region in which one can define the generators.
We will define a cosmology with a certain amount 
of SUSY by first looking for Euclidean
ten manifolds and three form field configurations which are solutions of the
equations of 11D SUGRA and have a certain number of Killing spinors.
The first approximation to cosmology will 
be to study motion on a moduli space of
such solutions.
The motivation for this is that at least 
in the semiclassical approximation we are guaranteed
to find arbitrarily slow motions of the moduli.  
In fact, in many cases, SUSY
nonrenormalization theorems guarantee that the semiclassical 
approximation becomes valid for
slow motions because the low energy effective Lagrangian of the
moduli is to a large
extent determined by SUSY.  There are however a number of 
pitfalls inherent in our approach.
We know that for some SUSY algebras, the moduli space of 
compactifications to four or six 
dimensions is not a manifold.  New moduli can appear at 
singular points in moduli space
and a new branch of the space, attached to the old one at the 
singular point, must
be added.  There may be cosmologies which traverse from one branch to 
the other in the
course of their evolution.  If that occurs, there will be a point at
which the moduli space approximation breaks down.
Furthermore, there are many examples of SUSY vacua of M-theory which 
have not yet been 
continuously connected on to the 11D limit, even through 
a series of ``conifold'' 
transitions such as those described above \cite{islands}.   
In particular, it has been suggested that there
might be a completely isolated vacuum state of M-theory \cite{evadine}.
Thus it might not be possible to imagine that all cosmological solutions
which preserve a given amount of SUSY are continuously connected to the
11D SUGRA regime.

Despite these potential problems, we think it is worthwhile to 
begin a study of compact, 
SUSY preserving, ten manifolds.  In this chapter we will only study 
examples where the
three form field vanishes.  The well known local condition for a 
Killing spinor, 
$D_{\mu} \epsilon = 0$, has as a condition for local integrability 
the vanishing
curvature condition
\eqn{killspin}{R_{\mu\nu}^{ab} \gamma_{ab} \epsilon = 0}
Thus, locally the curvature must lie in a subalgebra of the Lie 
algebra of $Spin (10)$ which
annihilates a spinor.  The global condition is that the holonomy 
around any 
closed path must lie in a subgroup which preserves a spinor.  
Since we are dealing with
11D SUGRA, we always have both the ${\bf 16}$  and ${\bf\overline{16}}$
representations
of 
$Spin (10)$ so SUSYs 
come in pairs.

For maximal SUSY the curvature must vanish identically and the space 
must be a torus.
The next possibility is to preserve half the spinors and this is
achieved 
by manifolds 
of the form $K3 \times T^7$ or orbifolds of them by freely acting
discrete 
symmetries. 

We now jump to the case of 4 SUSYs.   To find examples, it is convenient to
consider the decompositions $Spin (10) \supseteq
 Spin (k) \times Spin (10-k) $.

The ${\bf 16}$ is then a tensor product of two lower dimensional spinors.
For
$k=2$, the holonomy must be contained in $SU(4) \subseteq Spin (8)$ in
order to preserve a spinor, and it then preserves two (four once the
complex conjugate representation is taken into account). The corresponding
manifolds are products of Calabi-Yau fourfolds with two tori, perhaps
identified by the action of a freely acting discrete group.  This moduli
space is closely related to that of F-theory compactifications to four
dimensions with minimal four dimensional SUSY. The three spatial
dimensions are then compactified on a torus. For $k=3$ the holonomy must
be in $G_2 \subseteq Spin (7)$.  The manifolds are, up to discrete
identifications, products of Joyce manifolds and three tori.  For $k=4$
the holonomy is in $SU(2) \times SU(3)$.  The manifolds are free orbifolds
of products of Calabi-Yau threefolds and K3 manifolds.  This moduli space
is that of the heterotic string compactified on a three torus and
Calabi-Yau three fold. The case $k=5$ does not lead to any more examples
with precisely 4 SUSYs.

It is possible that M-theory contains U-duality transformations which map
us between these classes.  For example, there are at least some examples
of F-theory compactifications to four dimensional Minkowski space which
are dual to heterotic compactifications on threefolds.  After further
compactification on three tori we expect to find a map between the $k=2$
and $k=4$ moduli spaces.

To begin the study of the cosmology of these moduli spaces we restrict the
Einstein Lagrangian to metrics of the form 
\eqn{modmet}{ds^2 = - dt^2 +
g_{AB} (t) dx^A dx^B} 
where the euclidean signature metric $g_{AB}$ lies
in one of the moduli spaces.  Since all of these are spaces of solutions
of the Einstein equations they are closed under constant rescaling of the
metric.  Since they are spaces of restricted holonomy, this is the only
Weyl transformation which relates two metrics in a moduli space. Therefore
the equations (\ref{cwde}) and (\ref{nlsigma}) remain valid, where
$G_{ij}$ is now the De~Witt metric on the restricted moduli space of unit
volume metrics.

It is clear that the metric on the full moduli space still has Lorentzian
signature in the SUGRA approximation.  In some of these cases of lower
SUSY, we expect the metric to be corrected in the quantum theory.  
However, we do not expect these corrections to alter the signature of the
metric.  To see this note that each of the cases we have described has a
two torus factor.  If we decompactify the two torus, we expect a low
energy field theoretic description as three dimensional gravity coupled to
scalar fields and we can perform a Weyl transformation so that the
coefficient of the Einstein action is constant.  The scalar fields must
have positive kinetic energy and the Einstein term must have its
conventional sign if the theory is to be unitary.  Thus, the
decompactified moduli space has a positive metric.  In further
compactifying on the two torus, the only new moduli are those contained in
gravity, and the metric on the full moduli space has Lorentzian signature.

Note that as in the case of maximal SUSY, the region of the moduli space
with large ten volume and all other moduli held fixed, is in the future
light cone of any finite point in the moduli space.  Thus we suspect that
much of the general structure that we uncovered in the toroidal moduli
space, will survive in these less supersymmetric settings.

The most serious obstacle to this generalization appears in the case
of 4 (or fewer) supercharges.  In that case, general arguments do not
forbid the appearance of a potential in the Lagrangian for the moduli.
Furthermore, at generic points in the moduli space one would expect
the energy density associated with that potential to be of order the
fundamental scales in the theory.  In such a situation, it is difficult
to justify the Born-Oppenheimer separation between moduli and high
energy degrees of freedom.  Typical motions of the moduli on their
potential have frequencies of the same order as those of 
the ultraviolet degrees of freedom.

We do not really have a good answer to this question.  Perhaps the
approximation only makes sense in regions where the potential is small.
We know that this is true in extreme regions of moduli space in which
SUSYs are approximately restored.
However, it is notoriously difficult to stabilize the system
asymptotically far into such a region.  This difficulty is particularly
vexing in the context of currently popular ideas
\cite{horwita}-\cite{horwitc} in
which the fundamental scale of M-theory is taken to be orders of
magnitude smaller than the Planck scale.

\section{Discussion}

We have demonstrated that the modular group of toroidally compactified
M-theory prescribes a Lorentzian structure in the moduli space which
precisely mirrors that found in the low energy effective Einstein
action.
We argued that a similar structure will be found for moduli
spaces of lower SUSY, although the precise details could not be worked
out because the moduli spaces and metrics on them generally receive
quantum corrections.  As a consequence the mathematical structure of the
modular group is unknown.  Nonetheless we were able to argue that it
will be a group of isometries of a Lorentzian manifold.
Thus, we argue that the generic mathematical structure discussed in   
minisuperspace\footnote{A term which we have avoided up to this point
because it is confusing in a supersymmetric theory.}
approximations to quantum cosmology based on the low
energy field equations actually has an exact meaning in M-theory.
We note however that the detailed structure of the equations will be  
different in M-theory, since the correct minisuperspace is a moduli
space of static, SUSY preserving static solutions.

The Lorentzian structure prescribes a natural arrow of time, with a
general cosmological solution interpolating between a singular past   
where the holographic principle is violated and a future described by
11D SUGRA or weakly coupled string theory where low energy effective
field theory is a good approximation to the gross features of the
universe.   Note that it is {\it not} the naive arrow of time of any given
low
energy gravity theory, which points from small volume to large volume.
Many of the safe regions of moduli space are singular from the point of
view of any given low energy effective theory.  We briefly described how
those singularities are avoided in the presence of matter.

We believe that the connections we have uncovered are important and   
suggest that there are crucial things to be learned from cosmological
M-theory even if we are only interested in microphysics.  We 
realize that we have only made a small beginning in
understanding the import of these observations.  

Finally, we want to emphasize that our identification of moduli spaces
of SUSY preserving static solutions of SUGRA (which perhaps deserve
a more exact, M-theoretical characterization) as the appropriate
arena for early universe cosmology, provides a new starting point for 
investigating old cosmological puzzles.  We hope to report some progress
in the investigation of these modular cosmologies in the near future.


\section{Appendix A: The appearance of $\Gamma_8$ and $\Gamma_{9,1}$}

In this appendix we will explain the appearance of the lattices $\Gamma_8$
and $\Gamma_{9,1}$ in our framework. For the
group $\Rut_k$
we have found a bilinear invariant 
\eqn{scpr}{\vec u\cdot \vec v=(9-k)\sum_{i=1}^k (u_iv_i)
+(\sum_{i=1}^k u_i)(\sum_{i=1}^k v_i).}
Now let us take a vector
\eqn{vdef}{\vec v\equiv
\vec v_{123}=(-\frac23,-\frac23,-\frac23,+\frac13,+\frac13,\dots)}
and calculate its scalar product with a vector $\vec p$ according to
(\ref{scpr}). The result is
\eqn{scprp}{\vec v\cdot \vec p=(9-k)\sum_{i=1}^k (v_ip_i)
+\frac{k-9}3 (\sum_{i=1}^k p_i)=(k-9)(p_1+p_2+p_3).}
Thus for $k=9$ the product vanishes and for $k=10$ and $k=8$
the product equals $\pm (p_1+p_2+p_3)$. If the entries $(-2/3)$ are
at the positions $i,j,k$ instead of $1,2,3$, we get $\pm(p_i+p_j+p_k)$.
Substituting $\vec v_{ijk}$ also for $\vec p$, we obtain
\eqn{normv}{\vec v_{ijk}\cdot \vec v_{ijk} = 2(9-k).}
So this squared norm equals $\pm 2$ for $k=8,10$. More generally, we can
calculate the scalar products of any two $v_{ijk}$'s and the result is
\eqn{anysc}{\!\frac{\vec v_{ijk}\!\cdot\! \vec
v_{lmn}}{9-k}\!=\!\left\lbrace
\begin{array}{ll}
+\frac23+\frac23+\frac23=+2&\mbox{if $\{i,j,k\}$ and $\{l,m,n\}$ have
3 elements in common}\\
+\frac23+\frac23-\frac13=+1&\mbox{if $\{i,j,k\}$ and $\{l,m,n\}$ have 
2 elements in common}\\
+\frac23-\frac13-\frac13=0&\mbox{if $\{i,j,k\}$ and $\{l,m,n\}$ have 
1 element in common}\\
-\frac13-\frac13-\frac13=-1&\mbox{if $\{i,j,k\}$ and $\{l,m,n\}$ have 
no elements in common}
\end{array}
\right.
}
so the corresponding angles are $0^\circ$, $60^\circ$, $90^\circ$
and $120^\circ$.
Now a reflection with respect to the hyperplane perpendicular to
$\vec v\equiv \vec v_{ijk}$
is given by
\eqn{reflp}{\vec p\mapsto \vec p' = \vec p-2\vec v\,\,\frac{\vec
p\cdot\vec
v}{\vec v\cdot\vec v}}
and we see that for $k\ne 9$ it precisely reproduces the 2/5
transformation (\ref{rutdef}).

Let us define the lattice $\Lambda_k$ to be the lattice of all integer
combinations of the vectors $\vec v_{ijk}$. We will concentrate on the
cases $k=8$ and $k=10$ since $\Lambda_k$ will be shown to be even
self-dual lattices.
Thus they are isometric to the unique even self-dual lattices in these
dimensions.

It is easy to see that for $k=8,10$, the lattice $\Lambda_k$ contains
exactly those
vectors whose entries are multiples of $1/3$ 
equal modulo 1.
The reason is that all possible multiples of $1/3$ modulo 1
i.e. $-1/3,0,+1/3$ are realized by $-\vec v_{ijk},0,\vec v_{ijk}$
and we can also change any coordinate by one (or any integer) because
for instance
\eqn{byone}{\begin{array}{rcll}
\vec v_{123}+\vec v_{456}+\vec v_{789}&=&(0,0,0,0,0,0,
0,0,0,1)&\mbox{for }k=10\\
\vec v_{812}+\vec v_{345}+\vec v_{678}&=&(0,0,0,0,0,0,
0,-1)&\mbox{for }k=8
\end{array}}
Since the scalar products of any two $\vec v_{ijk}$ are integers
and the squared norms of $\vec v$'s are even, $\Lambda_8$ and
$\Lambda_{10}$ are even lattices. Finally we prove that they are
self-dual.

The dual lattice is defined as the lattice of all vectors whose scalar
products with any elements of the original lattice 
(or with any of its generators $\vec v_{ijk}$) are integers. Because of
(\ref{scprp}) it means that the sum of any three coordinates should be
an integer. But it is easy to see that this condition is identical to the
condition that the entries are multiples of $1/3$ equal modulo 1.
The reason is that the difference
$(p_1+p_2+p_3)-(p_1+p_2+p_4)=p_3-p_4$ must be also an integer -- so
all the coordinates are equal modulo one. But if they are equal modulo 1,
they must be equal to a multiple of $1/3$ because the sum of three such
numbers must be integer.

We think that this construction of $\Gamma_8$ and $\Gamma_{9,1}$ is 
more natural in the
context of U-dualities than the construction using the root lattice of
$E_8$ with the $SO(16)$ sublattice generated by $\pm e_i\pm e_j$.

\vspace{3mm}

Let us finally mention that the orders of $\Rut_3,$
$\Rut_4,$\dots $\Rut_8$ are equal to $2\cdot 3!$,
$5\cdot 4!$, $16\cdot 5!$, $72\cdot 6!$, $576\cdot 7!$ and
$17280\cdot 8!$. For instance the group $\Rut_4$ is isomorphic
to $S_5$. We can see it explicitly. Defining
\eqn{fourfromfive}{p_i=R_i-\frac 13(R_1+R_2+R_3+R_4-R_5),\qquad
i=1,2,3,4}
which leaves $p_i$ invariant under the transformation
$R_i\mapsto R_i+\lambda$, the permutation of $R_4$ and $R_5$ is easily
seen to generate the 2/5 transformation on $p_1,p_2,p_3$. Note that
$p_1+p_2+p_3=-R_4+R_5$.

%% file: chap7-bfmhetero.tex
\chapter{Hyperbolic structure of the heterotic moduli space}

In analogy with the previous chapter, we study the asymptotic limits of
the heterotic string theories compactified on tori \cite{BANXMOTL}. We
find a bilinear form uniquely determined by dualities which becomes
Lorentzian in the case of one spacetime dimension.  For the case of the
${\rm SO}(32)$ theory, the limiting descriptions include ${\rm SO}(32)$
heterotic strings, type~I, type~IA and other T-duals, M-theory on K3, type
IIA theory on K3 and type IIB theory on K3 and possibly new limits not
understood yet.

\section{Introduction}

In the previous chapter we showed that the space of asymptotic directions
in the moduli space of toroidally compactified M-theory had a hyperbolic
metric, related to the hyperbolic structure of the $E_{10}$ duality group.  
We pointed out that this could have been anticipated from the hyperbolic
nature of metric on moduli space in low energy SUGRA, which ultimately
derives from the negative kinetic term for the conformal factor.

An important consequence of this claim is that there are asymptotic
regions of the moduli space which cannot be mapped onto either 11D
SUGRA (on a large smooth manifold) or weakly coupled Type II string
theory.  These regions represent true singularities of M-theory at
which no known description of the theory is applicable.
Interestingly, the classical solutions of the theory all follow
trajectories which interpolate between the mysterious singular region
and the regions which are amenable to a semiclassical description.
This introduces a natural arrow of time into the theory.  We suggested
that moduli were the natural semiclassical variables that define
cosmological time in M-theory and that ``the Universe began'' in the
mysterious singular region.

We note that many of the singularities of the classical solutions {\it
can} be removed by duality transformations.  This makes the special
nature of the singular region all the more striking.\footnote{For
reference, we note that there are actually two different types of
singular region: neither the exterior of the light cone in the space
of asymptotic directions, nor the past light cone, can be mapped into
the safe domain.  Classical solutions do not visit the exterior of the
light cone.}

In view of the connection to the properties of the low energy SUGRA
Lagrangian, we conjectured in~\cite{BFM} that the same sort of
hyperbolic structure would characterize moduli spaces of M-theory with
less SUSY than the toroidal background.  In this chapter, we verify this
conjecture for 11D SUGRA backgrounds of the form $K3 \times T^6$,
which is the same as the moduli space of heterotic strings
compactified on $T^9$.  A notable difference is the absence of a
completely satisfactory description of the safe domains of asymptotic
moduli space.  This is not surprising.  The moduli space is known to
have an F-theory limit in which there is no complete semiclassical
description of the physics.  Rather, there are different semiclassical
limits valid in different regions of a large spacetime.

Another difference is the appearance of asymptotic domains with
different internal symmetry groups.  11D SUGRA on $K3 \times T^3$
exhibits a U$(1)^{28}$ gauge group in four noncompact dimensions.  At
certain singularities, this is enhanced to a nonabelian group, but
these singularities have finite codimension in the moduli space.
Nonetheless, there are asymptotic limits in the full moduli space
(i.e. generic asymptotic directions) in which the full heterotic
symmetry group is restored.  From the heterotic point of view, the
singularity removing, symmetry breaking, parameters are Wilson lines
on $T^9$.  In the infinite (heterotic torus) volume limit, these
become irrelevant.  In this chapter we will only describe the subspace
of asymptotic moduli space with full ${\rm SO}(32)$ symmetry.  We will
call this the HO moduli space from now on.  The points of the moduli
space will be parametrized by the dimensionless heterotic string
coupling constant $\gh=\exp{p_0}$ and the radii $R_i=\lh\exp{p_i}$
where $i=1,\dots, 10-d$ with $d$ being the number of large spacetime
dimensions and $\lh$ denoting the heterotic string length. Throughout
the chapter we will neglect factors of order one.

Apart from these, more or less expected, differences, our results are
quite similar to those of~\cite{BFM}.  The modular group of the
completely compactified theory preserves a Lorentzian bilinear form
with one timelike direction.  The (more or less) well understood
regimes correspond to the future light cone of this bilinear form,
while all classical solutions interpolate between the past and future
light cones.\linebreak We interpret this as evidence for a new
hyperbolic algebra ${\cal O}$, whose infinite momentum frame Galilean
subalgebra is precisely the affine algebra $\hat{{\rm o}} (8,24)$
of~\cite{sentwod}--\cite{schwtwod}.  This would precisely mirror the
relation between $E_{10}$ and $E_9$.  Recently, Ganor~\cite{ori} has
suggested the $DE_{18}$ Dynkin diagram as the definition of the basic
algebra of toroidally compactified heterotic
strings.  This is indeed a hyperbolic algebra in the sense that it
preserves a nondegenerate bilinear form with precisely one negative
eigenvalue.\footnote{Kac' definition of a hyperbolic algebra requires
it to turn into an affine or finite dimensional algebra when one root
of the Dynkin diagram is cut.  We believe that this is too restrictive
and that the name hyperbolic should be based solely on the signature
of the Cartan metric.  We thank O. Ganor for discussions of this
point.}

\subsection{The bilinear form}

We adopt the result of~\cite{BFM} with a few changes in
notation. First we will use $d=11-k$ instead of $k$ because now we
start in ten dimensions instead of eleven.  The parameter that makes
the parallel between toroidal M-theory and heterotic compactifications
most obvious is the number of large spacetime dimensions
$d$. In~\cite{BFM}, the bilinear form was
\eqn{bilstart}{I=\left(\sum_{i=1}^{k}P_i\right)^2 +
(d-2)\sum_{i=1}^k(P_i^2)\,.}  
where $P_i$ (denoted $p_i$ in~\cite{BFM}) are the logarithms of the
radii in 11-dimensional Planck units.

Now let us employ the last logarithm $P_k$ as the M-theoretical circle
of a type IIA description. For the HE theory, which can be understood
as M-theory on a line interval, we expect the same bilinear form where
$P_k$ is the logarithm of the length of the Ho\hacek rava-Witten line
interval.  Now we convert~(\ref{bilstart}) to the heterotic units
according to the formulae ($k-1=10-d$)
\eqn{bildva}{P_k=\frac 23 p_0\,,\qquad 
P_i=p_i-\frac 13 p_0\,, \qquad i=1,\dots, 10-d\,,}
where $p_0=\ln{\gh}$ and $p_i=\ln(R_i/ \lh)$ for $i=1,\dots 10-d$.  To
simplify things, we use natural logarithms instead of the logarithms
with a large base $t$ like in~\cite{BFM}. This corresponds to a simple
rescaling of $p$'s but the directions are finally the only thing that
we study.  In obtaining~(\ref{bildva}) we have used the well-known
formulae $R_{11}=\lple \gh^{2/3}$ and
$\lple=\gh^{1/3}\lh$. Substituing~(\ref{bildva}) into~(\ref{bilstart})
we obtain 
\eqn{biltri}{I=\left(-2p_0+\sum_{i=1}^{10-d}p_i\right)^2
+(d-2)\sum_{i=1}^{10-d}(p_i^2).}
This bilinear form encodes the kinetic terms for the moduli in the
$E_8\times E_8$ heterotic theory (HE) in the Einstein frame for the
large coordinates.

We can see very easily that~(\ref{biltri}) is conserved by
T-dualities.  A simple T-duality (without Wilson lines) takes HE
theory to HE theory with $R_1$ inverted and acts on the parameters
like
\eqn{tdu}{(p_0,p_1,p_2,\dots)\to(p_0-p_1,-p_1,p_2, \dots)\,.}
The change of the coupling constant keeps the effective 9-dimensional
gravitational constant $\gh^2/R_1=\gh'^2/R'_1$ (in units of $\lh$)
fixed.  In any number of dimensions~(\ref{tdu}) conserves the quantity
\eqn{pten}{p_{10}=-2p_0+\sum_{i=1}^{10-d}p_i}
and therefore also the first term in~(\ref{biltri}). The second term
in (\ref{biltri}) is fixed trivially since only the sign of $p_1$ was
changed. Sometimes we will use $p_{10}$ instead of $p_0$ as the extra
parameter apart from $p_1,\dots, p_{10-d}$.

In fact those two terms in~(\ref{biltri}) are the only terms conserved
by T-dualities and only the relative ratio between them is
undetermined.  However it is determined by S-dualities, which exist
for $d\leq 4$. For the moment, we ask the reader to take this claim on
faith.  Since the HE and HO moduli spaces are the same on a torus, the
same bilinear form can be viewed in the ${\rm SO}(32)$ language.  It
takes the form~(\ref{biltri}) in ${\rm SO}(32)$ variables as well.

Let us note also another interesting invariance 
of~(\ref{biltri}), which is useful for the ${\rm SO}(32)$ case.
Let us express the parameters in the terms of the natural
parameters of the S-dual type~I theory
\eqn{sduone}{p_0=-q_0=-\ln(\gi)\,, \qquad
p_i=q_i-\frac 12 q_0\,,\qquad i=1,\dots, 10-d\,,}
where $q_i=\ln(R_i/ \li)$. We used $\gi=1/ \gh$ and
$\lh=\gi^{1/2}\li$, the latter expresses that the tension of the
D1-brane and the heterotic strings are equal. Substituing this
into~(\ref{biltri}) we get the same formula with $q$'s.
\eqn{qbiltri}{I=\left(-2q_0+\sum_{i=1}^{10-d}q_i\right)^2
+(d-2)\sum_{i=1}^{10-d}(q_i^2)}

\subsection{Moduli spaces and heterotic S-duality}

Let us recall a few well-known facts about the moduli space of
heterotic strings toroidally compactified to $d$ dimensions. For $d>4$
the moduli space is
\eqn{modhet}{{\cal M}_d=\IR^+ \times
({\rm SO}(26-d,10-d,\IZ) \backslash {\rm SO}(26-d,10-d,\IR) / {\rm SO}(26-d,\IR)\times
{\rm SO}(10-d,\IR))\,.}
The factor $\IR^+$ determines the coupling constant $\lh$. For $d=8$
the second factor can be understood as the moduli space of
elliptically fibered K3's (with unit fiber volume), giving the duality
with the F-theory. For $d=7$ the second factor also corresponds to the
Einstein metrics on a K3 manifold with unit volume which expresses the
duality with M-theory on K3. In this context, the factor $\IR^+$ can
be understood as the volume of the K3. Similarly for $d=5,6,7$ the
second factor describes conformal field theory of type~II string
theories on K3, the factor $\IR^+$ is related to the type~IIA coupling
constant.

For $d=4$, i.e.\ compactification on $T^6$, there is a new
surprise. The field strength $H_{\kappa\lambda\mu}$ of the $B$-field
can be Hodge-dualized to a 1-form which is the exterior derivative of
a dual 0-form potential, the axion field.  The dilaton and axion are
combined in the $S$-field which means that in four noncompact
dimensions, toroidally compactified heterotic strings exhibit the
${\rm SL}(2,\IZ)$ S-duality.
\eqn{modhetfour}{{\cal M}_4={\rm SL}(2,\IZ)\backslash
{\rm SL}(2,\IR)/{\rm SO}(2,\IR) \,\times\,
({\rm SO}(22,6,\IZ) \backslash {\rm SO}(22,6,\IR) / {\rm SO}(22)\times
{\rm SO}(6)).}

Let us find how our parameters $p_i$ transform under S-duality.  The
S-duality is a kind of electromagnetic duality. Therefore an
electrically charged state must be mapped to a magnetically charged
state. The ${\rm U}(1)$ symmetry expressing rotations of one of the
six toroidal coordinate is just one of the 28 ${\rm U}(1)$'s in the
Cartan subalgebra of the full gauge group. It means that the
electrically charged states, the momentum modes in the given direction
of the six torus, must be mapped to the magnetically charged objects
which are the KK-monopoles.

The strings wrapped on the $T^6$ must be therefore mapped to the only
remaining point-like\footnote{Macroscopic strings (and
higher-dimensional objects) in $d=4$ have at least logarithmic IR
divergence of the dilaton and other fields and therefore their tension
becomes infinite.}  BPS objects available, i.e.\ to wrapped
NS5-branes.  We know that NS5-branes are magnetically charged with
respect to the $B$-field so this action of the electromagnetic duality
should not surprise us. We find it convenient to combine this
S-duality with T-dualities on all six coordinates of the torus. The
combined symmetry $ST^6$ exchanges the point-like BPS objects in the
following way:

\begin{equation}
\begin{array}[b]{rcl}
\mbox{momentum modes} & ~\leftrightarrow ~& 
\mbox{wrapped NS5-branes}\\[1ex]
\mbox{wrapped strings} & ~\leftrightarrow~ & \mbox{KK-monopoles}\\
\end{array}\label{tabulka}
\end{equation}

Of course, the distinguished direction inside the $T^6$ on both sides
is the same. The tension of the NS5-brane is equal to
$1/(\gh^2\lh^6)$. Now consider the tension of the KK-monopole. In 11
dimensions, a KK-monopole is reinterpreted as the D6-brane so its
tension must be
\eqn{tdsix}{T_{D6}=\frac{1}{g_{IIA}L_{IIA}^7}=\frac{R_{11}^2}{(\lple)^9}\,,}
where we have used $g_{IIA}=R_{11}^{3/2}\lple^{-3/2}$ and
$L_{IIA}=\lple^{3/2}R_{11}^{-1/2}$ (from the tension of the fundamental
string).

The KK-monopole must always be a $(d-5)$-brane where $d$ is the
dimension of the spacetime. Since it is a gravitational object and the
dimensions along its worldvolume play no role, the tension must be
always of order $(R_1)^2$ in appropriate Planck units where $R_1$ is
the radius of the circle under whose ${\rm U}(1)$ the monopole is
magnetically charged. Namely in the case of the heterotic string in
$d=4$, the KK-monopole must be another fivebrane whose tension is
equal to 
\eqn{kkhet}{T_{KK5}=\frac{R_1^2}{(\lt)^8}=
\frac{{R_1}^2}{\gh^2\lh^8}\,,}
where the denominators express the ten-dimensional Newton's constant.

Knowing this, we can find the transformation laws for $p$'s with
respect to the $ST^6$ symmetry. Here $V_6=R_1R_2R_3R_4R_5R_6$ denotes
the volume of the six-torus. Identifying the tensions
in~(\ref{tabulka}) we get
\eqn{vypocet}{\frac{1}{R'_1}=\frac{V_6}{\gh^2 R_1\lh^6}\,,\qquad
\frac{R'_1}{(\lh')^2}= \frac{V_6 R_1}{\gh^2\lh^8}\,.}

Dividing and multiplying these two equations we get respectively
\eqn{podil}{\frac{R'_1}{\lh'}=\frac{R_1}{\lh},\qquad
\frac{1}{\lh'}=\frac{V_6}{\gh^2\lh^7}\,.}

It means that the radii of the six-torus are fixed in string units
i.e.\ $p_1,\dots,p_6$ are fixed. Now it is straightforward to see that
the effective four-dimensional ${\rm SO}(32)$ coupling constant $\gh^2
\lh^6/V_6$ is inverted and the four-dimensional Newton's constant must
remain unchanged.  The induced transformation on the $p$'s is
\eqn{indtr}{(p_0,\,p_1,\dots, p_6,\, p_7,\,p_8 \dots)\to
(p_0+m,\,p_1,\dots,\, p_6,\, p_7+m,\,p_8+m\dots)}
where $m=(p_1+p_2+p_3+p_4+p_5+p_6-2p_0)$ and the form~(\ref{biltri})
can be checked to be constant.  It is also easy to see that such an
invariance uniquely determines the form up to an overall normalization
i.e.\ it determines the relative magnitude of two terms
in~(\ref{biltri}).

For $d=4$ this $ST^6$ symmetry can be expressed as $p_{10}\to -p_{10}$
with $p_1,\dots, p_6$ fixed which gives the $\IZ_2$ subgroup of the
${\rm SL}(2,\IZ)$. For $d=3$ the transformation~(\ref{indtr}) acts as
$p_7\leftrightarrow p_{10}$ so $p_{10}$ becomes one of eight
parameters that can be permuted with each other. It is a trivial
consequence of the more general fact that in three dimensions, the
dilaton-axion field unifies with the other moduli and the total space
becomes~\cite{senthreed} 
\eqn{modth}{{\cal M}_3= {\rm SO}(24,8,\IZ)\,\backslash\, {\rm
SO}(24,8,\IR)\, /\, {\rm SO}(24,\IR)\times {\rm SO}(8,\IR)\,.}

We have thus repaid our debt to the indulgent reader, and verified
that the bilinear form~(\ref{biltri}) is indeed invariant under the
dualities of the heterotic moduli space for $d \geq 3$.  For $d=2$ the
bilinear form is degenerate and is the Cartan form of the affine
algebra $\hat{o}(8,24)$ studied by~\cite{sentwod}. For $d=1$ it is the
Cartan form of $DE_{18}$~\cite{ori}.  The consequences of this for the
structure of the extremes of moduli space are nearly identical to
those of~\cite{BFM}.  The major difference is our relative lack of
understanding of the safe domain.  We believe that this is a
consequence of the existence of regimes like F-theory or 11D SUGRA on
a large smooth K3 with isolated singularities, where much of the
physics is accessible but there is no systematic expansion of all
scattering amplitudes.  In the next section we make some remarks about
different extreme regions of the restricted moduli space that
preserves the full ${\rm SO}(32)$ symmetry.

\section{Covering the ${\rm SO}(32)$ moduli space}

\subsection{Heterotic strings, type~I, type~IA and $d\geq 9$}

One new feature of heterotic moduli spaces is the apparent possibility
of having asymptotic domains with enhanced gauge symmetry.  For
example, if we consider the description of heterotic string theory on
a torus from the usual weak coupling point of view, there are domains
with asymptotically large heterotic radii and weak coupling, where the
the full nonabelian rank $16$ Lie groups are restored.  All other
parameters are held fixed at what appears from the weak coupling point
of view to be \lq\lq generic\rq\rq values.  This includes Wilson
lines.  In the large volume limit, local physics is not sensitive to
the Wilson line symmetry breaking.

Now, consider the limit described by weakly coupled Type IA string
theory on a large orbifold.  In this limit, the theory consists of
D-branes and orientifolds, placed along a line interval.  There is no
way to restore the $E_8\times E_8$ symmetry in this regime.  Thus,
even the safe domain of asymptotic moduli space appears to be divided
into regimes in which different nonabelian symmetries are restored.
Apart from sets of measure zero (e.g.\ partial decompactifications) we
either have one of the full rank $16$ nonabelian groups, or no
nonabelian symmetry at all.  The example of F-theory tells us that the
abelian portion of asymptotic moduli space has regions without a
systematic semiclassical expansion.

In a similar manner, consider the moduli space of the $E_8\times E_8$
heterotic strings on rectilinear tori.  We have only two semiclassical
descriptions with manifest $E_8\times E_8$ symmetry, namely HE strings
and the Ho\hacek rava-Witten (HW) domain walls. Already for $d=9$ (and
any $d<9$) we would find limits that are described neither by HE nor
by HW. For example, consider a limit of M-theory on a cylinder with
very large $\gh$ but the radius of the circle, $R$, in the domain $L_P
\gg R \gg \lh^2/ \lple $, and unbroken $E_8\times E_8$.  We do not know
how to describe this limit with any known semiclassical expansion.  We
will find that we can get a more systematic description of asymptotic
domains in the HO case, and will restrict attention to that regime for
the rest of this chapter.

\begin{figure} [t]
\qquad\qquad\qquad\epsfig{file=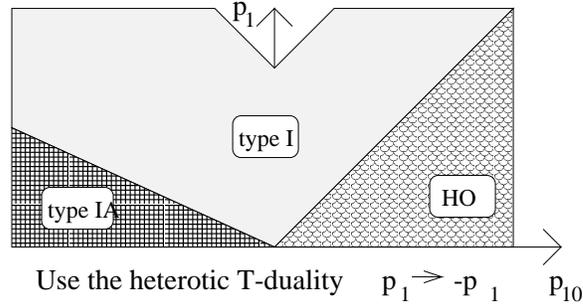}
\caption{The limits in $d=9$.}
\label{ninefigure}
\end{figure}

For $d=10$ there are only two limits.  $p_0<0$ gives the heterotic
strings and $p_0>0$ is the type I theory. However already for $d=9$ we
have a more interesting picture analogous to the
figure~\ref{ninefigure} in~\cite{BFM}. Let us make a counterclockwise
trip around the figure.  We start at a HO point with $p_1=0$ which is
a weakly coupled heterotic string theory with radii of order $\lh$
(therefore it is adjacent to its T-dual region). When we go around the
circle, the radius and also the coupling increases and we reach the
line $p_0\equiv (p_1-p_{10})/2=0$ where we must switch to the type I
description. Then the radius decreases again so that we must perform a
T-duality and switch to the type~IA description. This happens for
$p_1-(p_0/2)=(3p_1+p_{10})/4=0$; we had to convert $R_1$ to the units
of $\li=\gh^{1/2}\lh$. Then we go on and the coupling $g_{IA}$ and/or
the size of the line interval increases. The most interesting is the
final boundary given by $p_1=0$ which guarantees that each of the
point of the $p$-space is covered precisely by one limit.

We can show that $p_1>0$ is precisely the condition that the dilaton
in the type~IA theory is not divergent. Rou\-g\-h\-ly speaking, in
units of $\li=L_{IA}$ the ``gravitational potential'' is linear in
$x_1$ and proportional to $g_{IA}^2 / g_{IA}$. Here $g_{IA}^2$ comes
from the gravitational constant and $1/g_{IA}$ comes from the tension
of the D8-branes. Therefore we require not only $g_{IA}<1$ but also
$g_{IA}<\li / R_{line\,interval}$. Performing the T-duality $\li /
R_{line\,interval}=R_{circle} / \li$ and converting to $\lh$ the
condition becomes precisely $R_{circle}>\lh$.

In all the text we adopt (and slightly modify) the standard definition
\cite{BFM} for an asymptotic description to be viable: dimensionless
coupling constants should be smaller than one, but in cases without
translational invariance, the dilaton should not diverge anywhere, and
the sizes of the effective geometry should be greater than the
appropriate typical scale (the string length for string theories or
the Planck length for M-theory). It is important to realize that in
the asymptotic regions we can distinguish between e.g.\ type~I and
type~IA because their physics is different. We cannot distinguish
between them in case the T-dualized circle is of order $\li$ but such
vacua are of measure zero in our investigation and form various
boundaries in the parameter space.  This is the analog of the
distinction we made between the IIA and IIB asymptotic toroidal moduli
spaces in~\cite{BFM}

\subsection{Type IA${}^2$ and $d=8$}

In $d=8$ we will have to use a new desciption to cover the parameter
space, namely the double T-dual of type I which we call type~IA${}^2$.
Generally, type~IA${}^k$ contains 16 D-$(9-k)$-branes, their images
and $2^k$ orientifold $(9-k)$-planes. We find it also useful to
perform heterotic T-dualities to make $p_i$ positive for $i=1,\dots,
10-d$ and sort $p$'s so that our interest is (without a loss of
generality) only in configurations~with
\eqn{sort}{0\leq p_1\leq p_2\leq \cdots \leq p_{10-d}\,.}
We need positive $p$'s for the heterotic description to be valid but
such a transformation can only improve the situation also for type~I
and its T-dual descriptions since when we turn $p$'s from negative to
positive values, $\gh$ increases and therefore $\gi$ decreases. For
type~I we also need large radii. For its T-duals we need a very small
string coupling and if we make a T-duality to convert $R>\li$ into $R<
\li$, the coupling $g_{IA}$ still decreases; therefore it is good to
have as large radii in the type~I limit as possible.

In $d=8$ our parameters are $p_0,p_1,p_2$ or $p_{10},p_1,p_2$ where
$p_{10}=-2p_0+p_1+p_2$ and we will assume $0< p_1< p_2$ as we have
explained (sets of measure zero such as the boundaries between regions
will be neglected). If $p_0<0$, the HO description is good. Otherwise
$p_0>0$. If furthermore $2p_1-p_0>0$ (and therefore also
$2p_2-p_0>0$), the radii are large in the type~I units and we can use
the (weakly coupled) type~I description. Otherwise $2p_1-p_0<0$. If
furthermore $2p_2-p_0>0$, we can use type~IA strings. Otherwise
$2p_2-p_0<0$ and the type~IA${}^2$ description is valid. Therefore we
cover all the parameter space. Note that the F-theory on K3 did not
appear here.  In asymptotic moduli space, the F-theory regime
generically has no enhanced nonabelian symmetries.

In describing the boundaries of the moduli space, we used the
relations $\lh=\gi^{1/2}\li$, $\gh=1/\gi$.  The condition for the
dilaton not to diverge is still $p_1>0$ for any type~IA${}^k$
description. The longest direction of the $T^k/ \IZ_2$ of this theory
is still the most dangerous for the dilaton divergence and is not
affected by the T-dualities on the shorter directions of the $T^k/
\IZ_2$ orientifold. For $d=9$ (and fortunately also for $d=8$) the
finiteness of the dilaton field automatically implied that $g_{IA^k} <
1$. However this is not true for general $d$. After a short chase
through a sequence of S and T-dualities we find that the condition
$g_{IA^k}<1$ can be written as
\eqn{karkulka}{(k-2)p_0 -2\sum_{i=1}^k p_i<0\,.}
We used the trivial requirement that the T-dualities must be performed
on the shortest radii (if $R_j< \li$, also $R_{j-1} < \li$ and
therefore it must be also T-dualized). Note that for $k=1$ the
relation is $-p_0-2p_1<0$ which is a trivial consequence of $p_1>0$
and $p_0>0$.  Also for $k=2$ we get a trivial condition
$-2(p_1+p_2)<0$. However for $k>2$ this condition starts to be
nontrivial. This is neccessary for consistency: otherwise IA${}^k$
theories would be sufficient to cover the whole asymptotic moduli
space, and because of S-dualities we would cover the space several
times. It would be also surprising not to encounter regimes described
by large K3 geometries.

\subsection{Type IA${}^3$, M-theory on K3 and $d=7$}

This happens already for $d=7$ where the type~IA${}^3$ description
must be added. The reasoning starts in the same way: for $p_0<0$ HO,
for $2p_1-p_0>0$ type~I, for $2p_2-p_0>0$ type~IA, for $2p_3-p_0>0$
type~IA${}^2$.

However, when we have $2p_3-p_0<0$ we cannot deduce that the
conditions for type~IA${}^3$ are obeyed because also~(\ref{karkulka})
must be imposed: 
\eqn{kremilek}{p_0-2(p_1+p_2+p_3)<0\,.}
It is easy to see that this condition is the weakest one i.e.\ that it
is implied by any of the conditions $p_0<0$, $2p_1-p_0>0$,
$2p_2-p_0>0$ or $2p_3-p_0>0$.  Therefore the region that we have not
covered yet is given by the opposite equation
\eqn{vochomurka}{2p_0-4(p_1+p_2+p_3)=-p_{10}-3(p_1+p_2+p_3)>0} 
The natural hypothesis is that this part of the asymptotic parameter
space is the limit where we can use the description of M-theory on a
K3 manifold. However things are not so easy: the condition that
$V_{K3}> (\lple)^4$ gives just $p_{10}<0$ which is a weaker requirement
than~(\ref{vochomurka}).

The K3 manifold has a $D_{16}$ singularity but this is not the real
source of the troubles. A more serious issue is that the various
typical sizes of such a K3 are very different and we should require
that each of them is greater than $\lple$ (which means that the
shortest one is). In an analogous situation with $T^4$ instead of K3
the condition $V_{T^4}>\lple^4$ would be also insufficient: all the
radii of the four-torus must be greater than $\lple$.

Now we would like to argue that the region defined
by~(\ref{vochomurka}) with our gauge $0<p_1<p_2<p_3$ can indeed be
described by the 11D SUGRA on K3, except near the $D_{16}$
singularity.  Therefore, all of the asymptotic moduli space is covered
by regions which have a reasonable semiclassical description.

While the fourth root of the volume of K3 equals
\eqn{arabela}{\frac{V_{K3}^{1/4}}{\lple}=\frac{\gh^{1/3}\lh^{1/2}}{V_3^{1/6}}
=\exp\left({p_0\over 3}-{(p_1+p_2+p_3)\over 6}\right)=\exp\left({-p_{10}\over 6}\right),}
the minimal typical distance in K3 must be corrected to agree
with~(\ref{vochomurka}). We must correct it only by a factor depending
on the three radii in heterotic units (because only those are the
parameters in the moduli space of metric on the K3) so the distance
equals (confirming~(\ref{vochomurka}))
\eqn{rumburak}{\frac{L_{min.K3}}{\lple}=\exp\left({-p_{10}\over 6}
-{(p_1+p_2+p_3)\over 2}\right).}
Evidence that~(\ref{rumburak}) is really correct and thus that we
understand the limits for $d=7$ is the following.  We must first
realize that 16 independent two-cycles are shrunk to zero size because
of the $D_{16}$ singularity present in the K3 manifold. This
singularity implies a lack of understanding of the physics in a
vicinity of this point but it does not prevent us from describing the
physics in the rest of K3 by 11D SUGRA. So we allow the 16 two-cycles
to shrink. The remaining 6 two-cycles generate a space of signature
3+3 in the cohomology lattice: the intersection numbers are identical
to the second cohomology of $T^4$. We can compute the areas of those 6
two-cycles because the M2-brane wrapped on the 6-cycles are dual to
the wrapped heterotic strings and their momentum modes. Now let us
imagine that the geometry of the two-cycles of K3 can be replaced by
the 6 two-cycles of a $T^4$ which have the same intersection number.

It means that the areas can be written as $a_1a_2,a_1a_3,a_1a_4$,
$a_2a_3,a_2a_4,a_3a_4$ where $a_1,a_2,a_3,a_4$ are the radii of the
four-torus and correspond to some typical distances of the K3. If we
order the $a$'s so that $a_1<a_2<a_3<a_4$, we see that the smallest of
the six areas is $a_1a_2$ (the largest two-cycle is the dual $a_3a_4$)
and similarly the second smallest area is $a_1a_3$ (the second largest
two-cycle is the dual $a_2a_4$). On the heterotic side we have radii
$\lh<R_1<R_2<R_3$ (thus also $\lh^2/R_3< \lh^2/R_2< \lh^2/R_1<\lh$)
and therefore the correspondence between the membranes and the
wrapping and momentum modes of heterotic strings tells us that
\eqn{budulinek}{\frac{a_1a_2}{\lple^3}=\frac{1}{R_3}\,,
\qquad \frac{a_3a_4}{\lple^3}=\frac{R_3}{\lh^2}\,,
\qquad
\frac{a_1a_3}{\lple^3}=\frac{1}{R_2}\,,
\qquad \frac{a_2a_4}{\lple^3}=\frac{R_2}{\lh^2}\,.}
As a check, note that $V_{K3}=a_1a_2a_3a_4$ gives us $\lple^6/ \lh^2$
as expected (since heterotic strings are M5-branes wrapped on $K3$).
We will also assume that
\eqn{liska}{\frac{a_1a_4}{\lple^3}=\frac{1}{R_1}\,, 
\qquad \frac{a_2a_3}{\lple^3}=\frac{R_1}{\lh^2}\,.}
Now we can calculate the smallest typical distance on the K3.
\eqn{jezinka}{a_1=\sqrt{\frac{a_1a_2\cdot a_1a_3}{a_2a_3}}=
\frac{\lple^{3/2}\lh}{\sqrt{R_1R_2R_3}}\,,}
which can be seen to coincide with~(\ref{rumburak}). There is a
subtlety that we should mention. It is not completely clear whether
$a_1a_4<a_2a_3$ as we assumed in~(\ref{liska}). The opposite
possibility is obtained by exchanging $a_1a_4$ and $a_2a_3$
in~(\ref{liska}) and leads to $a_1$ greater than~(\ref{jezinka}) which
would imply an overlap with the other regions. Therefore we believe
that the calculation in~(\ref{liska}) and~(\ref{jezinka}) is the
correct way to find the condition for the K3 manifold to be large
enough for the 11-dimensional supergravity (as a limit of M-theory) to
be a good description.

\subsection{Type IA${}^{4,5}$, type~IIA/B on K3 and $d=6,5$}

Before we will study new phenomena in lower dimensions, it is useful
to note that in any dimension we add new descriptions of the physics.
The last added limit always corresponds to the ``true'' S-dual of the
original heterotic string theory -- defined by keeping the radii fixed
in the heterotic string units (i.e.\ also keeping the shape of the K3
geometry) and sending the coupling to infinity -- because this last
limit always contains the direction with $p_0$ large and positive (or
$p_{10}$ large and negative) and other $p_i$'s much smaller.

\begin{itemize}
\item In 10 dimensions, the true S-dual of
heterotic strings is the type~I theory.
\item In 9 dimensions it is type~IA.
\item In 8 dimensions type~IA${}^2$.
\item In 7 dimensions we get M-theory on K3.
\item In 6 dimensions type IIA strings on K3.
\item In 5 dimensions type IIB strings on K3$\times S^1$
where the circle decompactifies as the coupling goes to infinity. The
limit is therefore a six-dimensional theory.
\item In 4 dimensions we observe a mirror copy of the region $p_{10}<0$
to arise for $p_{10}>0$. The strong coupling limit is the heterotic
string itself.
\item In 3 dimensions the dilaton-axion is already unified with the other
moduli so it becomes clear that we studied an overly specialized
 direction in the
examples above. Nevertheless the same claim as in $d=4$ can be made.
\item In 2 dimensions only positive values of $p_{10}$ are possible
therefore the strong coupling limit does not exist in the safe domain 
of moduli space. 
\item In 1 dimension the Lorentzian structure of the parameter space
emerges. Only the future light cone corresponds to semiclassical physics
which is reasonably well understood.
The strong coupling limit defined above would lie inside the unphysical
past light cone.
\end{itemize}

Now let us return to the discussion of how to separate the parameter
space into regions where different semiclassical descriptions are
valid. We may repeat the same inequalities as in $d=7$ to define the
limits HO, I, IA, IA${}^2$, IA${}^3$. But for M-theory on K3 we must
add one more condition to the constraint~(\ref{vochomurka}): a new
circle has been added and its size should be also greater than
$\lple$. For the new limit of the type~IIA strings on K3 we encounter
similar problems as in the case of the M-theory on K3. Furthermore if
we use the definition~(\ref{rumburak}) and postulate this shortest
distance to be greater than the type~IIA string length, we do not seem
to get a consistent picture covering the whole moduli space. Similarly
for $d=5$, there appear two new asymptotic descriptions, namely
type~IA${}^5$ theory and type~IIB strings on $K3\times S^1$. It is
clear that the condition $g_{IA^5}<1$ means part of the parameter
space is not understood and another description, most probably
type~IIB strings on $K3\times S^1$, must be used.  Unfortunately at
this moment we are not able to show that the condition for the IIB
theory on K3 to be valid is complementary to the condition
$g_{IA^5}<1$. A straightforward application of~(\ref{jezinka}) already
for the type~IIA theory on a K3 gives us a different inequality. Our
lack of understanding of the limits for $d<7$ might be solved by
employing a correct T-duality of the type~IIA on K3 but we do not have
a complete and consistent picture at this time.

\subsection{Type IA${}^6$ and S-duality in $d=4$}

Let us turn to the questions that we understand better.  As we have
already said, in $d=4$ we see the $\IZ_2$ subgroup of the ${\rm
SL}(2,\IZ)$ S-duality which acts as $p_{10}\to -p_{10}$ and
$p_1,\dots,p_6$ fixed in our formalism. This reflection divides the
$p$-space to subregions $p_{10}>0$ and $p_{10}<0$ which will be
exchanged by the S-duality. This implies that a new description should
require $p_{10}>0$. Fortunately this is precisely what happens: in
$d=4$ we have one new limit, namely the type~IA${}^6$ strings and the
condition (\ref{karkulka}) for $g_{IA^6}<1$ gives
\eqn{myslivec}{4p_0-2\sum_{i=1}^6 p_i =-2p_{10}<0} or $p_{10}>0$.

In the case of $d=3$ we find also a fundamental domain that is copied
several times by S-dualities. This fundamental region is again bounded
by the condition $g_{gauge}^{eff.4-dim}<1$ which is the same like
$g_{IA^6}<1$ and the internal structure has been partly described: the
fundamental region is divided into several subregions HO, type~I,
type~IA${}^k$, M/K3, IIA/K3, IIB/K3. As we have said, we do not
understand the limits with a K3 geometry well enough to separate the
fundamental region into the subregions enumerated above. We are not
even sure whether those limits are sufficient to cover the whole
parameter space. In the case of $E_8\times E_8$ theory, we are pretty
sure that there are some limits that we do not understand already for
$d=9$ and similar claim can be true in the case of the ${\rm SO}(32)$
vacua for $d<7$.  We understand much better how the entire parameter
space can be divided into the copies of the fundamental region and we
want to concentrate on this question.

The inequality $g_{gauge}^{eff.4-dim}<1$ should hold independently of
which of the six radii are chosen to be the radii of the six-torus. In
other words, it must hold for the smallest radii and the condition is
again~(\ref{myslivec}) which can be for $d=3$ reexpressed as
$p_{7}<p_{10}$.

So the ``last'' limit at the boundary of the fundamental region is
again type~IA${}^6$ and not type~IA${}^7$, for instance. It is easy to
show that the condition $g_{IA^6}<1$ is implied by any of the
conditions for the other limits so this condition is the weakest of
all: all the regions are inside $g_{IA^6}<1$.

This should not be surprising, since
$g_{gauge}^{eff.4-dim}=(g_{IA^6})^{1/2} =g_{IA^6}^{open}$; the
heterotic S-duality in this type~IA${}^6$ limit can be identified with
the S-duality of the effective low-energy description of the D3-branes
of the type~IA${}^6$ theory.  As we have already said, this inequality
reads for $d=3$ 
\eqn{cipisek}{2p_0-\sum_{i=1}^6 p_i =-p_{10}+p_7<0} 
or $p_{10}>p_7$. We know that precisely in $d=3$ the S-duality (more
precisely the $ST^6$ transformation) acts as the permutation of $p_7$
and $p_{10}$. Therefore it is not hard to see what to do if we want to
reach the fundamental domain: we change all signs to pluses by
T-dualities and sort all {\it eight} numbers $p_1,\dots, p_7; p_{10}$
in the ascending order.  The inequality~(\ref{cipisek}) will be then
satisfied. The condition $g_{gauge}^{eff.4-dim}<1$ or~(\ref{myslivec})
will define the fundamental region also for the case of one or two
dimensions.

\subsection{The infinite groups in $d\leq 2$}

In the dimensions $d>2$ the bilinear form is positive definite and the
group of dualities conserves the lattice $\IZ^{11-d}$ in the
$p$-space.  Therefore the groups are finite. However for $d=2$ (and
\emph{a fortiori} for $d=1$ because the $d=2$ group is isomorphic to a
subgroup of the $d=1$ group) the group becomes infinite. In this
dimension $p_{10}$ is unchanged by T-dualities and S-dualities. The
regions with $p_{10}\leq 0$ again correspond to mysterious regions
where the holographic principle appears to be violated, as
in~\cite{BFM}. Thus we may assume that $p_{10}=1$; the overall
normalization does not matter.

Start for instance with $p_{10}=1$ and \eqn{rumcajs}{(p_1,p_2,\dots,
p_8)=(0,0,0,0,0,0,0,0)} and perform the S-duality ($ST^6$ from the
formula~(\ref{indtr})) with $p_7$ and $p_8$ understood as the large
dimensions (and $p_1,\dots, p_6$ as the 6-torus). This transformation
maps $p_7\mapsto p_{10}-p_8$ and $p_8\mapsto p_{10}-p_7$. So if we
repeat $ST^6$ on $p_7,p_8$, T-duality of $p_7,p_8$, $ST^6$, $T^2$ and
so on, $p_{1}\dots, p_6$ will be still zero and the values of
$p_7,p_8$ are
\eqn{rakosnicek}{(p_7,p_8)=(1,1)\to(-1,-1)\to(2,2)\to(-2,-2)\to(3,3)
\to\cdots}
and thus grow linearly to infinity, proving the infinite order of the
group. The equation for $g_{IA^6}<1$ now gives
\eqn{bobek}{2p_0-\sum_{i=1}^6 p_i =-p_{10}+p_7+p_8<0}
or $p_{10}>p_7+p_8$. Now it is clear how to get to such a fundamental
region with~(\ref{bobek}) and $0<p_1<\cdots< p_8$. We repeat the
$ST^6$ transformation with the two largest radii ($p_7,p_8$) as the
large coordinates. After each step we turn the signs to $+$ by
T-dualities and order $p_1<\cdots< p_8$ by permutations of radii.  A
bilinear quantity decreases assuming $p_{10}>0$ and $p_{10}<p_7+p_8$
much like in~\cite{BFM}, the case $k=9$ ($d=2$):
\eqn{pokuston}{C_{d=2}=\sum_{i=1}^8 (p_i)^2\to \sum_{i=1}^8
(p_i)^2+2p_{10}(p_{10}-(p_7+p_8))}
In the same way as in~\cite{BFM}, starting with a rational
approximation of a vector $\vec p$, the quantity $C_{d=2}$ cannot
decrease indefinitely and therefore finally we must get to a point
with $p_{10}>p_7+p_8$.

In the case $d=1$ the bilinear form has a Minkowski signature. The
fundamental region is now limited by 
\vspace*{-1ex}
\eqn{hurvinek}{2p_0-\sum_{i=1}^6 p_i =-p_{10}+p_7+p_8+p_9<0}
and it is easy to see that under the $ST^6$ transformation on radii
$p_1,\dots, p_6$, $p_{10}$ transforms as
\vspace*{-1ex}
\eqn{spejbl}{p_{10}\to 2p_{10}-(p_7+p_8+p_9)\,.}
Since the $ST^6$ transformation is a reflection of a spatial coordinate in
all cases, it keeps us inside the future light cone if we start there.
Furthermore, after each step we make such T-dualities and permutations
to ensure $0<p_1<\cdots< p_9$.

If the initial $p_{10}$ is greater than $[(p_1)^2+ \cdots+
(p_9)^2]^{1/2}$ (and therefore positive), it remains positive and
assuming $p_{10}<p_7+p_8 +p_9$, it decreases according
to~(\ref{spejbl}). But it cannot decrease indefinitely (if we
approximate $p$'s by rational numbers or integers after a scale
transformation). So at some point the assumption $p_{10}<p_7+p_8+p_9$
must break down and we reach the conclusion that fundamental domain is
characterized by $p_{10}>p_7+p_8+p_9$.

\subsection{The lattices}

In the maximally supersymmetric case~\cite{BFM}, we encountered
exceptional algebras and their corresponding lattices. We were able to
see some properties of the Weyl group of the exceptional algebra
$E_{10}$ and define its fundamental domain in the Cartan
subalgebra. In the present case with 16 supersymmetries, the structure
of lattices for $d>2$ is not as rich. The dualities always map integer
vectors $p_i$ onto integer vectors.

For $d>4$, there are no S-dualities and our T-dualities know about the
group $O(26-d,10-d,\IZ)$. For $d=4$ our group contains an extra
$\IZ_2$ factor from the single S-duality. For $d=3$ they unify to a
larger group $O(8,24,\IZ)$. We have seen the semidirect product of
$(\IZ_2)^8$ and $S_8$ related to its Weyl group in our formalism.  For
$d=2$ the equations of motion exhibit a larger affine $\hat{\rm
o}(8,24)$ algebra whose discrete duality group has been studied
in~\cite{sentwod}.

In $d=1$ our bilinear form has Minkowski signature. The S-duality can
be interpreted as a reflection with respect to the vector
\eqn{refsl}{(p_1,p_2,\dots, p_9,p_{10})=(0,0,0,0,0,0,-1,-1,-1,+1).}
This is a spatial vector with length-squared equal to minus two (the
form~(\ref{biltri}) has a time-like signature). As we have seen, such
reflections generate together with T-dualities an infinite group which
is an evidence for an underlying hyperbolic algebra analogous to
$E_{10}$.\pagebreak[3] Indeed, Ganor~\cite{ori} has argued that the
$DE_{18}$ ``hyperbolic'' algebra underlies the nonperturbative duality
group of maximally compactified heterotic string theory.  The Cartan
algebra of this Dynkin diagram unifies the asymptotic directions which
we have studied with compact internal symmetry directions.  Its Cartan
metric has one negative signature direction.

\section{Conclusions}

The parallel structure of the moduli spaces with 32 and 16 SUSYs gives
us reassurance that the features uncovered in~\cite{BFM} are general
properties of M-theory.  It would be interesting to extend these
arguments to moduli spaces with less SUSY.  Unfortunately, we know of
no algebraic characterization of the moduli space of M-theory on a
Calabi Yau threefold.  Furthermore, this moduli space is no longer an
orbifold.  It is stratified, with moduli spaces of different
dimensions connecting to each other via extremal transitions.
Furthermore, in general the metric on moduli space is no longer
protected by nonrenormalization theorems, and we are far from a
characterization of all the extreme regions.  For the case of four
SUSYs the situation is even worse, for most of what we usually think
of as the moduli space actually has a superpotential on it, which
generically is of order the fundamental scale of the
theory.\footnote{Apart from certain extreme regions, where the
superpotential asymptotes to zero, the only known loci on which it
vanishes are rather low dimensional subspaces of the classical moduli
space,~\cite{bdw}.}

There are thus many hurdles to be jumped before we can claim that the
concepts discussed here and in~\cite{BFM} have a practical application
to realistic cosmologies.

%% file: references.tex

%% file: vita.tex
\begin{vita}  
\heading{Lubo\v{s} Motl}
\vspace{15pt}


\begin{descriptionlist}{1994 July 29} 

\item[1995] Milo\v{s} Zahradn\'\i{}k, Lubo\v{s} Motl,
{\it P\v{e}stujeme line\'arn\'\i{} algebru} (We Grow Linear Algebra),
a textbook, Charles University, Prague, Czech Republic.

\item[1995] L.\,Motl, {\it Two parametric zeta function regularization in
superstring theory,} \hepth{9510105}.

\item[1996] L.\,Motl, {\it Quaternions and M(atrix) theory in spaces with
boundaries,} 
\newline\hepth{9612198}.

\item[1997] L.\,Motl, {\it Proposals on nonperturbative superstring
interactions,} 
\newline\hepth{9701025}.

\item[1997]     Mgr. (equivalent of MSc.) in Physics from the Charles
University, Prague, Czech Republic.

\item[1997-2001]        Graduate work in Physics and Astronomy, Rutgers,
The State University of New Jersey, New Brunswick, New Jersey.

\item[1997] T.\,Banks, L.\,Motl, {\it Heterotic strings from matrices,}
\jhep{9712}{1997}{004} [\hepth{9703218}].

\item[1997] L.\,Motl, L.\,Susskind, {\it Finite $N$ heterotic matrix
models and discrete light cone quantization,} \hepth{9708083}.

\item[1998] O.\,Ganor, L.\,Motl, {\it Equations of the (2,0) theory and
knitted fivebranes,} \jhep{9805}{1998}{009} [\hepth{9803108}].

\item[1998] T.\,Banks, W.\,Fischler, L.\,Motl, 
{\it Dualities versus singularities,}
\jhep{9901}{1999}{019} [\hepth{9811194}].

\item[1999] L.\,Motl, T.\,Banks,
{\it On the hyperbolic structure of moduli space with 16 SUSYs,}
\jhep{9905}{1999}{015} [\hepth{9904008}].

\item[1999] T.\,Banks, L.\,Motl,
{\it A nonsupersymmetric matrix orbifold,}
\jhep{0003}{2000}{027} [\hepth{9910164}].

\item[2000] T.\,Banks, M.\,Dine, L.\,Motl,
{\it On anthropic solutions to the cosmological constant problem,}
\hepth{0007206}.

\item[2001]	Ph.D. in Physics and Astronomy.

\end{descriptionlist} 
\end{vita}

%% file: lumo-phd.bbl
\begin{thebibliography}{99}        

\bibitem{gswitten} M.B.\,Green, J.H.\,Schwarz, E.\,Witten,
{\it Superstring theory,} two volumes, Cambridge University Press 1987.

\bibitem{GANORMOTL} O.\,Ganor, L.\,Motl,
{\it Equations of the (2,0) theory and knitted fivebranes,}
\jhep{9805}{1998}{009} [\hepth{9803108}].
  
\bibitem{somecom} E.\,Witten, {\it Some comments on string dynamics},
\hepth{9507121},
    published in {\em Future perspectives in string theory}, Los
Angeles 1995, p. 501-523.
\bibitem{openp} A.\,Strominger,  {\it Open P-branes},
\plb{383}{1996}{44} [\hepth{9512059}].
\bibitem{rozali} M.\,Rozali, {\it Matrix theory and U-duality
in seven dimensions}
  \plb{400}{1997}{260}, \hepth{9702136}.
\bibitem{berozaliseiberg} M.\,Berkooz, M.\,Rozali, N.\,Seiberg,
  {\it On transverse fivebranes in M(atrix) theory on $T^5$},
\plb{408}{1997}{105}
[\hepth{9704089}].
\bibitem{berozali} M.\,Berkooz, M.\,Rozali,
  {\it String dualities from matrix theory}, \npb{516}{1998}{229}
[\hepth{9705175}].
\bibitem{dkps} M.R.\,Douglas, D.\,Kabat, P.\,Pouliot,
S.\,Shenker,
{\it D-branes and short distances in string theory,}
\npb{485}{1997}{85-127} [\hepth{9608024}].
\bibitem{bfss} T.\,Banks, W.\,Fischler,\,S.H.\,Shenker, L.\,Susskind, 
  {\it M theory as a matrix model: a conjecture}, \prd{55}{1997}{5112}
[\hepth{9610043}].
\bibitem{taylorreview} W.\,Taylor,
{\it M(atrix) Theory: Matrix Quantum Mechanics as a Fundamental Theory,}
\hepth{0101126}.
\bibitem{prem} O.\,Aharony, M.\,Berkooz,
               S.\,Kachru, N.\,Seiberg, E.\,Silverstein,
  {\it Matrix description of interacting theories in six dimensions},
{\it Adv. Theor. Math. Phys.} {\bf 1} (1998) 148 [\hepth{9707079}].
\bibitem{WitQHB} E.\,Witten,
  {\it On the conformal field theory of the Higgs branch},
\jhep{07}{1997}{003} [\hepth{9707093}].
\bibitem{Lowe} D.\,Lowe,
  {\it $E_8 \times E_8$ small instantons in matrix theory},
\hepth{9709015}.
\bibitem{KSAB} O.\,Aharony, M.\,Berkooz, S.\,Kachru, E.\,Silverstein,
  {\it Matrix description of (1,0) theories in six dimensions},
\plb{420}{1998}{55} [\hepth{9709118}]. 
\bibitem{ABS} O.\,Aharony, M.\,Berkooz, N.\,Seiberg,
  {\it Light-cone description of (2,0) superconformal theories in six
dimensions},
 \hepth{9712117}.
\bibitem{APS} M.\,Aganagic, C.\,Popescu, and J.H.\,Schwarz,
  {\it D-brane actions with local 
kappa symmetry}, \plb{393}{1997}{311} [\hepth{9610249}].
\bibitem{APPS} M.\,Aganagic, J.\,Park, C.\,Popescu, and J.H.\,Schwarz,
  {\it World-volume action of the M theory five-brane},
 \npb{496}{1997}{191} [\hepth{9701166}].
\bibitem{VWSCT} C.\,Vafa and E.\,Witten,
  {\it A strong coupling test of S-duality}, \npb{431}{1994}{3}
[\hepth{9408074}].
\bibitem{DLM} M.J.\,Duff, J.T.\,Liu and R.\,Minasian,
  {\it Eleven dimensional origin of string/string duality: a one loop
  test}, \npb{452}{1995}{261} [\hepth{9506126}].
\bibitem{WTHR} A.\,Hanany and E.\,Witten,
  {\it Type II B superstrings, BPS monopoles, and three-dimensional gauge
  dynamics}, \npb{492}{1997}{152} [\hepth{9611230}].

\bibitem{Lambert}
  P.S.\,Howe, E.\,Sezgin and P.C.\,West,
  {\it Covariant field equations of the M-theory five-brane}
  \plb{399}{97}{49} [\hepth{9702008}];\\
  P.S.\,Howe, E.\,Lambert and P.C.\,West,
  {\it The selfdual string soliton}, \npb{515}{1998}{203}
[\hepth{9709014}];\\
  E.\,Lambert and P.C.\,West,
  {\it Gauge fields and M five-brane dynamics}, \hepth{9712040}.

\bibitem{SeiSXN} N.\,Seiberg,
  {\it Notes on theories with 16 supercharges}, \hepth{9705117}.
\bibitem{LGRU} U.\,Lindstrom, F.\,Gonzalez-Rey, M.\,Ro{\hacek c}ek,
R.\,von Unge
  {\it On $N=2$ low-energy effective actions}, \plb{388}{1996}{581}
[\hepth{9607089}].
\bibitem{SWSIXD} N.\,Seiberg and E.\,Witten,
  {\it Comments on string dynamics in six-dimensions},
\npb{471}{1996}{121} [\hepth{9603003}].
\bibitem{DS} M.\,Dine and N.\,Seiberg,
  {\it Comments on higher derivative operators in some SUSY field
theories},
  \plb{409}{1997}{239} [\hepth{9705057}].
               
\bibitem{CRRU} G.\,Chalmers, M.\,Ro{\hacek c}ek, and R.\,von Unge, 
  {\it Quantum corrections to monopoles}, \hepth{9801043}.
  
\bibitem{PSS} S.\,Paban, S.\,Sethi and M.\,Stern,
{\it Constraints from extended supersymmetry in quantum mechanics,}
\npb{534}{1998}{137-154} [\hepth{9805018}].
\bibitem{Juan} J.\,Maldacena,
  {\it The large N limit of superconformal field theories and
supergravity}, \hepth{9711200}.

\bibitem{WitFBE} E.\,Witten,
  {\it Five-brane effective action in M-theory}
  \jgp{22}{1997}{103} [\hepth{9610234}].


\bibitem{whycorrect} N.\,Seiberg,
  {\it Why is the matrix model correct?}, \prl{79}{1997}{3577}
[\hepth{9710009}].
\bibitem{taylor}
I.W.~Taylor, {\it D-brane field theory on compact spaces},
\plb{394}{1997}{283} [\hepth{9611042}]. 
\bibitem{seibwit}
N.~Seiberg and E.~Witten, {\it Spin structures in string theory},
\npb{276}{1986}{272}. 
\bibitem{dixonhar}
L.J. Dixon and J.A. Harvey, {\it String theories in ten-dimensions
without space-time supersymmetry}, \npb{274}{1986}{93}. 
\bibitem{sesva}
J.~Scherk and J.H. Schwarz, {\it Spontaneous breaking of
  supersymmetry through dimensional reduction}, \plb{82}{1979}{60}.
\bibitem{rohm}
R.~Rohm, {\it Spontaneous supersymmetry breaking in supersymmetric string
theories}, \npb{237}{1984}{553}.
\bibitem{cdl}
S.~Coleman and F.D. Luccia, {\it Gravitational effects on and of vacuum
decay},  \prd{21}{1980}{3305}.
\bibitem{ikkt} N.\,Ishibashi, H.\,Kawai, Y.\,Kitazawa,
A.\,Tsuchiya,
{\it A large $N$ reduced model as superstring,}
\npb{498}{1997}{467-491} [\hepth{9612115}].
\bibitem{pericross}
V.\,Periwal, {\it Antibranes and the crossing symmetry,}
\hepth{9612215}.
\bibitem{quiver}
M.R. Douglas and G.~Moore, {\it D-branes, quivers and ALE instantons},
\hepth{9603167}.
\bibitem{bg}
O.~Bergman and M.R. Gaberdiel, {\it Dualities of type-0 strings},
\jhep{07}{1999}{022} [\hepth{9906055}].
\bibitem{imamura}
Y.~Imamura, {\it Branes in type-0/type II duality},
\ptp{102}{1999}{859} [\hepth{9906090}].

\bibitem{oritensionless} O.\,Ganor,
  {\it Compactification of tensionless string theories}, \hepth{9607092}.
\bibitem{dhn} B.\,de Wit, J.\, Hoppe, H.\,Nicolai,
  {\it On the quantum mechanics of supermembranes}, \npb{305}{1988}{545}.
\bibitem{towns} P.K.\,Townsend, {\it D-branes from M-branes},
  \plb{373}{1996}{68} [\hepth{9512062}].
\bibitem{newcon} L.\,Susskind,
  {\it Another conjecture about M(atrix) theory}, \hepth{9704080}.  
\bibitem{cds} A.\,Connes, M.R.\,Douglas, A.\,Schwarz,
  {\it Noncommutative geometry and matrix theory: compactification on
tori}, \hepth{9711162}.
\bibitem{sussmotl} L.\,Motl, L.\,Susskind,
  {\it Finite N heterotic matrix models and discrete light
  cone quantization}, \hepth{9708083}.
\bibitem{howitten} P.\,Ho\hacek rava, E.\,Witten,
  {\it Heterotic and type I string dynamics from eleven dimensions}
  \npb{460}{1996}{506} [\hepth{9510209}].
  
\bibitem{orbi} U.\,Danielsson, G.\,Ferretti,
  {\it The heterotic life of the D-particle}, \ijmpa{12}{1997}{4581}
[\hepth{9610082}];\\
  N.\,Kim, S.-J.\,Rey,
  {\it M(atrix) theory on an orbifold and twisted membrane},
\npb{504}{1997}{189} [\hepth{9701139}];\\
  D.\,Lowe,
  {\it Bound states of type I' D-particles and enhanced gauge
  symmetry}, \npb{01}{1997}{134} [\hepth{9702006}];\\
  P.\,Ho\hacek rava, {\it Matrix theory and heterotic strings on tori},
  \npb{505}{1997}{84} [\hepth{9705055}].
\bibitem{bss}
T.\,Banks, N.\,Seiberg, E.\,Silverstein,
{\it Zero and one-dimensional probes with N=8 supersymmetry},
\plb{401}{1997}{30} [\hepth{9703052}].
\bibitem{evashamit}
S.\,Kachru, E.\,Silverstein,
  {\it On gauge bosons in the matrix model approach to
  M theory}, \plb{396}{1997}{70} [\hepth{9612162}].
\bibitem{lmztwo}
L.~Motl, {\it Quaternions and M(atrix) theory in spaces with boundaries},
\hepth{9612198}.
\bibitem{bamo}
T.~Banks and L.~Motl, {\it Heterotic strings from matrices},
\jhep{12}{1997}{004} [\hepth{9703218}].

\bibitem{jopo} J.\,Polchinski, E.\,Witten,
  {\it Evidence for heterotic -- type I string duality},
\npb{460}{1996}{525}
[\hepth{9510169}].
\bibitem{BPS} J.\,Maldacena, L.\,Susskind,
  {\it D-branes and fat black holes}, \npb{475}{1996}{679}
[\hepth{9604042}];\\
  S.\,Das, S.\,Mathur,  {\it Comparing decay rates for black holes and
D-branes},
\npb{478}{1996}{561} [\hepth{9606185}];\\
  G.\,Moore, R.\,Dijkgraaf, E.\,Verlinde, H.\,Verlinde, \quad
  {\it Elliptic Genera of Symmetric Products and Second Quantized
  Strings}, \cmp{185}{1997}{197} [\hepth{9608096}].
\bibitem{motlb} L.\,Motl,
  {\it Proposals on nonperturbative superstring interactions},
\hepth{9701025}.
\bibitem{bs} T.\,Banks, N.\,Seiberg,
  {\it Strings from matrices}, \npb{497}{1997}{41} [\hepth{9702187}].
\bibitem{dvv} R.\,Dijkgraaf, E.\,Verlinde, H.\,Verlinde,
  {\it Matrix string theory}, \npb{500}{1997}{43} [\hepth{9703030}].
\bibitem{jesus}
J.P. Penalba, {\it A matrix model for type-0 strings},
\jhep{08}{1999}{005} [\hepth{9907083}]. 
\bibitem{lust}
R.~Blumenhagen, C.~Kounnas and D.~Lust, {\it Continuous gauge and
supersymmetry breaking for open strings on D-branes},
\jhep{01}{2000}{036} [\hepth{9910094}].
  
\bibitem{GGV}{M.B. Green, M. Gutperle, P. Vanhove,
  {\it One loop in eleven dimensions}, \plb{409}{1997}{177}
[\hepth{9706175}].}
  
\bibitem{GG}{M.B. Green and M. Gutperle,
  {\it $\lambda^{16}$ and related terms in M-theory on $T^2$},
  \plb{421}{1998}{149} [\hepth{9710151}].}
  
\bibitem{SeiWHY}{N. Seiberg,
  {\it Why is the matrix model correct?}, \prl{79}{1997}{3577}
[\hepth{9710009}].}
  
\bibitem{SenWHY}{A. Sen,
  {\it D0 branes on $T^n$ and matrix theory}, \hepth{9709220}.}

\bibitem{SeiSO}{N. Seiberg,
  {\it Matrix description of M-theory on $T^5$ and $T^5/Z_2$},
 \plb{408}{1997}{98} [\hepth{9705221}].}
  
\bibitem{BK}{I. Brunner and A. Karch,
  {\it Matrix description of M-theory on $T^6$}, \plb{416}{1998}{67}
[\hepth{9707259}].}

\bibitem{LMS}{A. Losev, G. Moore and S.L. Shatashvili,
  {\it M\&m's}, \hepth{9707250}.}

\bibitem{HL}{A. Hanany and G. Lifschytz,
  {\it M(atrix) theory on $T^6$ and a m(atrix) theory description of KK
  monopoles}, \hepth{9708037}.}

\bibitem{PolPou}{J. Polchinski and P. Pouliot,
  {\it Membrane scattering with M-momentum transfer}, \prd{56}{1997}{6601}
[\hepth{9704029}].}
  
\bibitem{BFSeS}{T. Banks, W.Fischler, N. Seiberg and L. Susskind,
  {\it Instantons, scale invariance and Lorentz invariance
  in matrix theory}, \plb{408}{1997}{111} [\hepth{9705190}].}
  
\bibitem{thomp}{C. Boulahouache, G. Thompson,
  {\it One loop effects in various dimensions and D-branes},
\hepth{9801083}.}












\bibitem{BFM} T.\,Banks, W.\,Fischler, L.\,Motl,
{\it Dualities versus Singularities,}
\jhep{9905}{1999}{015} [\hepth{9811194}].

\bibitem{cosmoa} G.\,Veneziano,
{\it Scale Factor Duality For Classical And Quantum Strings,}
\plb{265}{1991}{287-294}.
\bibitem{cosmovf} A.A.\,Tseytlin, C.\,Vafa,
{\it Elements of String Cosmology,} \npb{372}{1992}{443-466},
\hepth{9109048}.
\bibitem{cosmob} M.\,Gasperini, G.\,Veneziano,
{\it Pre -- Big Bang In String Cosmology,}
{\it Astropart. Phys. }{\bf 1} (1993) 317-339.
\bibitem{cosmobtwo} A.\,Buonanno, T.\,Damour, G.\,Veneziano,
{\it Prebig Bang Bubbles From The Gravitational Instability Of Generic
String Vacua,} \hepth{9806230} and references therein;
see also {\tt http://www.to.infn.it/{}$\widetilde{\,\,\,}$gasperin/}.
\bibitem{cosmoc} P.\,Binetruy, M.K.\,Gaillard,
{\it Candidates For The Inflaton Field in Superstring Models,}
\prd{34}{1986}{3069-3083}.
\bibitem{cosmomoore} G.\,Moore, J.H.\,Horne,
{\it Chaotic Coupling Constants,}
\npb{432}{1994}{109-126}, \hepth{9403058}.
\bibitem{cosmod} T.\,Banks, M.\,Berkooz, S.H.\,Shenker,
G.\,Moore, P.J.\,Steinhardt,
{\it Modular Cosmology,}, \prd{52}{1995}{3548-3562}, \hepth{9503114}.
\bibitem{cosmoe} T.\,Banks, M.\,Berkooz, P.J.\,Steinhardt,
{\it The Cosmological Moduli Problem, Supersymmetry Breaking, And
Stability
In Postinflationary Cosmology,} \prd{52}{1995}{705-716},
\hepth{9501053}.
\bibitem{cosmofold} A.\,Lukas, B.A.\,Ovrut, {\it U Duality Covariant
M~Theory Cosmology,} \plb{437}{1998}{291-302}, \hepth{9709030},   
and references therein.
\bibitem{cosmof} S.-J.\,Rey, D.\,Bak,
{\it Holographic Principle and String Cosmology,}
\hepth{9811008}.
\bibitem{cosmog} O.\,Bertolami, R.\,Schiappa,
{\it Modular Quantum Cosmology,} \grqc{9810013}.
%
\bibitem{witten} E.\,Witten, {\it String Theory Dynamics In Various
                 Dimensions,} \npb{443}{1995}{85-126}, \hepth{9503124}.

\bibitem{duffrevo} M.J.\,Duff,
{\it Strong\,{}/{}\,Weak Coupling Duality from the Dual String,}
\npb{442}{1995}{47-63} [\hepth{9501030}].
\bibitem{hulltownsend} C.M.\,Hull, P.K.\,Townsend,
{\it Unity of Superstring Dualities,}
\npb{438}{1995}{109-137} [\hepth{9410167}].
\bibitem{lenwilly} W.\,Fischler, L.\,Susskind,
                 {\it Holography and Cosmology,} \hepth{9806039}.
\bibitem{bfstbruba} T.\,Banks, W.\,Fischler, L.\,Susskind,
  {\it Quantum Cosmology In $(2+1)$-Dimensions And $(3+1)$-Dimensions,}  
  \npb{262}{1985}{159}.
\bibitem{bfstbrubb} T.\,Banks, {\it TCP, Quantum Gravity, The Cosmological
Constant And All That\dots,} \npb{249}{1985}{332}.
\bibitem{bfstbrubc} V.A.\,Rubakov,
{\it Quantum Mechanics In The Tunneling Universe,}
\plb{148}{1984}{280-286}.
\bibitem{riotto} M.\,Maggiore, A.\,Riotto, {\it D-Branes and Cosmology,}
\hepth{9811089}.
\bibitem{rama} S.K.\,Rama, {\it Can String Theory Avoid Cosmological
Singularities?} \plb{408}{1997}{91-97}, \hepth{9701154}.
\bibitem{ofer} O.\,Aharony,
{\it String Theory Dualities from M-Theory,}
\npb{476}{1996}{470-483}, \hepth{9604103}.
\bibitem{elitzur} S.\,Elitzur, A.\,Giveon, D.\,Kutasov, E.\,Rabinovici,
{\it Algebraic Aspects of Matrix Theory on $T^d$,}
\npb{509}{1998}{122-144}, \hepth{9707217}.
\bibitem{pioline} N.A.\,Obers, B.\,Pioline,
{\it U-Duality and M-Theory,}
invited review for {\it Phys. Rept,}\newline
\hepth{9809039}.
\bibitem{islands} A.\,Dabholkar, J.A.\,Harvey,
{\it String Islands,} \hepth{9809122}.
\bibitem{evadine} M.\,Dine, E.\,Silverstein,
{\it New M-theory Backgrounds with Frozen Moduli,}\newline
\hepth{9712166}.
\bibitem{horwita} P.\,Ho\hacek rava, E.\,Witten,
{\it Heterotic and Type I String Dynamics from Eleven Dimensions,}
\npb{460}{1996}{506-524}, \hepth{9510209}.
\bibitem{horwitb} E.\,Witten, {\it Strong Coupling Expansion Of Calabi-Yau
Compactification,} \npb{471}{1996}{135-158}, \hepth{9602070}.
\bibitem{horwitc} T.\,Banks, M.\,Dine,
{\it Couplings And Scales In Strongly Coupled Heterotic String Theory,}
\npb{479}{1996}{173-196}, \hepth{9605136}.
\bibitem{horwitd} N.\,Arkani-Hamed, S.\,Dimopoulos, G.\,Dvali,
{\it The Hierarchy Problem and New Dimensions at a Millimeter,}
\plb{429}{1998}{263-272}, \hepph{9803315}. 






\bibitem{BANXMOTL} L.\,Motl, T.\,Banks,
{\it On the hyperbolic structure of moduli space with 16 SUSYs,}
\jhep{9905}{1999}{015} [\hepth{9904008}].

\bibitem{senthreed} A.~Sen,
{\it Strong-weak coupling duality in three-dimensional string theory,}
\npb{434}{1995}{179} [\hepth{9408083}].

\bibitem{sentwod} A.~Sen,  
{\it Duality symmetry group of two-dimensional heterotic string theory,}
\npb{447}{1995}{62} [\hepth{9503057}].

\bibitem{schwtwod} J.H.~Schwarz,
{\it Classical duality symmetries in two dimensions,}
\hepth{9505170}.

\bibitem{ori} O.~Ganor,
{\it Two conjectures on gauge theories, gravity, and infinite dimensional
Kac-Moody groups,} \hepth{9903110}.

\bibitem{bdw} T.~Banks and M.~Dine,   
{\it Quantum moduli spaces of $N=1$ string theories,}
\prd{53}{1996}{5790} [\hepth{9508071}].
\bibitem{tbcosmo} T.\,Banks,
{\it SUSY breaking, cosmology, vacuum selection and the cosmological
constant in string theory,} \hepth{9601151}.


  
\end{thebibliography}
